\documentclass[twocolumn,showpacs,aps,amsmath,amssymb,prx,superscriptaddress,nofootinbib,10pt]{revtex4}

\usepackage{standalone}
\usepackage[utf8]{inputenc}
\usepackage{amsmath}
\usepackage{soul}
\usepackage{cancel}
\usepackage{graphicx}
\usepackage{color}
\usepackage{comment}
\usepackage{amssymb}
\usepackage{amsmath}
\usepackage{amsthm}
\usepackage{commath}
\usepackage{float}
\usepackage{graphicx}
\usepackage{mathtools}
\usepackage{hyperref}
\usepackage{xcolor}
\usepackage{xargs}        
\usepackage{cancel}
\usepackage{arydshln}

\usepackage{subfigure}
\usepackage{scalerel}
\usepackage{fixltx2e} 
\usepackage{url} 

\newcommand{\Ron}{R_{\text{on}}}                     
\newcommand{\Roff}{R_{\text{off}}}                   

\begin{document}
\title{Phases of memristive circuits via an interacting disorder approach}

\author{F. Caravelli}
\author{F. C. Sheldon}
\affiliation{Theoretical Division (T4) and Center for Nonlinear Studies,\\
Los Alamos National Laboratory, Los Alamos, New Mexico 87545, USA}

\begin{abstract}
    We study the phase diagram of memristive circuit models in the replica-symmetric case using a novel Lyapunov function for the dynamics of these devices. Effectively, the model we propose is an Ising model with \textit{interacting} quenched disorder, which we study at the first order in a control parameter.  Notwithstanding these limitations, we find a complex phase diagram and a glass-ferromagnetic transition in the parameter space which generalizes earlier mean field theory results for a simpler model. Our results suggest a non-trivial landscape of asymptotic states for memristive circuits.
\end{abstract}
\maketitle

\section{Introduction}

There are several intriguing features, still unexplored, in the 
dynamical analysis of circuits with memory, and in particular memristors \cite{reviewCarCar,Rev1,Rev2}. 
An essential property of a memristor is its (pinched  at the origin) hysteretic behavior in the voltage-current diagram and the non-linearity of the component. Physical memristors \cite{chua71,chua76,stru8,Valov,stru13} have rather non-trivial voltage-current curves, but some core features are captured by a simple description which we adopt in this paper. The state of the resistance between two limiting values can be parametrized by a parameter $w$, which is constrained between 0 and 1 and can be thought as the first order term in a polynomial expansion for the resistance in terms of a an adimensional parameter. We will refer to $w$ as the \emph{internal memory parameter}. For the case of titanium dioxide devices, a rather simple toy model for the evolution of the resistance is the following:
$$ R(w)=\Ron (1-w) +w \Roff\equiv \Ron(1+\xi w),$$
\begin{equation}
     \frac{\dif}{\dif t} w(t)=\alpha w- \frac{\Ron}{\beta} i(t),
     \label{eq:memr1}
\end{equation}
initially studied for $\alpha=0$, and where $0\leq w\leq 1$, $\xi=\frac{\Roff-\Ron}{\Ron}$; in the equation above $i(t)$ is the current flowing in the device at time $t$. Physically, $w$ can be interpreted as the level of internal doping of the device, but this is a crude description. The constants $\alpha,\beta$ and $\xi$ control the decay and reinforcement time scales and the nonlinearity in the equation respectively, and can be measured experimentally. While $\xi$ is adimensional and depends only on the resistance boundaries, $\alpha$ has the dimension of an inverse time, while $\beta$ a time by a voltage. Aside from applications to memory devices, there is interest in these components also because memristors can serve for  neuromorphic computing devices \cite{indiveri}. 

The purpose of this paper is further understanding the asymptotic dynamics of a circuit of memristors, and specifically the statistics of the resistive states. While circuit laws impose correlations beween the currents across the circuits, for the case of memristors of the type above the Kirchhoff constraints can be solved exactly. In fact it possible to derive a set of coupled differential equation which incorporate the Kirchhoff laws, derived in \cite{Caravelli2016rl} and whose solution describes the statistics of asymptotic states and which we analyze in the following.   The equilibrium currents of a resistive circuit can be written in the vectorial form \cite{Zegarac} as 
    \begin{equation}
        \vec i(t)=- \Omega_{R/A} R^{-1} \vec S(t),
    \end{equation}
where $\Omega_{R/A}=A^t (A R A^t) A R$ is a non-orthogonal projector on the cycle space of the graph.
    
For a memristor (and in particular a circuit) the resulting dynamical equation(s) which we discuss below is (are) nonlinear and hard to solve analytically. An analysis for the asymptotic states can however be done via Lyapunov functions. After a first attempt at deriving a Lyapunov function in \cite{Caravelli2019Ent}, plagued by constraints on the external fields, here we provide a novel yet similar Lyapunov function free of these requirements.

For this purpose, we first focus on the derivation of the Lyapunov function and a correspondence between the Ising model and memristors. With such mapping at hand, the next step is understanding the nature of the asymptotic states using the Lyapunov function. One way of tackling this problem has been first proposed in \cite{mean-field} (and then in \cite{Caravelli2019Ent}). In that case a toy model of memristor dynamics corresponds a Lyapunov function equivalent to a Curie-Wei\ss\ model, the statistics of the asymptotic states have been understood via standard statistical mechanics methods. In particular, it has been noted in \cite{mean-field} that via the evaluation of the partition function and the mean magnetization of an equivalent Ising model, the asymptotic states $w=\sum_i w_i(\infty)/N$ could be predicted with a fair precision statistically, given random initial conditions.

Here we follow the same prescription, but for the case in which the interaction matrix is a projector operator on the cycle basis of the circuit.
For random circuits, and since the Lyapunov is being minimized, it make sense to study the problem via a quenched average over the projection matrix $\Omega$, and analyze the resulting phases.  This is because $\frac{dL}{dt}=\vec \nabla_w L \cdot \frac{d}{dt} \vec w=0$ implies $\frac{d\vec w}{dt}=0$, and thus asymptotic states are characterized by the local minima of the system.

The standard procedure commonly used in the study of disordered systems of this type is to analyze the phase diagram of the system and understand if a glass phase is present. For this purpose we perform the average using the replica trick in order to calculate the free energy,
    $f= \langle \log Z \rangle_{P(\Omega)}.$
Clearly, because one has the constraint $\Omega^2=\Omega$, the distribution $P(\Omega)$ is non-trivial. In this paper we introduce the constraint as interaction over i.i.d. Gau\ss ian variables, whose parameters can be fixed by the condition above. 
Our attempt is motivated by the fact that it was noticed numerically in \cite{Caravelli2019Ent} that the distribution of elements $P(\Omega_{ij})$ can be approximated by a short-tailed distribution.

Using the approach described above, we study the phases of the model as a function of the mean field adimensional parameters $\phi=\frac{s}{\alpha \beta}$ and $\xi$, and predict a ferromagnetic-spin glass transition which is robust with respect to the introduction of the interactions in the disorder model.

\section{The memristive circuit-spin model correspondence}
\subsection{Circuit dynamics}

    The extension of eqn. (\ref{eq:memr1}) to a circuit can be done, and is given by \cite{Caravelli2016rl}
\begin{equation}
\frac{d}{dt}\vec{w}(t)=\alpha\vec{w}(t)-\frac{1}{\beta} \left(I+\xi \Omega W(t)\right)^{-1} \Omega \vec S(t),
\label{eq:diffeq}
\end{equation}
with the constraints $0 \leq w_i\leq 1$. In a circuit, we can think of the variables $w_i$ as living on the edges of a graph,  and  $\Omega_{ij}$ contains the information about the topology of the graph. Each edge of the graph represents a resistive component. 
We note that because $\Omega$ is a projector operator, $\Omega=\Omega^2$ we can always write $\vec S=\Omega \vec S+(I-\Omega) \vec S$, it is easy to see that we can add to $\vec S$ any vector $\tilde S=(I-\Omega) \vec k$, which will not affect the dynamics. This form of freedom arises from the Kirchhoff (current and voltage) constraints from which the differential equation has been derived. 

\subsection{Dynamical aspects and Lyapunov function}
A possible way to understand the dynamics of memristive circuits is via the analysis of Lyapunov functions. 
We consider the following energy function which memristors minimize along their dynamics when controlled in direct current. In terms of the memristors asymptotic variables $w_i=\{1,0\}$, the first proposal for a Lyapunov function for memristor was given in \cite{mean-field,Caravelli2019Ent}, but these are not valid for arbitrary values either of the graph or the external voltage. Here we show that another Lyapunov function can be obtained, which is not plagued by these deficiencies.  We begin with the equations of motion,
\begin{equation}
    (I + \xi \Omega W)\dot{\vec{w}} = \alpha \vec{w} + \alpha \xi \Omega W \vec{w} - \frac{1}{\beta} \vec{x}.
\end{equation}Consider

\begin{equation}
    L = -\frac{\alpha}{3} \vec{w}^T W \vec{w} - \frac{\alpha\xi}{4} \vec{w}^T W\Omega W \vec{w}
    +\frac{1}{2\beta} \vec{w}^T W\vec{x}.
\end{equation}
In this case, we have
\begin{align}
    \frac{dL}{dt} &= \dot{\vec{w}}^T\left(-\alpha W\vec{w} - \alpha\xi W\Omega W\vec{w} + \frac{1}{\beta} W\vec{x} \right)\nonumber \\
    {} &= -\dot{\vec{w}}^T (W + \xi W\Omega W)\dot{\vec{w}}\nonumber\\
    &= -\dot{\vec{w} }^T \sqrt{W}(I + \xi \sqrt{W}\Omega \sqrt{W})\sqrt{W}\dot{\vec{w}} \nonumber\\
    &= -||\sqrt{W}\dot {\vec{w}}||^2_{ (I + \xi \sqrt{W}\Omega \sqrt{W})} 
\end{align}
and we have that $\frac{dL}{dt} \le 0$. 
\begin{equation}
   L = -\frac{\alpha}{3} \vec{w}^T W \vec{w} - \frac{\alpha\xi}{4} \vec{w}^T W\Omega W \vec{w}
    +\frac{1}{2\beta} \vec{w}^T W\vec{x}.
\end{equation}
An asymptotic form can be obtained by replacing $w_i^k=w_i$ for integer $k$, as asymptotically one has $w_i=\{1,0\}$.
Thus, the asymptotic Lyapunov function form is given by
\begin{eqnarray}
    L(\vec w)&=& = -\frac{\alpha}{3} \vec{w}^T  \vec{w} - \frac{\alpha\xi}{4} \vec{w}^T \Omega  \vec{w}
    +\frac{1}{2\beta} \vec{w}^T \vec{x}\nonumber \\ 
    L(\vec \sigma)&=& -\frac{\alpha}{3} \frac{\vec{s}+1}{2}\cdot \frac{\vec{\sigma}+1}{2} - \frac{\alpha\xi}{4} \frac{\vec{\sigma}+1}{2} \Omega  \frac{\vec{\sigma}+1}{2}
    +\frac{1}{2\beta} \frac{\vec{\sigma}+1}{2}\vec{x} \nonumber \\
    &\equiv&  -\frac{\alpha}{6} \vec{\sigma}\cdot \vec 1 - \frac{\alpha\xi}{4} \frac{\vec{\sigma}}{2} \Omega  \frac{\vec{\sigma}}{2}- \frac{\alpha\xi}{4}  \frac{\vec{\sigma}}{2} \Omega  \vec 1
    +\frac{1}{2\beta} \frac{\vec{\sigma}}{2}\vec{x}
\end{eqnarray}
from which we obtain
\begin{equation}
    \tilde L=\frac{8L(\vec \sigma)}{\alpha}=\vec \sigma\cdot (\frac{2\vec{x}}{\alpha \beta} -\frac{4}{3}  \vec 1- \xi   \Omega  \vec 1) - \frac{\xi}{2} \vec{\sigma}\ \tilde \Omega \  \vec{\sigma}
\end{equation}
where $\tilde \Omega$ has only the off-diagonal terms of $\Omega$.
The structure of the Lyapunov function above is very similar to the one described before, but only requires the spectral condition $I+\xi \Omega W\geq 0$, which is naturally satisfied by the network.

For the purpose of the calculation which follows, it is worth writing the Lyapunov function as
\begin{equation}
    \tilde L(s)=- \sum _{i=1}^n \sigma_i h_i-  \sum _{i=1}^n \sum _{j=1}^n \Omega_{ij} \pmb{b}_{ij}(s),
\end{equation}
up to a constant $-N \frac{\xi}{2}$.
We report in bold functions of the state variables $\sigma_{i}$. In the case of the corrected Lyapunov function we have $\pmb{b}_{ij}=\frac{\xi}{2}( \sigma_i \sigma_j-2 \sigma_i)$ and $\vec h=\frac{2}{\alpha \beta} \vec x-\frac{8}{3} \vec 1$. In this latter formulation of the Lyapunov function we  thus have the control over the external field and the absence of constraints.  In what follow we will however perform calculations with $\pmb{b}_{ij}=\frac{\xi}{2} \sigma_i \sigma_j-\sigma_i t_j$ for a generic $t_i$. We can however also define $\pmb{b}_{ij}=\frac{\xi}{2}( \sigma_i \sigma_j- \sigma_i t_j - \sigma_j t_i)$ and $\vec h=\frac{2}{\alpha \beta} \vec x-\frac{4}{3} \vec 1$. \footnote{We note that $\pmb{b}_{ij}$ is not symmetric by construction. However, the tensor will always be contracted with symmetric tensors or summed over both indices, and thus automatically symmetrized in the calculation.  Thus, it can be thought of being symmetrized in what follows, as $\pmb{b}_{ij}=\frac{1}{2}(\xi \sigma_i \sigma_j-\sigma_i t_j-\sigma_j t_i)$ without loss of generality.}  In the latter formulation we have thus the control over the external field and the and the absence of Lyapunov boundary. 

We can gain a little intuition by assuming that $\Omega$ is diagonal. In this case we have
\begin{equation}
    \tilde L=\frac{8L(\vec \sigma)}{\alpha}=\vec \sigma\cdot (\frac{2\vec{x}}{\alpha \beta} -\frac{4}{3}  \vec 1- \xi   \Omega \vec 1) 
\end{equation}
and we see that the asymptotic value of $\sigma_i$ is determined via the strength of $\xi$. 

\section{Constraints on the disorder average}

In the case of the memristor Lyapunov function there are a few additional properties that one needs to consider.
For a comparison, while in the Sherrington-Hamiltonian one has that self coupling terms $\Omega_{ii}s_i^2$ reduce to an additive constant and may be neglected, the binary memristor variables introduce a linear dependence on the diagonal terms $\Omega_{ii}$.
However, since the coupling matrix over which we will perform the average multiplies a term which is of the form $\pmb{b}_{ij}=\xi(\frac{1}{2}\sigma_i \sigma_j - \sigma_i)$, one has to make multiple assumptions about the distribution of the disorder, in particular the properties of the off-diagonal and diagonal elements. These assumptions must be compatible with the underlying assumption of a projector matrix.
In the following, we will call \textit{symmetric disorder} the case in which there is no difference between the fluctuations of the diagonal and off-diagonal elements, not to be confused with the matrix $\Omega_{ij}$ which is always symmetric with respect to transposition. 

\subsection{Gau\ss ian matrix disorder with soft constraints}
\subsubsection{Symmetric disorder}
Let us now discuss the fluctuations of the elements $\Omega_{ij}$. At the simplest level of description, we can assume that 
\begin{equation}
    P_{sym}(\Omega_{ij})\propto \exp\left(\frac{N (\Omega_{ij})^2}{2 J^2}\right),
\end{equation}
meaning that we assume completely symmetric disorder. The distribution of the elements $P(\Omega_{ij})$ was studied numerically in \cite{Caravelli2019Ent} for a variety of random graphs, showing that most of the probability can be roughly approximated by a Gau\ss ian distribution when averaging over various (random) graphs. The evaluation of the disorder average is rather simple in this case (leading to a model similar to the Sherrington-Kirkpatrick model \cite{SherringtonKirkpatrick}), and we will use it as a baseline of our calculations.

 Below we provide a refined distribution which considers the fact that $\Omega_{ij}$ is a projector operator.

\subsubsection{Asymmetric disorder}
The symmetric disorder above does not assume any particular properties for the matrix $\Omega_{ij}$. In order to make a more realistic ansatz about the distribution of the couplings, we assume that our distribution is of the form
\begin{eqnarray}
    P_{as}(\Omega_{ij})&\propto& \exp\left({\frac{(N-1) (\Omega_{ij}-\mu_{off})^2}{2 J^2}(1-\delta_{ij})}\right)\nonumber \\
    &\cdot &\ \ \ \exp\left(\delta_{ij}\frac{ (\Omega_{ii}-\mu)^2}{2 \rho^2}\right).
\end{eqnarray}
The Ansatz above must be made consistent with the fact that $\Omega^2=\Omega$, and thus need to find relationships between $\mu_{off},\mu,J$ and $\rho$.
In order for the system to be reminiscent of a projector operator, we must have that the off-diagonal elements and diagonal elements fluctuate differently, but with a well defined relation. Let us thus assume that for $i\neq j$, $\langle \Omega_{ij}^2\rangle=\frac{J^2}{N-1}$, $\langle \Omega_{ij}\rangle=\mu_{off}$, $\langle \Omega_{ii}^2\rangle=\rho^2$, $\langle \Omega_{ii}\rangle=\mu$. We must have
\begin{equation}
    \Omega_{ii}=\sum_{k} \Omega_{ik}\Omega_{ki} =\sum_{k\neq i} \Omega_{i k}\Omega_{ki}+\Omega_{ii}^2,
\end{equation}
from which
\begin{equation}
    \langle \Omega_{ii}\rangle= \sum_{k\neq i} \langle \Omega_{ik}^2 \rangle +\langle \Omega_{ii}^2\rangle,
    \label{eq:mistake}
\end{equation}
 %
which we can rewrite as
\begin{equation}
    \mu=(N-1) \frac{J^2}{N-1}+\rho^2=J^2+\rho^2
\end{equation}
Similarly, we must have other fluctuations relations between the mean of the off diagonal terms. In fact, for $i\neq j$,
\begin{eqnarray}
    \langle \Omega_{ij} \rangle &=& \langle \sum_{k\neq i\neq j} \Omega_{ik}\Omega_{kj}\rangle+\langle \Omega_{ii} \Omega_{ij}\rangle +\langle \Omega_{ij} \Omega_{jj}\rangle \nonumber \\
    &=&\sum_{k\neq i\neq j} \langle \Omega_{ik}\rangle \langle \Omega_{kj}\rangle+\langle \Omega_{ii}\rangle\langle \Omega_{ij}\rangle +\langle \Omega_{ij}\rangle \langle \Omega_{jj}\rangle\nonumber 
\end{eqnarray}
which, if we replace $\mu_{off}=\langle \Omega_{ij}\rangle$ can be rewritten as
\begin{equation}
    \mu_{off}=\frac{(N-2) \mu_{off}^2}{1-2\mu}.
\end{equation}
It is not hard to see that in the limit $N\rightarrow \infty$, we must have that if $\mu$ is constant (with respect to $N$),
\begin{equation}
    \mu_{off}\rightarrow 0.
\end{equation}
A similar argument applied to $\Omega_ii^2$ gives 
\begin{equation}
\mu=\frac{J^2}{1+J^2}.
\label{eq:scaling1}
\end{equation}
However, we will see that this constraint will come from the enforcement of a projector condition on the measure.

Let us now discuss how to fix $\mu$. Since we aim to use such approach for a projector operator, we must have $\text{Trace}(\Omega)=L$, where $L$ is the dimension of the vector space Span($\Omega$). We stress that $N$ here is not the number of nodes of the graph, but the edges. Thus, we can use
\begin{equation}
    \langle \text{Trace}(\Omega) \rangle = L= N \mu,
\end{equation}
where $L$ is the number of fundamental cycles in the graph, and from which $\mu=\frac{L}{N}$. In the case of the projector operator on the cycle space of a graph, we have that for a circuit $L=N-V+\chi$, where $\chi$ is the Euler characteristic (and is typically finite), and $V$ is the number of vertices in the circuit. For a random circuit, one must have that $N=\frac{p}{2} V (V-1) $, where $p$ is the probability of having a certain memristor (edge in the circuit). It follows that $V=\frac{1}{2}+\sqrt{\frac{1}{4}+\frac{2 N}{p}}\approx  \sqrt{\frac{2}{p}N}$ for large $N$. Thus, 

\begin{equation}
    \mu=\frac{L}{N}\approx 1-\sqrt{\frac{2}{p}} \frac{1}{\sqrt{N}},
\end{equation}
now, if $p$ is constant, in the thermodynamic limit $\mu\approx 1$. On the other hand, if $p=\frac{c}{N}$, where $c$ is a constant, then we have
\begin{equation}
    \mu=1-\sqrt{\frac{2}{c}}
\end{equation}
which is less than one for $c>2$, which is easy to see it must be the case for the graph to be connected. We will use this parametrization for the non-interacting disorder in what follows.
The scaling of the diagonal elements that the matrix $\Omega_{ij}$ is consistent with the observation obtained numerically in earlier works, in which $\Omega_{ii}\approx 1-\frac{1}{\sqrt{N}}$ and $P(\Omega_{ij})\propto e^{-\frac{\Omega_{ij}^2}{2 J^2} N}$. It is easy to see that we can always rewrite
\begin{equation}
    1=(\frac{J}{\sqrt{\mu}})^2+ (\frac{\rho}{\sqrt{\mu}})^2
\label{eq:projectioneq}
\end{equation}
in terms of an angle $\theta$, with $J=\sqrt{\mu} \cos(\theta)$ and $\rho=\sqrt{\mu} \sin(\theta)$. We will see later that we can fix the value of $\theta$ \textit{a posteriori} with assumptions on the scaling of the disorder average.
\subsection{Quenched dynamics approximation}
We wish to first gain some intuition on the effect of the quenched disorder on the dynamics of memristors.
Consider the differential equation for a memristive circuit
\begin{equation}
    \frac{d}{dt} \vec w=\alpha \vec w-\frac{1}{\beta}(I+\xi \Omega W)^{-1} \Omega \vec s
\end{equation}
where $\chi=\frac{R_{off}-R_{on}}{R_{on}}>1$.  Let us consider a first approximation.
If $\langle \Omega_{ij} \rangle=0$ for $i\neq j$, and $\langle \Omega_{ii}\rangle= \mu$.
Let us assume first that we can write at the first approximation
\begin{equation}
    P(\Omega)=\frac{1}{\mathcal N}
    \prod_{i\neq j} e^{-\frac{\Omega_{ij}^2}{2 J^2} } \prod_i e^{-\frac{(\Omega_{ii}-\mu)^2}{2 \rho^2}}
\end{equation}
where $\mathcal N=\frac{1}{2\pi \rho J}$. Consider $\chi\ll 1$, we can write at the first order in $\chi$ the following expansion
\begin{eqnarray}
    \langle (I+\xi \Omega W)^{-1} \Omega\rangle\approx \langle \Omega-\xi \Omega W \Omega\rangle.
\end{eqnarray}
Now, consider
\begin{eqnarray}
\sum_{k} w_k \langle \Omega_{ik}  \Omega_{kj}  \rangle =\sum_k w_k \delta_{ij}\big(\delta_{ik}(\mu^2+\rho^2) +(1-\delta_{ik}) J^2 \big)
\end{eqnarray}
It follows that
\begin{eqnarray}
\frac{d}{dt} \langle w_i\rangle &=& \langle w_i \rangle \big(-\xi (\mu^2+\rho^2)s_i+\xi J^2 s_i+\alpha) \big) \nonumber \\
&-&s_i (\frac{\mu}{\beta} +\xi  J^2 \sum_{i=1}^N \langle w_i\rangle  )
\end{eqnarray}
If $s_i$ is homogeneous, and $J^2$ scale as the inverse of $N$, then
\begin{eqnarray}
\frac{d}{dt} \langle w_i\rangle &=& \langle w_i \rangle \big(-\xi (c^2+\rho^2)s+\xi J^2 \frac{s}{N}+\alpha) \big)\nonumber \\
&-&s (\frac{\mu}{\beta} +\xi  J^2 \langle \langle w_i\rangle \rangle )
\end{eqnarray}
and in the limit $N\rightarrow \infty$, we have
\begin{eqnarray}
\frac{d}{dt} \langle w_i\rangle &=& \langle w_i \rangle \big(-\xi (\mu^2+\rho^2)s+\alpha) \big)\nonumber \\
& &-s (\frac{\mu}{\beta} +\xi  J^2 \langle \langle w_i\rangle \rangle )
\end{eqnarray}
which is an equation of the form
\begin{equation}
    \frac{d}{dt} \langle w_i \rangle=a \langle w_i \rangle+b \langle \langle w\rangle\rangle, 
\end{equation}
where $\langle \rangle$ is the quenched average, and $\langle\langle\rangle\rangle$ the mean field and quenched average. The dynamical equations above were shown to be equivalent to mean field Ising model, and were obtained in \cite{mean-field}. We know that such equation has a transition of the Curie-Wei\ss\ type, e.g. a ferromagnetic-paramagnetic asymptotic dynamics. Here we study a similar system using the Lyapunov function of the full system without a first order approximation, and show that the paramagnetic phase becomes a glass phase in what follows.

\subsection{Projection matrix measure}
At this point we wonder whether the parametrization for a non-interacting disorder is sufficient. The key point is that the matrix $\Omega_{ij}$ cannot be  identically and independently distributed, as we need to take into account the constraints $\Omega_{ij}^2=\Omega_{ij}$. Such constraint is problematic as we will discuss in a second. One possibility is to introduce a series of delta function in the average. We have
\begin{equation}
    dP(\Omega)\propto \prod_{ij} d\Omega_{ij} 
    P(\Omega_{ij})\delta(\sum_k \Omega_{ik}\Omega_{kj}-\Omega_{ij}).
\end{equation}
which can be written, writing $\delta(x)=\frac{1}{2\pi}\int _{\mathbb R}d\lambda e^{i\lambda x}$ as
\begin{eqnarray}
    dP(\Omega)\propto \prod_{ij}  d\Omega_{ij} d\lambda_{ij}& &e^{-\sum_{(ij),(kt)} \Omega_{ij} C_{(ij)(kt)}(\lambda) \Omega_{kt} }\nonumber \\
    &\cdot& \ \ \ \ \ e^{\sum_{ij} \Omega_{ij} P_{ij}(\lambda)}
\end{eqnarray}
where $C_{ijkt}=i \lambda_{it} \delta_{jk}+\delta_{ik}\delta_{jt} \frac{N}{2J^2}$, and $P_{ij}=i \lambda_{kt}$. However, we could not find an exact inverse for the matrix $C$ for arbitrary $\lambda$. Thus in this paper we focus on a simpler problem in which the fact the $\Omega_{ij}$ is a projector is applied to first order in perturbation theory.
We write the delta function as
\begin{equation}
    \delta(x)=\lim_{\gamma \rightarrow \infty} e^{-\frac{\gamma}{2} x^2 }
\end{equation}
as is commonly done for the Landau-Ginzburg problem, and we write

\begin{eqnarray}
    dP(\Omega)&\propto&\prod_{ij, i\neq j }  d\Omega_{ij} e^{-\frac{(N-1) \Omega_{ij}^2}{2 J^2}}
    \prod_{i} d\Omega_{ii} e^{-\frac{(\Omega_{ii}-\mu)^2}{2 \rho^2}}\nonumber \\
&\cdot&  e^{-\frac{\tilde \gamma}{2}\sum_{ij}(\sum_k \Omega_{ik}\Omega_{kj}-\Omega_{ij})^2} \nonumber \\
    &=&\prod_{ij, i\neq j}  d\Omega_{ij} e^{-\Omega_{ij}^2(\frac{N}{2 J^2}+\tilde \gamma)}
    \prod_{i} d\Omega_{ii} e^{-(\Omega_{ii}-\tilde \mu)^2(\frac{1}{2 \rho^2}+\tilde \gamma)}\nonumber \\
    &\cdot& e^{\gamma\sum_{ijt}\Omega_{ij}\Omega_{it}\Omega_{tj}-\frac{\tilde \gamma}{2} \sum_{t,t^\prime} \Omega_{it}\Omega_{tj} \Omega_{it^\prime}\Omega_{t^\prime j}+const} \nonumber 
\end{eqnarray}
which we see immediately is quartic in $\Omega_{ij}$, and where we defined $\tilde \mu=\frac{\mu}{1+\rho^2 \tilde \gamma}$.
Also this integral cannot be evaluated exactly, but it can be done in perturbation theory.

It is interesting to note that the presence of $\tilde \gamma$ renormalizes the variables $\mu$, $J$ and $\sigma$. If we want to still satisfy the projector conditions we must have 
\begin{eqnarray}
    \frac{\mu}{1+\rho^2 \tilde \gamma }=\frac{J^2}{1+ J^2 \frac{\tilde \gamma}{N-1}}+\frac{\rho^2}{1+ \rho^2 \tilde \gamma}.
    \label{eq:scaling2}
\end{eqnarray}
We are interested in integrals of the form
\begin{equation}
    \langle e^{\sum_{ij} \Omega_{ij} b_{ij}}\rangle_{P(\Omega)}
\end{equation}
in the following, and in particular in the thermodynamic limit in which $N-1\approx N$. We evaluate these at the first order in $\lambda$, and obtain a correction to the average over disorder due as a function of the interaction strength. Specifically, we find that at the first order in $\lambda$ we can write:
\begin{eqnarray}
    \langle e^{\sum_{ij} \Omega_{ij} b_{ij}}\rangle_{P(\Omega)}&=&\prod_{\langle ij\rangle} e^{\frac{b_{ij}^2 J_1^2}{2 N}}\prod_{i}e^{b_{ii} \mu_1 +\frac{b_{ii}^2 \rho_1^2}{2}} \prod_{ij}e^{\gamma N F_{ij}} \nonumber \\
    &+&O(\gamma)
    \label{eq:disorder}
\end{eqnarray}
Now it is interesting to note that there are two possibilities for the scaling of $\tilde \gamma$. The first option is to keep it constant in $N$, in which case it does not affect $J$, but only $\mu$ and $\sigma$.
Alternatively, we can re-scale $\tilde \gamma\rightarrow \gamma N$ and define $J^\prime$ as $\frac{1}{2(J_1)^2}=\frac{1}{2J^2}+\gamma$ and defined $C_{ij}$, in which case $\sigma$ and $\mu$ go to zero.
As we will see later, unless $\tilde \gamma\rightarrow \gamma N$ the corrections to the disorder average will not scale with the size of the system. Moreover, we will see that unless $\rho=0$, the corrections due to the projection constraints scale with the wrong power in $N$ in the thermodynamic limit. If we combine eqns (\ref{eq:scaling1}) and eqn. (\ref{eq:scaling2}), we see that in order for the two constraints to be satisfied, we should have $\gamma= N$ in the thermodynamic limit. 
\section{Disorder average}
Let us first describe the notation for the replicated partition function, which we define for integer $n$, as
\begin{eqnarray}
    Z_n(\Omega)&=&\left(\sum_{\{s\}=\pm 1}e^{\beta(\sum_{i<j}^N \Omega_{ij} \pmb{b}_{ij}+h_0 \sum_{j=1}^N \sigma_j)}\right)^n \nonumber \\
    &=&\prod_{\alpha=1}^n \sum_{\{s^i\}=\pm 1}   e^{\beta(\sum_{i<j}^N \Omega_{ij} \pmb{b}_{ij}^\alpha+h_0 \sum_{j=1}^N \sigma_j^\alpha)}.
\end{eqnarray}
In what follows, superscripts $\alpha,\beta$ indicate replicas, while subscript indices the spin sites ($\sigma_i^\alpha$ is the spin at the site $i$ of replica $\alpha$).

We begin by performing the average over the disorder. This step is typically the easiest point of the calculation, but in our case (because of the interacting disorder), requires some attention. We have
\begin{eqnarray}
Z_{n}&=&\overline{ Z_n (\Omega)} =\sum_{\Omega }P[\Omega]Z^{n}(\Omega) \nonumber \\
&=&\sum_{\Omega}P(\Omega)\Big(\prod_{\alpha=1}^n\sum_{\{s^{\alpha}\}}\exp \{ \sum_{\alpha=1}^{n} \Big[\sum_{1\leq i<j\leq N}\Omega_{ij}\pmb{b}_{ij}^\alpha\}\nonumber \\
&&\ \ \ \ \ \ \ \ \ \ \ \ \ \ \ \ \ \ \ \cdot\exp\{h_0 \sum_{i=1}^{N} s_{i}^{\alpha} \Big] \}\Big)
\end{eqnarray}
and using eqn (\ref{eq:disorder}), we can define
\begin{equation}
    Z_n=\langle Z^n\rangle_{P(\Omega)}.
\end{equation}
Specifically, the sum over $J$ has to be substituted with an integral:
\begin{equation}
P[\Omega]=\prod_{1\leq i<j\leq N}P(\Omega_{ij})d\Omega_{ij}
\end{equation}
which we perform in the next section. In what follows, we define $\pmb{b}_{ij}=\sum_{\alpha=1}^n \pmb{b}_{ij}^\alpha$.
For the disorder average, we have the following measure
\begin{eqnarray}
    P(\Omega_{ij})&=&\lim_{\gamma\rightarrow \infty}\frac{1}{\mathcal N} \int_{-\infty}^\infty d\Omega_{ij} \prod_{\langle i j\rangle}e^{-N\frac{\Omega_{ij}^2}{2 J^2}} \nonumber \\ 
    & &\ \ \ \ \ \ \ \cdot\prod_{i} e^{-\frac{(\Omega_{ii}-\mu)^2}{2 \rho^2}}e^{-\gamma (\sum_t \Omega_{it} \Omega_{tj}-\Omega_{ij})^2} \nonumber
\end{eqnarray}
where $\mathcal N$ is a normalization factor, and would like to perform the following average
$ \langle e^{\sum_{ij} \pmb{b}_{ij} \Omega_{ij}}\rangle_{P(\Omega_{ij})}$.

Let us note that from our previous analysis of the Lyapunov function, for large values of $\xi$ or the disorder in $\Omega_{ij}$ we should expect an asymptotic random state.
Intuitively this is true when $\gamma=0$ as we will see.

\subsection{Average over the bond disorder}
Often in spin glass calculations the average over the disorder is somewhat trivial, given that the elements of the coupling matrix $J_{ij}$ (here $\Omega_{ij}$) are assumed to be i.i.d. distributed.  However, as mentioned before we assume that the disorder is interacting. The corrections are discussed below, albeit the details of the calculations are provided in the Appendices.
\subsubsection{Calculation of $F_{ij}$}
We would like to perform the following average over the disorder using a first order correction induced by the fact that the our exchange coupling matrix is a projector operator. The projector constraint is represented by the parameter $\gamma$, which multiplies $F_{ij}$ in eqn.( \ref{eq:disorder}).
Now note that if $\gamma=0$ we have the standard average over the disorder introduced by Sherrington and Kirkpatrick (with $\Omega_0=0$), while for $\gamma>0$ we have corrections due to the projector operator property. For this reason, here we consider only the first order corrections, e.g. we assume for the time being that $\gamma\ll 1$. We will use this as a probe of the effect of non-trivial correlations between the elements of of the matrix $\Omega_{ij}$. 
With this prescription we have an interacting disorder, rather than a free one, simply meaning integrals over disorder are not Gau\ss ian. 
While seemingly harmless, such modification complicates substantially the average over the disorder.  These calculations became soon very involved, and we had to soldier through them. 
We have
\begin{widetext}
\begin{eqnarray}
    \langle e^{\sum_{ij} \pmb{b}_{ij} \Omega_{ij}}\rangle_{P(\Omega_{ij})}&=&
    \lim_{\tilde \gamma\rightarrow \infty}\frac{1}{\mathcal N} \int_{-\infty}^\infty d\Omega_{ij} \prod_{\langle i j\rangle}e^{-N\frac{\Omega_{ij}^2}{2 J^2}} \prod_i  e^{-\frac{(\Omega_{ii}-\mu)^2}{2\rho^2}}e^{-\tilde \gamma (\sum_t \Omega_{it} \Omega_{tj}-\Omega_{ij})^2} e^{\sum_{ij} \pmb{b}_{ij} \Omega_{ij}} \nonumber \\
&\approx &\frac{1}{\mathcal N} \int_{-\infty}^\infty d\Omega_{ij} \prod_{\langle i j\rangle}e^{-\Omega_{ij}^2(N\frac{1}{2 J^2}+\tilde \gamma)}  e^{\sum_{ij} \pmb{b}_{ij} \Omega_{ij}}  \left(1-\tilde \gamma \frac{1}{2}\sum_{ij}(\sum_t \Omega_{it} \Omega_{tj}-\Omega_{ij})^2 +\frac{\tilde \gamma}{2} \sum_{ij} \Omega_{ij}^2\right)   \nonumber \\
&\approx &\frac{1}{\mathcal N} \int_{-\infty}^\infty d\Omega_{ij} \prod_{\langle i j\rangle}e^{-\Omega_{ij}^2(N\frac{1}{2 J^2}+\tilde \gamma)}  e^{\sum_{ij} \pmb{b}_{ij} \Omega_{ij}}  \left(1-\frac{\tilde \gamma}{2}\sum_{ij} \sum_{tt^\prime} \Omega_{it} \Omega_{tj}\Omega_{it^\prime} \Omega_{t^\prime j}+\tilde \gamma \sum_{ij} \sum_t \Omega_{ij}\Omega_{it}\Omega_{tj} \right)   \nonumber
\end{eqnarray}
\end{widetext}
We see in that in order for the average to be extensive, we must have $\tilde \gamma$ to scale as $N$. We thus simply rescale $\tilde \gamma\rightarrow \gamma N$, and define a new effective constant $J_1$ such that $\frac{1}{2J_1^2}=\frac{1}{2J^2}+\gamma$. With this prescription we have
\begin{widetext}
\begin{eqnarray}
    \langle e^{\sum_{ij} \pmb{b}_{ij} \Omega_{ij}}\rangle_{P(\Omega_{ij})}
&\approx &\frac{1}{\mathcal N} \int_{-\infty}^\infty d\Omega_{ij} \prod_{\langle i j\rangle}e^{-N\frac{\Omega_{ij}^2}{2 J_1^2}}  e^{\sum_{ij} \pmb{b}_{ij} \Omega_{ij}}  \left(1-N \frac{\gamma}{2}\sum_{ij} \sum_{tt^\prime} \Omega_{it} \Omega_{tj}\Omega_{it^\prime} \Omega_{t^\prime j}+\gamma N  \sum_{ij} \sum_t \Omega_{ij}\Omega_{it}\Omega_{tj} \right).   \nonumber
\end{eqnarray}
\end{widetext}
The calculation above is a little long but simple. We have
\begin{eqnarray}
    \langle 1 \rangle &=&\int_{-\infty}^\infty d\Omega_{ij} \prod_{\langle i j\rangle}e^{-N\frac{\Omega_{ij}^2}{2 J_1^2}} \prod_i  e^{-\frac{(\Omega_{ii}-\mu)^2}{2\rho^2}} e^{\sum_{ij} \pmb{b}_{ij} \Omega_{ij}}\nonumber \\
    &=&\mathcal N \prod_{i<j} e^{\frac{J_1^2 \pmb{b}_{ij}^2}{2 N}} \prod_{i} e^{\pmb{b}_{ii} \mu_1 +\frac{\rho_1^2 \pmb{b}_{ii}^2}{2 }}
\end{eqnarray}
It is faster in what follows to write
\begin{eqnarray}
\langle 1 \rangle&=&\prod_{ij} e^{C_{ij} \pmb{b}_{ij}+Q_{ij} \pmb{b}_{ij}^2}, \nonumber \\
C_{ij}&=&\delta_{ij} \mu_1  \nonumber \\
Q_{ij}&=&\delta_{ij} \frac{\rho_1^2}{2}+(1-\delta_{ij}) \frac{J_1^2}{2 (N-1)}.
\end{eqnarray}
Using the formulae above we can calculate
\begin{equation}
    \langle \Omega_{i_1 j_1}\cdots \Omega_{i_k j_k}\rangle=\frac{\partial}{\partial \pmb{b}_{i_1 j_1}}\cdots \frac{\partial}{\partial \pmb{b}_{i_k j_k}} \langle 1 \rangle.
\end{equation}
For the case of the first order correction, we can write
\begin{eqnarray}
    \langle e^{\sum_{ij} \pmb{b}_{ij} \Omega_{ij}}\rangle_{P(\Omega_{ij})}
&\approx& (1+\gamma N \sum_{i j} \sum_t \frac{\partial}{\partial \pmb{b}_{ij}}\frac{\partial}{\partial \pmb{b}_{it}}\frac{\partial}{\partial \pmb{b}_{tj}}\nonumber \\
&-&\frac{N \gamma}{2} \sum_{ij} \sum_{t,t^\prime} \frac{\partial}{\partial \pmb{b}_{it}}\frac{\partial}{\partial \pmb{b}_{tj}}\frac{\partial}{\partial \pmb{b}_{it^\prime}}\frac{\partial}{\partial \pmb{b}_{t^\prime j}})\langle 1 \rangle.\nonumber 
\end{eqnarray}
Let us now calculate the second term in the expansion.
In the symmetric toy model case, we have $\mu=0$ and $\rho_1^2=\frac{J_1^2}{N-1}$. In this case, 
\begin{equation}
    \langle 1 \rangle =\int_{-\infty}^\infty d\Omega_{ij} \prod_{ i, j}e^{-N\frac{\Omega_{ij}^2}{2 J_1^2}}  e^{\sum_{ij} \pmb{b}_{ij} \Omega_{ij}}=\mathcal N  \prod_{ij} e^{\frac{J_1^2 \pmb{b}_{ij}^2}{2 N}}.
\end{equation}
Once the average has been performed, at the first order in $\gamma$ we can re-insert the average in the exponential, as $1-\gamma x\approx e^{-\gamma x}$, as it is common in perturbation theory. While this is an approximation, it is not inconsistent with the assumption $\tilde \gamma \rightarrow \gamma$, as we will choose the parameters for the average such that at the exponent we have a quantity which is homogeneous in $N$. This said, this approximation will force us to consider only small but non-zero values of $\gamma$ in what follows.

\subsection{Spin glass mean field equations}
At this point, we can use standard techniques for the analysis of mean field spin glasses in the replica-ansatz approximation. Assuming the rescaling $\tilde \gamma\rightarrow \gamma N$,
\begin{eqnarray}
   \lim_{N\gg 1} \sum_{ij} N F_{ij} &=& N \sum_{ij} a_{ij}(n) q^i m^j.
\end{eqnarray}
The partition function can thus be written as 
\begin{eqnarray}
    \beta F&=&-\lim_{n\rightarrow 0} \frac{Z_\Omega-1}{n} \nonumber \\
    &=&\lim_{n\rightarrow 0} \frac{1}{n}\text{Tr}_{s^1\cdots s^n}\left( \prod_{\langle ij\rangle} e^{\frac{\pmb{b}_{ij}^2 J_1^2}{ 2N} + \gamma N F_{ij}} \right)-\frac{1}{n}
\end{eqnarray}
In order to perform the calculation, we can insert $1$ smartly into the partition function, with the technique introduced in \cite{CrisantiSommers}, both for the magnetization and the overlap parameter:
\begin{eqnarray}
    1&=&\int_{q>0} dq^{\alpha \beta}  \delta(N q^{\alpha \beta} - \sum_i \sigma_i^\alpha \sigma_i^\beta)\nonumber \\
    &=&  \int_{q>0} dq^{\alpha \beta} \int_{-\infty i}^{\infty i} \frac{ N d\lambda_{\alpha \beta}}{2\pi i}e^{-\frac{1}{2} \sum_{\alpha\neq \beta} \lambda_{\alpha \beta}(N q^{\alpha \beta} - \sum_i \sigma_i^\alpha \sigma_i^\beta)} \nonumber \\
    1&=&\int_{m>0} dm^\alpha \delta(N m^\alpha - \sum_i \sigma_i^\alpha)\nonumber \\
    &=&\int_{m>0} dm^\alpha\int_{-\infty i}^{i\infty} \frac{d\eta_\alpha}{2\pi i} e^{-\sum_\alpha \eta_\alpha (N m^\alpha - \sum_i \sigma_i^\alpha)}.
 \end{eqnarray}
 We now would like to take the trace over the spins.
 This can be done as follows.
 
First, we note that we can write
\begin{eqnarray}
    \sum_{\{s^1\}}\cdots \sum_{\{s^n\}} & &e^{\sum_{a\neq b}\lambda_{ab} \sum_i \sigma_i^a \sigma_i^b+\sum_a \eta_a \sum_i \sigma_i^a}\nonumber \\
    & &\ \ \ \  =(\sum_{s^1}\cdots \sum_{s^n} e^{\sum_{a\neq b}\lambda_{ab} \sigma^a \sigma^b+\sum_a \eta_a \sigma^a})^N.\nonumber 
\end{eqnarray}
In order to calculate this exactly, we assume the RSA for the overlap, $q^{ab}$, e.g. $q^{ab}=q$ for $a\neq b$ and $q^{ab}=0$ for $a=b$, and $m^{a}=m$. At this point we note that it is not necessary to introduce a Lagrange multiplier for every element of the matrix $q^{ab}$ and for the mean magnetization, but two only. Thus we can write
\begin{eqnarray}
    \sum_{s^1}\cdots& &\sum_{s^n} e^{\sum_{a\neq b}\lambda_{ab} \sigma^a \sigma^b+\sum_a \eta_a \sigma^a}
    \nonumber \\
    & &\ \ \ \ =\sum_{s^1}\cdots \sum_{s^n} e^{\sum_{a\neq b}\lambda \sigma^a \sigma^b+\sum_a \eta \sigma^a}.
\end{eqnarray}
We now see that this is the partition function of a Curie-Wei\ss\ model. We can thus write
\begin{eqnarray}
    Z_{CW}(\lambda,\eta)&=&\sum_{s^1}\cdots \sum_{s^n} e^{\sum_{a\neq b}\lambda \sigma^a \sigma^b+\sum_a \eta \sigma^a}\nonumber \\
    &=&\frac{1}{\sqrt{2\pi}}\int d\tilde x\ e^{-\frac{\tilde x^2}{2}+ n \log 2\cosh(\sqrt{2 \lambda } \tilde x+\eta)-n \lambda},\nonumber 
\end{eqnarray}
 This implies that we can write an effective term in the quenched partition function of the form     $e^{N \log Z_{CW}(\lambda,\eta)}$.
Since terms which are proportional to $n^2$ and higher are going to be suppressed in the limit $n\rightarrow 0$, it is important to know those that are linear in $n$. We thus see that factors that scale as $N^{-1}$ and proportional to $n^2$ can be suppressed.  We performed in the Appendices the steps that lead to the function $H_{RS}$. In Appendix A we calculate the perturbative corrections, while in B and C we calculate the corrections in terms of the parameters $q$ and $m$ in the symmetric replica Ansatz. Some assumptions here have been done. We first noticed that, unless $\rho_1^2=\frac{J_1^2}{N-1}$, the final result depends also on higher order overlaps such as $\rho^{\alpha \beta \gamma}=\frac{1}{N}\sum_i \sigma_i^\alpha \sigma_i^\beta \sigma_i^\gamma$ and $\eta^{\alpha \beta \gamma \delta}=\frac{1}{N}\sum_i \sigma_i^\alpha \sigma_i^\beta \sigma_i^\gamma \sigma_i^\delta$. Such analysis will be performed in following studies, but here we focus on the reduced set of parameters $J,\xi,\gamma,s=\frac{1}{N} \sum_{i} \frac{x_i}{2 \alpha \beta}$ and $\gamma$. Of these, we are interested in the phase diagram as a function of the two physical parameters $\xi$ and $s$. We note that since in this approximation $\rho$ now scales with $N$, necessarily because of eqn. (\ref{eq:projectioneq}) we must have $1=\frac{J}{\sqrt{\mu}}$. In terms of the angle of eqn. (\ref{eq:projectioneq}), we thus see that we choose $\theta=0$.
These results will be effectively be perturbed as function of the strength of $\gamma$. We can thus interpret the role of $\gamma$ as a measure of how robust are our phase diagrams with the introduction of non-trivial correlations in the disorder average. Given the prescriptions above, let us now show the dependence of $H_{RS}$ on the parameters above and $q$ and $m$.

In the replica symmetric ansatz, the final result of the calculation depends on the polynomial mean field functional:
\begin{eqnarray}
\tilde H_{RS}(q,m)&=&\Big(h-\mu_1\xi+\gamma \big(-2J_1^4(1-2\mu_1 ) \nonumber \\
& & \ \ \ \ \ -  J_1^2\xi\mu_1^2(3-2\mu_1)  \big)\Big) m \nonumber \\
   &-&\gamma \Big(\frac{1}{4}J_1^8 \xi ^4  \Big) q^3 m +\gamma \Big(-\frac{3\xi}{4}J_1^6(1-2\mu_1)  \Big) q^2m \nonumber \\
   &+&\left(\frac{J_1^2}{2}\xi^2+ \gamma\xi^2 \mu_1 J_1^4\big(\frac{3}{2} -\mu_1\big)\right) qm \nonumber \\
   &-&\gamma \Big(\frac{\xi^3}{2} J_1^6(1-2\mu_1) \Big) qm^2 \nonumber \\
   &+&\left(-\frac{J_1^2}{4} \xi^2+\gamma J_1^4\xi^2\big(- \mu_1 (1 -\mu_1)+\frac{J_1^2}{2} \big) \right)q \nonumber \\
   &+&\Big(-\frac{J_1^2}{2}\frac{\xi^2}{8}+\gamma J_1^4\xi \Big(- \frac{\mu_1}{8}   \nonumber \\
   & & \ \ \ \ \ \ -\mu_1 (1-\mu_1)\frac{\xi}{4}+\frac{\xi^2 }{2}   J_1^2(1-2\mu_1) \Big)\Big) q^2 \nonumber \\
   &+&\gamma J_1^6\frac{\xi^2 }{8}\Big(1+2(1-2\mu_1) \xi  \Big) q^3 \nonumber \\
   &+&\frac{ 7 \gamma J_1^8}{32} \xi ^4   q^4 \nonumber \\
\end{eqnarray}

which one has also to consider two terms of the form
\begin{equation}
    H_l(\lambda,\eta)=\lambda q+ \eta m
\end{equation}
and for small $n$ we have
\begin{eqnarray}
    \log Z_n(\lambda,\eta)&\approx&  n(\frac{1}{\sqrt{2\pi}} \int_{-\infty}^\infty d\tilde x\ e^{-\frac{\tilde x^2}{2} } \log 2\cosh(\sqrt{2\lambda} \tilde x+\eta) \nonumber \\
    &
    &\ \ \ \ \ -\lambda),
\end{eqnarray}
it follows that
\begin{eqnarray}
    -f&=&\lim_{n\rightarrow 0}\frac{Z_\Omega-1}{nN} \nonumber \\
    &=&   \tilde H_{RS}(q,m)+H_l\nonumber \\
    & &+\frac{1}{\sqrt{2\pi}} \int_{-\infty}^\infty d\tilde x\ e^{-\frac{\tilde x^2}{2}} \log 2\cosh(\sqrt{2\lambda} \tilde x+\eta)-\lambda, \nonumber \\
\end{eqnarray}
where $q,m,\lambda,\eta$ have to satisfy the saddle point equations (SPE's). The SPE's in the variables $\lambda,\eta,q$ and $m$, after a rapid calculation, given by 
\begin{eqnarray}
    1-q&=&\frac{1}{\sqrt{2\pi}} \int_{-\infty}^\infty d\tilde x\ e^{-\frac{\tilde x^2}{2}}   \frac{\tilde x \tanh \left(\eta +\sqrt{2} \sqrt{\lambda } \tilde x\right)}{\sqrt{2} \sqrt{\lambda }} \nonumber \\ 
m&=&\frac{1}{\sqrt{2\pi}} \int_{-\infty}^\infty d\tilde x\ e^{-\frac{\tilde x^2}{2}}   \tanh(\sqrt{2\lambda} \tilde x+\eta) \nonumber  \\
            \lambda&=&-\partial_q  \tilde H_{RS}(q,m), \nonumber \\
    \eta &=&- \partial_m  \tilde H_{RS}(q,m),
\end{eqnarray}
respectively.
We can thus substitute in the first two equations and get the final mean field equations
\begin{eqnarray}
q&=&\frac{1}{\sqrt{2\pi}} \int_{-\infty}^\infty d\tilde x\ e^{-\frac{\tilde x^2}{2}}  \tanh^2\Big(M(q,m)\Big)
    \nonumber \\
    m&=&\frac{1}{\sqrt{2\pi}} \int_{-\infty}^\infty d\tilde x\ e^{-\frac{\tilde x^2}{2}}   \tanh\Big(M(q,m)\Big)\nonumber 
\end{eqnarray}
where $M(q,m)=\sqrt{-2(\partial_q  \tilde H_{RS}(q,m))} \tilde x-\partial_m  \tilde H_{RS}(q,m)$.

which define the state of the system depending on the parameters obtained via the mean field equations.
\section{Numerical solutions of the mean field equations}
At this point we can, with the equations at hand, try to understand the behavior of the overlap and magnetization as a function of the parameters of the model in the RSA, in which the only relevant parameters are the mean magnetization $m=\frac{1}{n}\sum_\alpha m^\alpha=\frac{1}{n N}\sum_\alpha \sum_i m_i^\alpha$ and the replica overlap $q=\frac{2}{n(n-1) N}\sum_{\alpha \beta} \sum_i \sigma_i^\alpha \sigma_i^\beta $.
We recall that a ferromagnetic phase is equivalent to $m\neq0, q\neq0$ a paramagnetic phase $m=q=0$, and a glass phase $m=0$, $q\neq 0$.

We first study the case $\gamma=0$, 
\subsection{Case $\gamma=0$}
\begin{figure}
    \centering
    \includegraphics[scale=0.15]{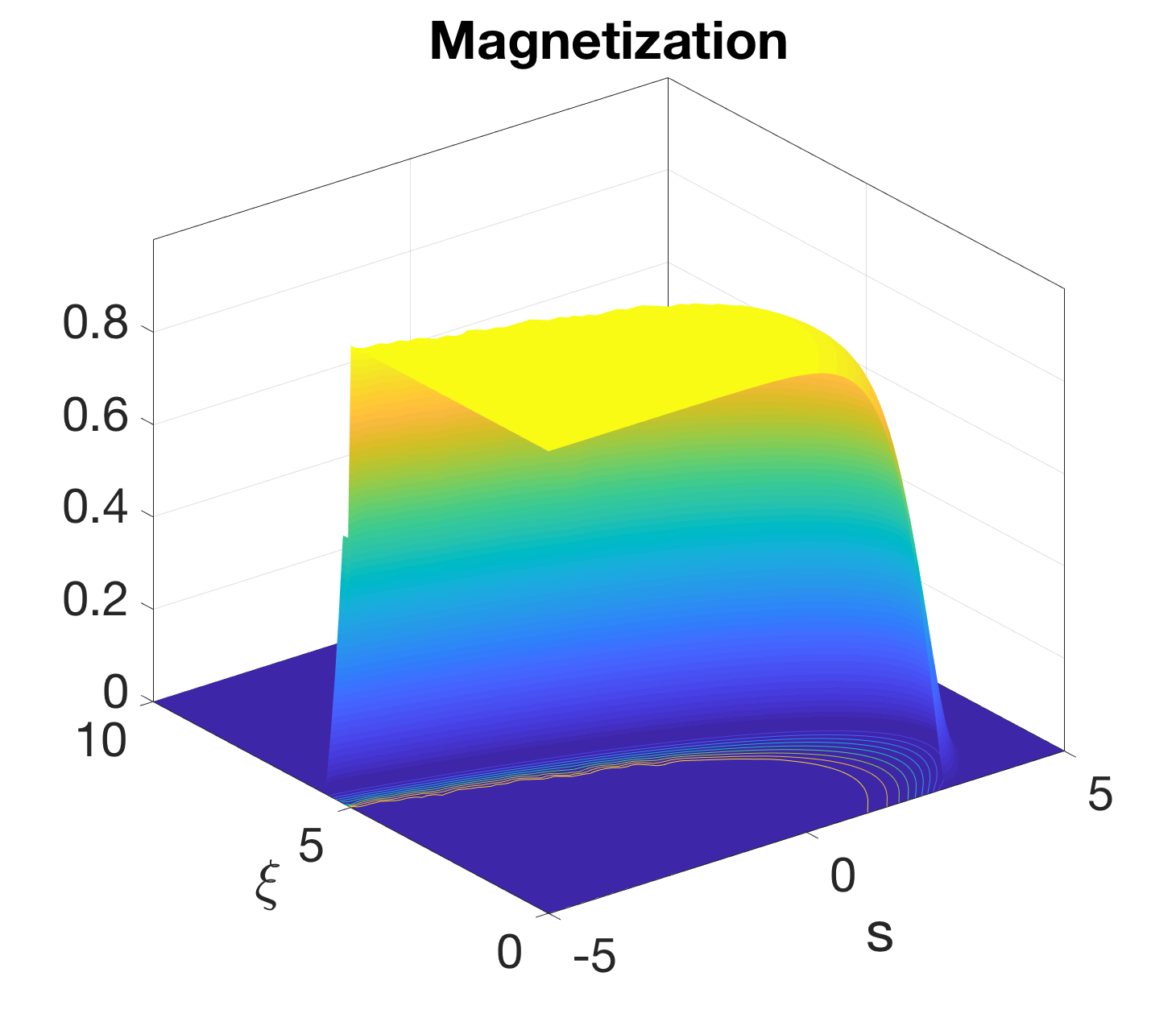}\\
    \includegraphics[scale=0.15]{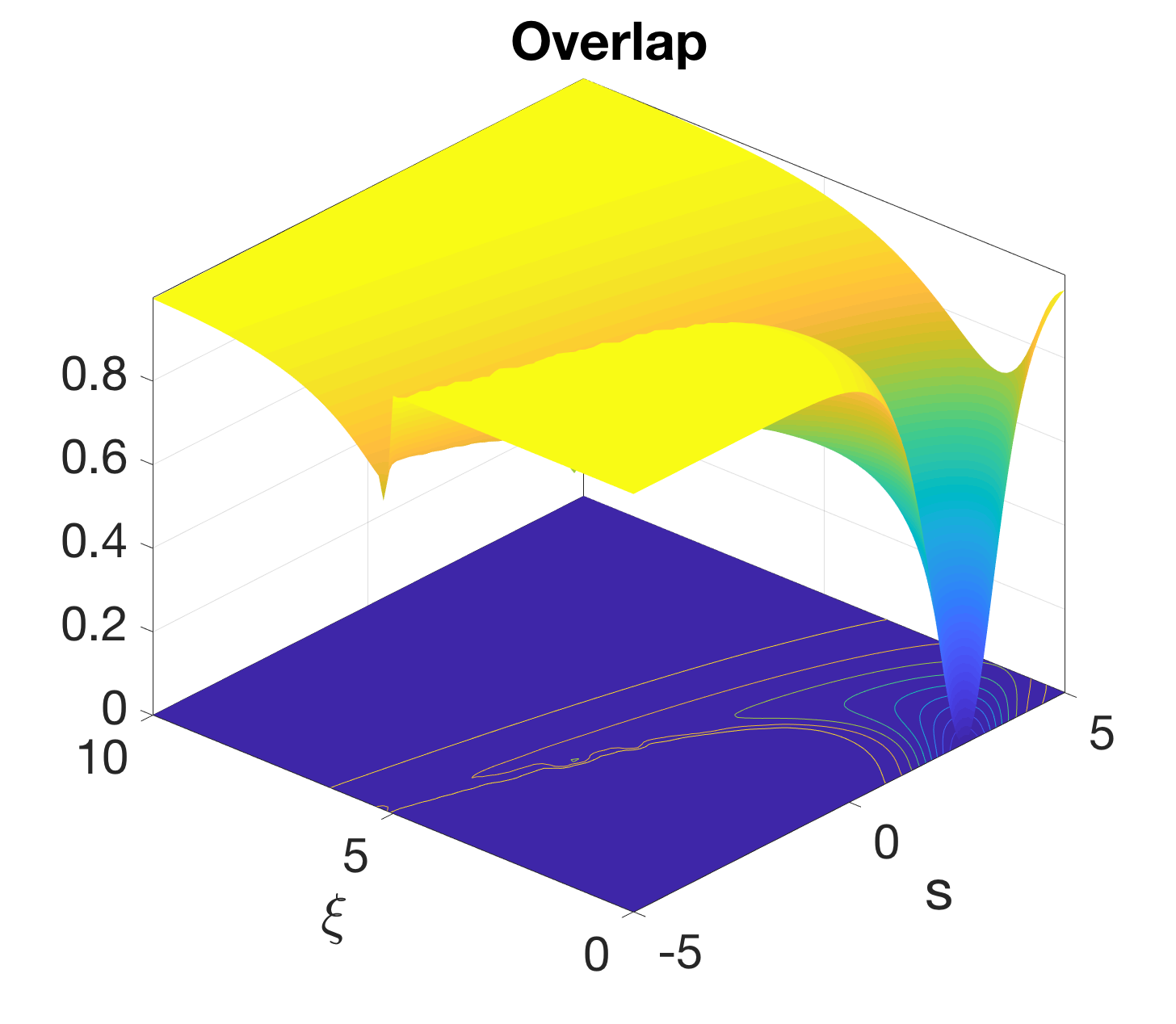}
    \caption{Magnetization and Overlaps as a function of $\xi$ and $s$ for $J=1$, $\gamma=0$. We observe a transition between a glass and a ferromagnetic phase.}
    \label{fig:ovmag}
\end{figure}

as a function of $\xi$ and the mean external field $s$. These are the two figures in Fig. \ref{fig:ovmag} in which we plot the magnetization and the overlap as a function of $\xi$ and $s$, for $J_1=J=\sqrt{\mu}=1$. We observe that there are two phases: a glass phase and a ferromagnetic phase, and that there is a special paramagnetic point along the $\xi=0$ line at $s=\frac{8}{3}$, which is where $m=q=0$.  Such point is special as we will discuss shortly, as it plays an important role also in the other phase diagrams. Numerically one can see that the the phase separation line does depend on the strength of the disorder.

\begin{figure}
    \centering
    \includegraphics[scale=0.15]{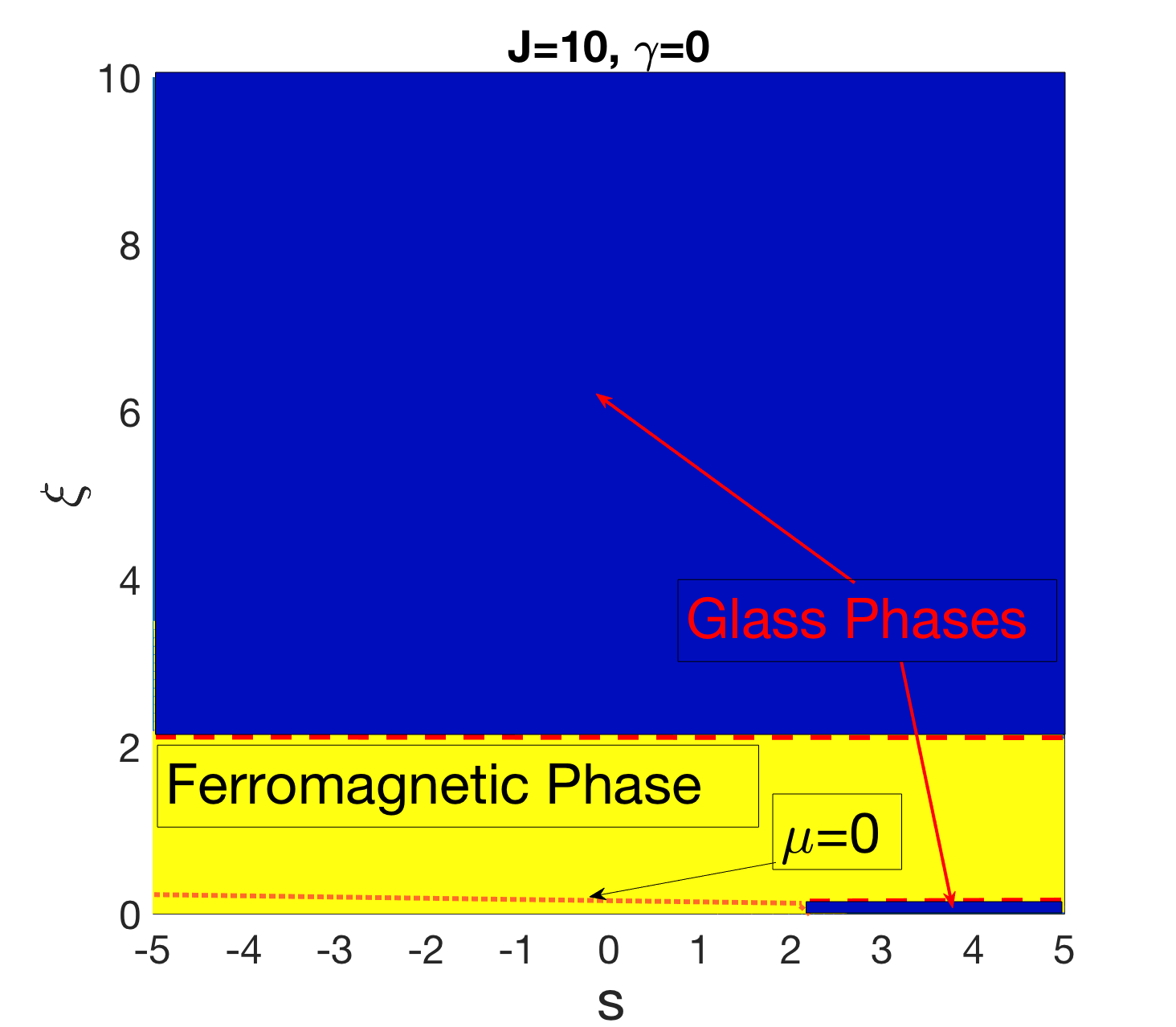}\\
    \includegraphics[scale=0.15]{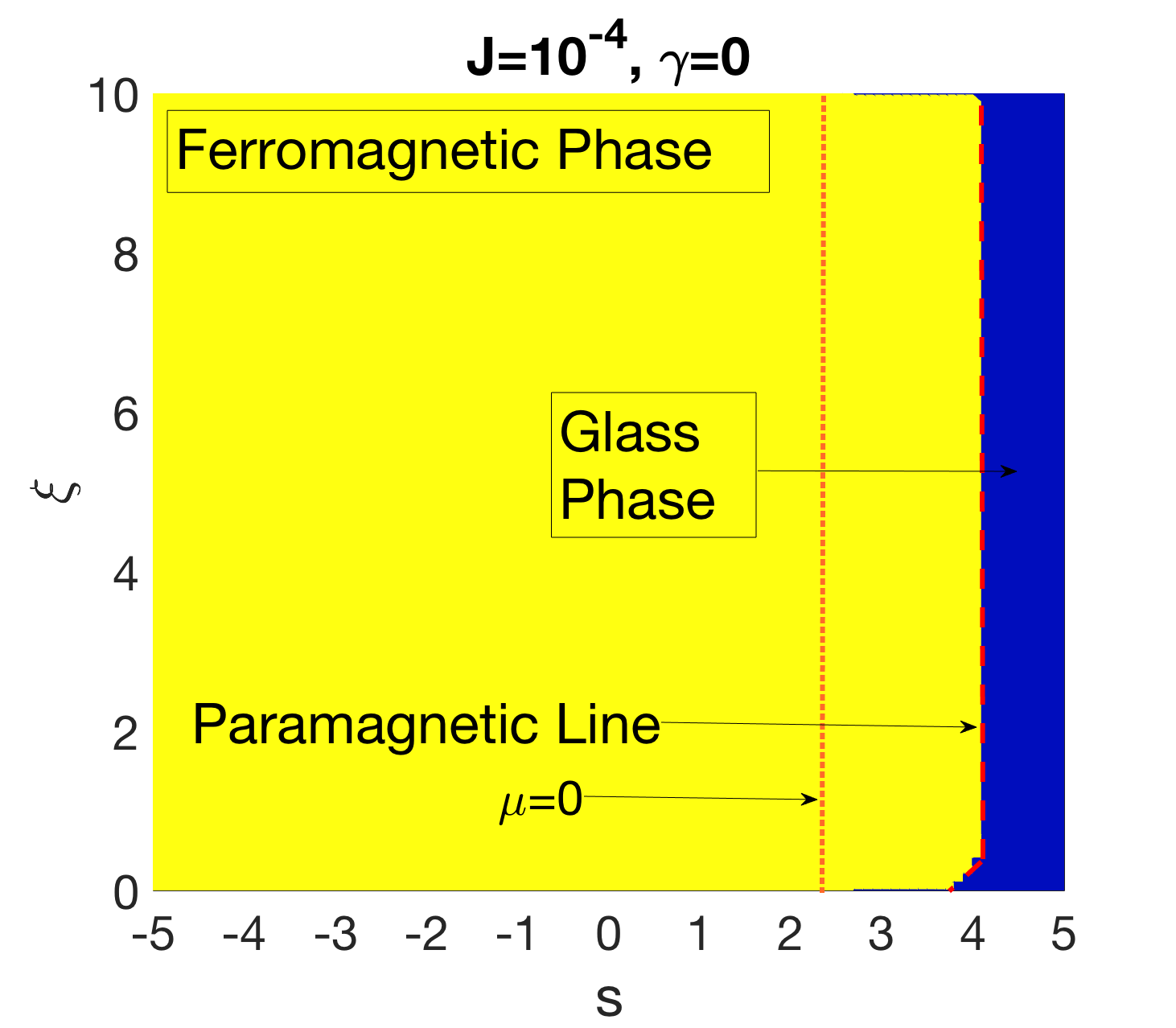}
    \caption{The diagrams above describe in the parameter region $\xi\in [0,10]$ and $s\in [-5,5]$, and for $\gamma=0$, the two limiting cases $J\ll 1$ and $J\gg 1$, with the condition $J=\sqrt{\mu}$. The dashed red line is the phase line between the ferromagnetic and glass phases. Yellow corresponds to a ferromagnetic phase, while blue to a glass phase.}
    \label{fig:highlowd}
\end{figure}

\begin{figure}
    \centering
    \includegraphics[scale=0.15]{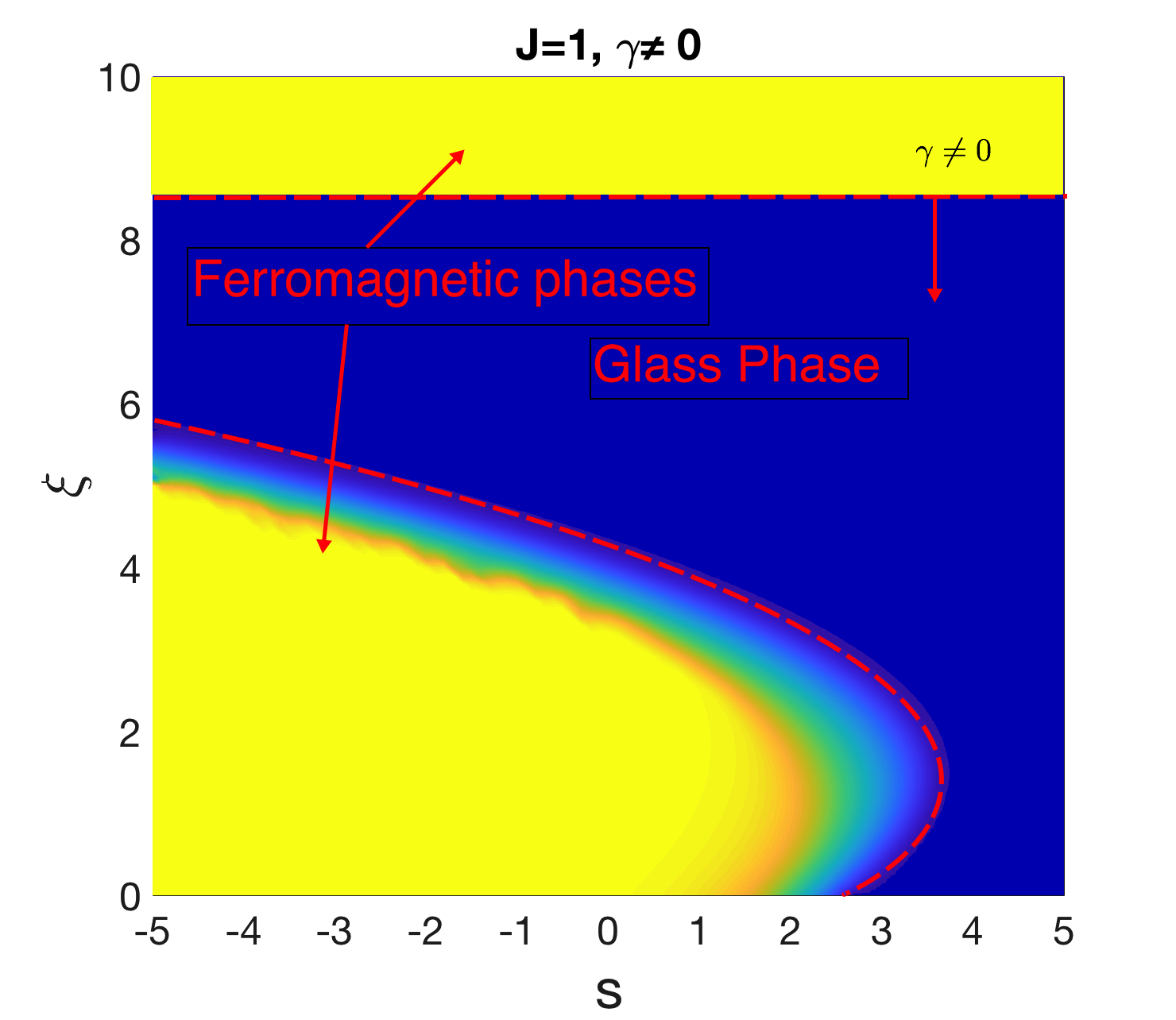}\\
    \caption{Phase diagram with the introduction of the interactions, for $\gamma=0.05$, and $J=\sqrt{\mu}=1$. Yellow corresponds to a ferromagnetic phase, while blue to a glass phase.}
    \label{fig:int}
\end{figure}
For $J\rightarrow 0$, the phase separation is characterized, for $J_1=\sqrt{\mu}=10^{-4}$, by a horizontal line $\xi=2$, for which above $\xi=2$ we have a glass phase, and below a ferromagnetic phase. We can see the phase diagram in Fig. \ref{fig:highlowd} (top). Also in this case, however, along the $\xi0$ line we have that there is a special point at $s=s_0\approx 2.2$ for which there is a secluded glass phase for $s>s_0$. If we allow for $J$ to remain constant and and $\mu\rightarrow 0$, the ferromagnetic phase shrinks, and the secluded glass phase joins with the glass phase which covers essentially the entire plane aside from a region below $s_0$.

On the other hand, the situation is inverted for large values of the disorder, $J_1\gg 1$, the ferromagnetic-glass line is separated along a vertical line $s^*=const(\mu)$ which for $\mu=0$ is approximately equal to $s_0$, but that moves further right for $J=\sqrt{\mu}$. We plot the phase diagram in Fig. \ref{fig:highlowd} (bottom). It is interesting to note however that along the $\xi=0$ line the ferromagnetic-glass transition is still at $s_0$, implying that there is a re-entrant glass region for $\xi\approx 0$ and $s\geq s_0$. For very large values of $J$, the critical line $s^*$ is paramagnetic, as on that specific line $q=0$.

\subsection{Case $\gamma\neq 0$}
In the case in which the higher order terms in $H_{RS}$ are present, and in particular with an interacting disorder, we find that some of the features of the previous phase diagrams are robust. Let us focus on Fig. \ref{fig:int} in which we plot for $J=\sqrt{\mu}=1$ the phase diagram of the model for $\gamma=0.05$. The bottom region of the $[s,\xi]$ diagram is very similar to the situation we had in Fig. \ref{fig:ovmag}, with the difference that for $\gamma>0$ we have a re-entrant glass phase within the ferromagnetic region (for $s\leq s_0\approx 8/3$). It is interesting to note  that for $\gamma>0$ a new ferromagnetic region above $\xi=\xi_0(\gamma)$ appears. For larger values of $\gamma$ such region moves down , and in the regimes in which we should include second order contributions in $\gamma$, it merges with the ferromagnetic region below. It follows that most of the $[s,\xi]$ plane is covered with a ferromagnetic regime. 

\section{Conclusions}
In this paper we have presented an attempt to characterize the asymptotic behavior of purely memristive circuits in terms of Lyapunov functions, via a disorder average on the projection matrix $\Omega$ which represents the underlying cycle space of the graph.
We have first introduced a new Lyapunov function in which no constraints on the parameters $\xi$ and $s$ must be satisfied. Since along the dynamics $\frac{d}{dt} L\leq 0$, the nature and complexity of the minima of $L$ characterizes the possible asymptotic states in which the memristor states can relax on. At long times, such Lyapunov function is well approximated by an Ising model. In the equivalent Ising model, we have a linear term in the state variables which depends on the circuit control (e.g. voltage or current generators, either in series or parallel),  via the adimensional parameter $s=\frac{\phi}{\alpha \beta}$. The parameters $\alpha$ and  $ \beta$ control the relaxation and excitation timescales of the memristors. A quadratic term  emerges in the Lyapunov function, which depends on $\Omega$, a matrix (a projector operator), in which the circuit topology enters, and the adimensional parameter $\xi$, which controls the degree of nonlinearity in the system.

Such memristors-Ising model correspondence allows for an analysis typical of disordered systems via the introduction of a disorder distribution over the parameters $\Omega_{ij}$, for random circuits.
 Such analysis can be performed via standard techniques applied to study of glasses and spin glasses, via a quenched average over the the possible $\Omega_{ij}$ values based on empirical evidence.
In previous papers it had been noticed that for random graphs the distributions of the elements of the projector operator, $P(\Omega_{ij})$ which characterizes both the dynamics and the Lyapunov function was short tailed, and well approximated by a Gau\ss ian. Such distribution cannot be i.i.d. for all the variables states,  random because of the constraints $\Omega^2=\Omega_{ij}$ element by element. We have thus introduced these constraints perturbatively into the measure for the disorder average in terms of a control parameter.

The phases of the model have been then analyzed in terms of the physical parameters of the model: $\xi$, which controls the nonlinearity in the system, and that corresponds to a quadratic term in the Lyapunov function, and $s=\frac{\phi}{\alpha \beta}$ which is an effective external field. What we obtained is evidence that memristors can be in a glass or ferromagnetic phase depending on the value of $\xi=\frac{R_{off}-R_{on}}{R_{on}}$ and the control parameter $s$, and which seem robust to the introduction of correlations into the disorder average.

From the point of view of memristor dynamics, the glass phase  seems to confirm many observations of strong dependence of the system dynamics on the initial conditions.
 The ferromagnetic phase is instead interpreted as a phase in which the memristors eventually end all in the $R_{on}$ or $R_{off}$ states, while a glass phase one in which depending on the initial conditions the memristors end up in some random states.
 This is interesting for the following reason. In \cite{mean-field} it was noticed that a paramagnetic-ferromagnetic transition was present in a memristor model equivalent to a Curie-Wei\ss\ Ising model. For random circuits, such transition seem to become glass-ferromagnetic.
 
 It is fair to say that in this paper we have performed however various approximations, both at the level of the disorder and in the spin glass calculations via the replica symmetric Ansatz and have chosen a set of parameters for which higher order overlaps can be neglected. The importance and relevance of these terms will be analyzed in future works.\ \\\ \\

\textit{Acknowledgements.}  We acknowledge the support of NNSA for the U.S. DoE at LANL under Contract No. DE-AC52-06NA25396.  FC was also financed via DOE-LDRD grants PRD20170660 and PRD20190195.

\clearpage
\onecolumngrid

\appendix
\section{The symmetric case}
Perturbatively, we should expect the following type of expansion in $N$ and $J$ for the terms of order $3$ and $4$:
\begin{eqnarray}
\frac{\int_{-\infty }^{\infty } x^3 e^{-\frac{N x^2}{2 J^2}+r x} \, dx}{\int_{-\infty
   }^{\infty } e^{-\frac{N x^2}{2 J^2}+r x} \, dx}&=&\frac{3 J^4 r}{N^2}+ \frac{J^6}{N^3} r^3 \nonumber \\
    \frac{\int_{-\infty }^{\infty } x^4 e^{-\frac{N x^2}{2 J^2}+r x} \, dx}{\int_{-\infty
   }^{\infty } e^{-\frac{ N x^2}{2 J^2}+r x} \, dx}&=&\frac{3 J^4 }{N^2}+\frac{6 J^6 r^2}{N^3}+\frac{J^8 r^4}{N^4}.
\end{eqnarray}
This expectation is confirmed for symmetric disorder, but it is less evident for asymmetric disorder. 
A long but easy calculation shows that
\begin{equation}
    \frac{\langle \Omega_{ij} \Omega_{kt} \Omega_{pq}\rangle}{\langle 1 \rangle} =\frac{J^4_1}{N^2} \left(\delta_{ip} \delta_{jq}\pmb{b}_{kt}+\delta_{ik}\delta_{jt} \pmb{b}_{pq}+\delta_{kp}\delta_{tq} \pmb{b}_{ij} \right)+ \frac{J_1^6}{N^3} \pmb{b}_{pq} \pmb{b}_{kt} \pmb{b}_{ij}
\end{equation}
while 
\begin{eqnarray}
    \frac{\langle \Omega_{ij} \Omega_{kt} \Omega_{pq} \Omega_{rs} \rangle}{\langle 1 \rangle}&=&\frac{J_1^4}{(N-1)^2} \left(\delta_{ip} \delta_{jq}\delta_{rk}\delta_{st}+\delta_{ik}\delta_{jt}\delta_{rp}\delta_{sq}+\delta_{kp}\delta_{tq}\delta_{ri}\delta_{sj}\right) \nonumber \\
    &+&\frac{J_1^6}{N^3} \left(\delta_{rp} \delta_{sq} \pmb{b}_{kt} \pmb{b}_{ij}+\pmb{b}_{pq} \delta_{rk} \delta_{st} \pmb{b}_{ij}+\pmb{b}_{pq} \pmb{b}_{kt} \delta_{ri}\delta_{sj}+ \pmb{b}_{rs}\delta_{ip} \delta_{jq}\pmb{b}_{kt}+ \pmb{b}_{rs}\delta_{ik}\delta_{jt} \pmb{b}_{pq}+ \pmb{b}_{rs}\delta_{kp}\delta_{tq} \pmb{b}_{ij}\right)\nonumber\\
    &+& \frac{J_1^8}{N^4} \pmb{b}_{pq}\pmb{b}_{kt} \pmb{b}_{ij} \pmb{b}_{rs}.
\end{eqnarray}

Those expansions are a generalization of the scalar cases discussed above.
The cubic term leads, once we sum over $t$ and we choose appropriately the indices, to

\begin{equation}
    F^{3}_{ij}= N \sum_t \frac{\langle \Omega_{it} \Omega_{tj} \Omega_{ij}\rangle}{\langle 1 \rangle}= \gamma N \left(\frac{J_1^4}{(N-1)^2}(\pmb{b}_{ii}+\pmb{b}_{jj}+\delta_{ij}\pmb{b}_{ij})+\frac{J_{1}^6}{N^3} \pmb{b}_{ij} \sum_t \pmb{b}_{it} \pmb{b}_{tj} \right)
\end{equation}

Thus, we have that the non-constant terms, dependent on $b_{ij}$ are
\begin{equation}
    \sum_{ij}F^{3}_{ij}=(\mu_1+\rho_1^2)^2 \sum_{i} \pmb{b}_{ii}+\sum_{i\neq j}(\mu_1+\rho_1^2) \frac{J_1^2}{N-1} \pmb{b}_{ij}
\end{equation}

\begin{eqnarray}
    F^{4}_{ij}=-\frac{ N}{2} \sum_{t,t^\prime}\frac{\langle \Omega_{it} \Omega_{tj} \Omega_{it^\prime} \Omega_{t^\prime j}\rangle}{\langle 1 \rangle}&=&-\frac{\gamma N}{2} \Big( \frac{J_1^4}{N^2}(N+2\delta_{ij})+\frac{J_1^6}{N^3} (\delta_{ii} \sum_{t} \pmb{b}_{tj}^2+\delta_{jj} \sum_{t} \pmb{b}_{it}^2+ 2 \pmb{b}_{ii} \pmb{b}_{jj}+2 \delta_{ij} \sum_{t} \pmb{b}_{it}\pmb{b}_{tj}) \nonumber \\
    &+& \frac{J_1^8}{N^4} (\sum_{t} \pmb{b}_{it} \pmb{b}_{tj})^2    \Big)\nonumber \\
\end{eqnarray}
In the symmetric case, we have $\mu=0$ and $\rho_1^2=\frac{J_1^2}{N-1}\approx \frac{J_1^2}{N}$, and adding these two results we have, up to some constants, 
\begin{eqnarray}
    N F_{ij}=N F^{3}_{ij}+NF^{4}_{ij}&=& N \left(\frac{J_1^4}{N^2}(\pmb{b}_{ii}+\pmb{b}_{jj}+\delta_{ij}\pmb{b}_{ij})+\frac{J_{1}^6}{N^3} \pmb{b}_{ij} \sum_t \pmb{b}_{it} \pmb{b}_{tj} \right) \nonumber \\
    & &-\frac{ N}{2} \Big( \frac{J_1^6}{N^3} (\delta_{ii} \sum_{t} \pmb{b}_{tj}^2+\delta_{jj} \sum_{t} \pmb{b}_{it}^2+ 2 \pmb{b}_{ii} \pmb{b}_{jj}+2 \delta_{ij} \sum_{t} \pmb{b}_{it}\pmb{b}_{tj}) + \frac{J_1^8}{N^4} (\sum_{t} \pmb{b}_{it} \pmb{b}_{tj})^2    \Big)\nonumber \\
    &=&\frac{J_1^4}{N}\Big(\pmb{b}_{ii}+\pmb{b}_{jj}+\delta_{ij}\pmb{b}_{ij} - \frac{J_1^2}{2N} (\delta_{ii} \sum_{t} \pmb{b}_{tj}^2+\delta_{jj} \sum_{t} \pmb{b}_{it}^2+ 2 \pmb{b}_{ii} \pmb{b}_{jj}+2 (\delta_{ij}-\pmb{b}_{ij}) \sum_{t} \pmb{b}_{it}\pmb{b}_{tj})\nonumber \\
    & &\ \ \ \ \ \ \ \ \ \ \ \ \ \ \  -\frac{J_1^4}{N^2}(\sum_{t} \pmb{b}_{it} \pmb{b}_{tj})^2  \Big)        \nonumber \\
\end{eqnarray}

Which is the basis of the calculation. Since the symmetric disorder case is the limit of the asymmetric disorder for which $C_{ij}=0$ and $Q_{ij}=\frac{J_1^2}{2 N}$, the calculation above is a reference to check that the the asymmetric calculation is correct in those limits.

\section{Asymmetric disorder case}
Let us now focus on the harder case.
\begin{eqnarray}
    \frac{\langle \Omega_{ij}\rangle}{\langle 1\rangle}&=&\frac{\partial_{\pmb{b}_{ij}} \langle 1\rangle}{\langle 1\rangle}=(C_{ij}+2 Q_{ij} \pmb{b}_{ij}) \nonumber \\
    \frac{\langle \Omega_{ij}\Omega_{kt}\rangle}{\langle 1\rangle}&=&2 Q_{ij} \delta_{ik} \delta_{tj} +(C_{ij}+2 Q_{ij}\pmb{b}_{ij})(C_{kt}+2 Q_{kt} \pmb{b}_{kt})\nonumber \\
    \frac{\langle \Omega_{ij} \Omega_{kt} \Omega_{pq}\rangle}{\langle 1 \rangle}&=& 2 Q_{ij} \delta_{ik} \delta_{tj}(C_{pq}+2 Q_{pq}\pmb{b}_{pq}) \nonumber \\
    &+&(2 Q_{ij} \delta_{ip} \delta_{qj})(C_{kt}+2 Q_{kt} \pmb{b}_{kt})+(C_{ij}+2 Q_{ij} \pmb{b}_{ij})2 Q_{kt} \delta_{kp}\delta_{tq} \nonumber \\
    &+&(C_{ij}+2 Q_{ij} \pmb{b}_{ij})(C_{kt}+2 Q_{kt} \pmb{b}_{kt})(C_{pq}+2 Q_{pq} \pmb{b}_{pq}) \nonumber \\
    \frac{\langle \Omega_{ij} \Omega_{kt} \Omega_{pq}\Omega_{rs} \rangle}{\langle 1 \rangle}&=&2 Q_{ij} \delta_{ik} \delta_{tj} 2 Q_{pq} \delta_{rp} \delta_{qs} \nonumber \\
        &+& 2 Q_{ij} \delta_{ip} \delta_{qj} 2 Q_{kt} \delta_{kr}\delta_{ts} \nonumber \\
    &+& 2 Q_{ij} \delta_{ir}\delta_{js}2 Q_{kt} \delta_{kp} \delta_{tq}\nonumber \\
    &+&2 Q_{ij} \delta_{ik}\delta_{tj}(C_{pq}+2 Q_{pq} \pmb{b}_{pq}) (C_{rs}+2 Q_{rs} \pmb{b}_{rs})\nonumber \\
    &+& 2 Q_{ij} \delta_{ip} \delta_{qj} (C_{kt}+2 Q_{kt}\pmb{b}_{kt})(C_{rs}+2 Q_{rs} \pmb{b}_{rs})\nonumber \\
    &+&(C_{ij}+2 Q_{ij} \pmb{b}_{ij})2 Q_{kt} \delta_{kp}\delta_{tq}(C_{rs}+2 Q_{rs} \pmb{b}_{rs})\nonumber \\
    &+&2 Q_{ij} \delta_{ir} \delta_{js} (C_{kt}+2 Q_{kt} \pmb{b}_{kt})(C_{pq}+2 Q_{pq} \pmb{b}_{pq}) \nonumber \\
    &+&(C_{ij}+2 Q_{ij} \pmb{b}_{ij} ) 2 Q_{kt} \delta_{kr}\delta_{ts}(C_{pq}+2 Q_{pq }\pmb{b}_{pq}) \nonumber \\
    &+&(C_{ij}+2 Q_{ij} \pmb{b}_{ij})(C_{kt}+2 Q_{kt} \pmb{b}_{kt})2 Q_{pq} \delta_{pr} \delta_{qs} \nonumber \\
    &+& (C_{ij}+2 Q_{ij} \pmb{b}_{ij})(C_{kt}+2 Q_{kt} \pmb{b}_{kt})(C_{pq}+2 Q_{pq }\pmb{b}_{pq})(C_{rs}+2 Q_{rs }\pmb{b}_{rs})
\end{eqnarray}

where $C_{ij}=\delta_{ij} \mu_1$ and $Q_{ij}=\delta_{ij} \frac{\rho_1^2}{2}+(1-\delta_{ij}) \frac{J_1^2}{2 (N-1)}=\frac{1}{2} \delta_{ij}(\rho_1^2-\frac{J_1^2}{N-1})+\frac{J^2_1}{2N}$.

\begin{eqnarray}
    \frac{F^{3}_{ij}}{\gamma N}=\sum_{t}\langle \Omega_{ij} \Omega_{it} \Omega_{tj}\rangle&=&\sum_{t}\Big( 2 Q_{ij} \delta_{ii} \delta_{tj}(C_{tj}+2 Q_{tj}\pmb{b}_{tj})+(2 Q_{ij} \delta_{it} \delta_{jj})(C_{it}+2 Q_{it} \pmb{b}_{it})+2 Q_{it} \delta_{it}\delta_{tj}(C_{it}+2 Q_{ij} \pmb{b}_{ij}) \nonumber \\
    &+&(C_{ij}+2 Q_{ij} \pmb{b}_{ij})(C_{it}+2 Q_{it} \pmb{b}_{it})(C_{tj}+2 Q_{tj} \pmb{b}_{tj})\Big)\nonumber \\
 \text{up to constant terms}  &=& \sum_{t} \Big( 2 Q_{ij} \delta_{ii} \delta_{tj}2 Q_{tj}\pmb{b}_{tj}+2 Q_{ij} \delta_{it} \delta_{jj}2 Q_{it} \pmb{b}_{it}+2 Q_{it} \delta_{it}\delta_{tj}2 Q_{ij} \pmb{b}_{ij} \nonumber \\
 &+& 2 Q_{ij} \pmb{b}_{ij} C_{it} C_{tj}+ C_{ij} C_{it} 2 Q_{tj} \pmb{b}_{tj}+C_{ij} 2 Q_{it} \pmb{b}_{it} C_{tj} \nonumber \\
 &+& C_{ij} 2 Q_{it} \pmb{b}_{it} 2 Q_{tj} \pmb{b}_{tj}+2 Q_{ij} \pmb{b}_{ij} C_{it} 2 Q_{tj} \pmb{b}_{tj}\nonumber \\
 &+& 2 Q_{ij} \pmb{b}_{ij} 2 Q_{it}\pmb{b}_{it} C_{tj}+ 2 Q_{ij} \pmb{b}_{ij} 2 Q_{it} \pmb{b}_{it} 2 Q_{tj} \pmb{b}_{tj}\Big) \nonumber \\
 \end{eqnarray}

And after a lengthy but easy calculation, we obtain the final result:
\begin{eqnarray}
  \frac{F^{3}_{ij}}{\gamma N}&=&   \rho_1^2 \big(\rho_1^2 \delta_{ij}+(1-\delta_{ij}) \frac{J_1^2}{N-1}\big) (\pmb{b}_{jj}+\pmb{b}_{ii}) \nonumber \\
 &+& \rho_1^2(3\mu_1^2+\rho_1^2)\delta_{ij} \pmb{b}_{ij}+3 \mu_1 \rho_1^2 \delta_{ij} \pmb{b}_{ij}^2  \nonumber \\
&+&   \mu_1  \frac{J_1^4}{(N-1)^2}\sum_{t}\big(\delta_{ij}(1-\delta_{it})(1-\delta_{jt}) \pmb{b}_{it}\pmb{b}_{tj}+(1-\delta_{tj})\delta_{it}   \pmb{b}_{tj}^2+
 (1-\delta_{it})\delta_{tj}  \pmb{b}_{it}^2\big)
  \nonumber \\
&+& \big(\rho_1^2 \delta_{ij}+(1-\delta_{ij}) \frac{J_1^2}{N-1}\big) \pmb{b}_{ij} \\
&\cdot& \Big(\sigma_1^4 \delta_{ij} \pmb{b}_{ii}\pmb{b}_{jj}
+ (1-\delta_{ij}) \frac{\rho_1^2J_1^2}{N}(\pmb{b}_{ij}\pmb{b}_{jj}+\pmb{b}_{ii}\pmb{b}_{ij})+\sum_t \frac{J_1^4}{(N-1)^2}(1-\delta_{it})(1-\delta_{tj})\pmb{b}_{it}\pmb{b}_{tj} \Big) 
\nonumber \\  
\end{eqnarray}

 For the case $\rho_1^2=\frac{J_1^2}{N-1}$, but $\mu_1\neq 0$, we have
 \begin{eqnarray}
    \frac{F^{3}_{ij}}{\gamma N} &=&  \frac{J_1^4}{(N-1)^2} (\pmb{b}_{jj}+ \pmb{b}_{ii}+)+\delta_{ij}\Big(\frac{J_1^4}{(N-1)^2}+3\frac{J_1^2\mu_1^2}{N-1}\Big)  \pmb{b}_{ij} \nonumber \\
 &+& \frac{\mu_1J_1^4}{(N-1)^2} \Big(\delta_{ij} \sum_t \pmb{b}_{it}  \pmb{b}_{tj}+2 \pmb{b}_{ij}^2 \Big)  \nonumber \\
 &+& \frac{J_1^6}{(N-1)^3} \pmb{b}_{ij} \sum_t \pmb{b}_{it}  \pmb{b}_{tj} \nonumber \\
 \end{eqnarray}
In order to double check that the result is correct, let us assume that $C_{ij}=0$, and $Q_{ij}=\frac{J_1^2}{2(N-1)}$.
Then we have
\begin{eqnarray}
   \frac{F_{ij}^3}{\gamma N}= \sum_{t}\langle \Omega_{ij} \Omega_{it} \Omega_{tj}\rangle&=&\sum_{t}\Big(  \frac{J_1^4}{(N-1)^2}( \delta_{ii} \delta_{tj} \pmb{b}_{tj}+ \delta_{it} \delta_{jj} \pmb{b}_{it}+ \delta_{it}\delta_{tj} \pmb{b}_{ij}) \nonumber \\
    &+&\frac{J_1^6}{N^3} \pmb{b}_{ij}\pmb{b}_{it} \pmb{b}_{tj}\Big) \nonumber \\
    &=&\frac{J_1^4}{(N-1)^2}(\pmb{b}_{ii}+\pmb{b}_{jj}+\pmb{b}_{ij})+\frac{J_1^6}{N^3} \pmb{b}_{ij}\sum_t \pmb{b}_{it} \pmb{b}_{tj}
\end{eqnarray}
which is exactly the result we had in the symmetric case. For the fourth order term we have
\begin{eqnarray}
     -\frac{ 2 F^4_{ij}}{\gamma N}=\sum_{t,t^\prime}\langle \Omega_{it} \Omega_{tj} \Omega_{it^\prime}\Omega_{t^\prime j} \rangle
&=&  \sum_{t,t^\prime}\Big(2 Q_{it} \delta_{it} \delta_{jt} 2 Q_{i t^\prime} \delta_{t^\prime i} \delta_{t^\prime j}\nonumber \\
& &\ \ \ \ \ + 2 Q_{it} \delta_{ii} \delta_{t^\prime t} 2 Q_{tj} \delta_{tt^\prime}\delta_{jj} \nonumber \\
& &\ \ \ \ \ + 2 Q_{it} \delta_{it^\prime} \delta_{tj} 2 Q_{tj} \delta_{ti} \delta_{jt^\prime} \nonumber \\
& &\ \ \ \ \  + (C_{it^\prime}+2 Q_{it^\prime} \pmb{b}_{it^\prime})(C_{it}+2 Q_{it} \pmb{b}_{it}) 2 Q_{tj} \delta_{t^\prime j} \delta_{t^\prime t} \delta_{jj} \nonumber \\
& &\ \ \ \ \ +(C_{it^\prime}+2 Q_{it^\prime} \pmb{b}_{it^\prime})2 Q_{it} \delta_{i t^\prime} \delta_{tj} (C_{tj}+2 Q_{tj} \pmb{b}_{tj}) \nonumber \\
& &\ \ \ \ \ + 2 Q_{it^\prime} \delta_{it^\prime} \delta_{t^\prime j} (C_{it}+2Q_{it}\pmb{b}_{it})(C_{tj}+2Q_{tj}\pmb{b}_{tj}) \nonumber \\
& &\ \ \ \ \ + (C_{t^\prime j}+2 Q_{t^\prime j}\pmb{b}_{t^\prime j})(C_{it}+2 Q_{it}\pmb{b}_{it}) 2 Q_{tj} \delta_{ti}\delta_{jt^\prime} \nonumber \\
& &\ \ \ \ \ +(C_{t^\prime j}+2 Q_{t^\prime j}\pmb{b}_{t^\prime j})2 Q_{it} \delta_{ii} \delta_{t t^\prime} (C_{t j}+2 Q_{tj}\pmb{b}_{t j}) \nonumber \\
& &\ \ \ \ \ +(C_{t^\prime j}+2 Q_{t^\prime j}\pmb{b}_{t^\prime j})(C_{i t^\prime}+2 Q_{i t^\prime}\pmb{b}_{i t^\prime})2 Q_{it} \delta_{it}\delta_{tj}\nonumber \\
& &\ \ \ \ \ +(C_{t^\prime j}+2 Q_{t^\prime j}\pmb{b}_{t^\prime j})(C_{i t^\prime}+2 Q_{i t^\prime}\pmb{b}_{i t^\prime})(C_{i t}+2 Q_{i t}\pmb{b}_{i t})(C_{t j}+2 Q_{t j}\pmb{b}_{t j})\Big) \nonumber \\
\end{eqnarray}

After a lengthy calculation, we find:

\begin{eqnarray}
-\frac{2F_{ij}^4}{\gamma N}&=&\Big(\ \ \ 
  4\mu_1 \rho_1^2   \big(\rho_1^2 \delta_{ij}+(1-\delta_{ij}) \frac{J_1^2}{N-1}\big)  (\pmb{b}_{ii}+\pmb{b}_{jj}) \nonumber \\
   & &\ \ \ \ \ +2\sigma_1^4  \big(\rho_1^2 \delta_{ij}+(1-\delta_{ij}) \frac{J_1^2}{N-1}\big)  \pmb{b}_{ii}\pmb{b}_{jj} \nonumber \\
& &\ \ \ \ \ +4(\mu_1 \sigma_1^4+\mu_1^3 \rho_1^2)\delta_{ij}   \pmb{b}_{ij}\nonumber \\
& &\ \ \ \ \ +\sum_t \big(\rho_1^2 \delta_{it}+(1-\delta_{it}) \frac{J_1^2}{N-1}\big)  \big(\sigma_1^4 \delta_{tj}+(1-\delta_{tj}) \frac{J_1^4}{(N-1)^2}\big)(\pmb{b}_{it}^2+\pmb{b}_{t j}^2) \nonumber \\
& &\ \ \ \ \ + \sum_t (2\mu_1^2+2\rho_1^2) \delta_{ij}\big(\sigma_1^4 \delta_{t j}+(1-\delta_{t j}) \frac{J_1^4}{(N-1)^2}\big)\pmb{b}_{i t}\pmb{b}_{t j} \nonumber \\
& &\ \ \ \ \ +4\mu_1^2\big(\sigma_1^4 \delta_{i j}+(1-\delta_{i j}) \frac{J_1^4}{(N-1)^2}\big) \pmb{b}_{i j}^2\nonumber \\
& &\ \ \ \ \ +\sum_{t^\prime} \mu_1\big(\rho_1^2 \delta_{it^\prime}+(1-\delta_{it^\prime}) \frac{J_1^2}{N-1}\big)\big(\rho_1^2 \delta_{t^\prime j}+(1-\delta_{t^\prime j}) \frac{J_1^2}{N-1}\big)\pmb{b}_{i t^\prime}\pmb{b}_{t^\prime j} \nonumber \\
& &\ \ \ \ \ +\sum_t 3\mu_1 \big(\rho_1^2 \delta_{ij}+(1-\delta_{ij}) \frac{J_1^2}{N-1}\big)\pmb{b}_{i j}\big(\rho_1^2 \delta_{it}+(1-\delta_{it}) \frac{J_1^2}{N-1}\big)\big(\rho_1^2 \delta_{t j}+(1-\delta_{t j}) \frac{J_1^2}{N-1}\big)\pmb{b}_{i t}\pmb{b}_{t j}\nonumber \\
& &\ \ \ \ \  +\sum_{t^\prime t}\big(\rho_1^2 \delta_{it^\prime}+(1-\delta_{it^\prime}) \frac{J_1^2}{N-1}\big)\big(\rho_1^2 \delta_{t^\prime j}+(1-\delta_{t^\prime j}) \frac{J_1^2}{N-1}\big)\pmb{b}_{i t^\prime}\pmb{b}_{t^\prime j} \nonumber \\
& &\ \ \ \ \ \ \ \ \ \ \cdot \big(\rho_1^2 \delta_{tj}+(1-\delta_{tj})\big(\rho_1^2 \delta_{it}+(1-\delta_{it}) \frac{J_1^2}{N-1}\big) \pmb{b}_{i t}\pmb{b}_{t j}\Big)
\end{eqnarray}
In the case in which $\rho_1^2=\frac{J_1^2}{N-1}$, but $\mu_1\neq 0$, we have

\begin{eqnarray} 
-\frac{2F^4_{ij}}{\gamma N}&=&
 4\mu_1 \frac{J_1^4}{(N-1)^2} (\pmb{b}_{ii}+\pmb{b}_{j j})  +  \frac{J_1^6}{(N-1)^3} \sum_t(\pmb{b}_{it}^2 +\pmb{b}_{t j}^2)  +2\frac{J_1^6}{(N-1)^3} \pmb{b}_{ii}    \pmb{b}_{jj} \nonumber \\
& &\ \ \ \ \  +4\Big(\mu_1\frac{J_1^4}{(N-1)^2}+\mu_1^3\frac{J_1^2}{N-1}\Big) \delta_{ij}  \pmb{b}_{ij}+2\Big(\frac{J_1^6}{(N-1)^3}+\mu_1^2\frac{J_1^4}{(N-1)^2}\Big) \delta_{ij} \sum_t \pmb{b}_{it}\pmb{b}_{tj}\nonumber \\  
     & &\ \ \ \ \ +4\mu_1^2\frac{J_1^4}{(N-1)^2}\pmb{b}_{i j}^2+4\mu_1\frac{J_1^6}{(N-1)^3}\pmb{b}_{i j} \sum_t \pmb{b}_{i t}\pmb{b}_{t j}\nonumber \\ 
& &\ \ \ \ \ +\frac{J_1^8}{(N-1)^4}\sum_t \sum_{t^\prime}\pmb{b}_{t^\prime j}\pmb{b}_{i t^\prime}\pmb{b}_{i t}\pmb{b}_{t j}
\end{eqnarray}

 Below we double checked that the results are consistent with the symmetric ones. If we now set $C_{ij}=0$ and $Q_{ij}=\frac{J_1^2}{2(N-1)}$ we obtain:
\begin{eqnarray}
-\frac{2F_{ij}^4}{\gamma N}&=&2 \frac{J_1^6}{(N-1)^3}  \pmb{b}_{ii}\pmb{b}_{jj}  +\frac{J_1^6}{(N-1)^3}(\pmb{b}_{it}^2+\pmb{b}_{t j}^2)  + 2\frac{J_1^6}{(N-1)^3} \delta_{ij}\sum_t \pmb{b}_{i t}\pmb{b}_{t j}  +\frac{J_1^8}{(N-1)^4}(\sum_t \pmb{b}_{i t}\pmb{b}_{t j})^2
\end{eqnarray}
which is exactly the result obtained for the symmetric case.
The correction due to the disorder is thus
\begin{equation}
    F_{ij}=\gamma N(F_{ij}^3-\frac{1}{2} F_{ij}^4)
\end{equation}
and for $\sigma_1=\frac{J_1^2}{N-1}$ we have
 \begin{eqnarray}
    \frac{F_{ij}}{\gamma N} &=&  \frac{J_1^4}{(N-1)^2}(1-2\mu_1) (\pmb{b}_{jj}+ \pmb{b}_{ii}) \nonumber \\
    &+&\Big(\frac{J_1^4}{(N-1)^2}+3\frac{J_1^2\mu_1^2}{N-1}-2\mu_1\frac{J_1^4}{(N-1)^2}-2\mu_1^3\frac{J_1^2}{N-1}-\frac{J_1^6}{(N-1)^3}-\mu_1^2\frac{J_1^4}{(N-1)^2}\Big) \delta_{ij} \pmb{b}_{ij} \nonumber \\
 &+&  \frac{\mu_1J_1^4}{(N-1)^2}\delta_{ij} \sum_t \pmb{b}_{it}  \pmb{b}_{tj}+2\frac{J_1^4}{(N-1)^2}(\mu_1 -\mu_1^2)\pmb{b}_{ij}^2   \nonumber \\
 &+& \frac{J_1^6}{(N-1)^3}(1-2\mu_1) \pmb{b}_{ij} \sum_t \pmb{b}_{it}  \pmb{b}_{tj} \nonumber \\
&&
   -  \frac{J_1^6}{2(N-1)^3} \sum_t(\pmb{b}_{it}^2 +\pmb{b}_{t j}^2)  -\frac{J_1^6}{(N-1)^3} \pmb{b}_{ii}    \pmb{b}_{jj} \nonumber \\
& &\ \ \ \ \ -\frac{1}{2}\frac{J_1^8}{(N-1)^4}\sum_t \sum_{t^\prime}\pmb{b}_{t^\prime j}\pmb{b}_{i t^\prime}\pmb{b}_{i t}\pmb{b}_{t j}
\end{eqnarray}
Each term has been calculated separately. We have
 
 \begin{eqnarray}
    \sum_{ij}\frac{F_{ij}}{\gamma N} &=&  \frac{J_1^4}{(N-1)^2}(1-2\mu_1 ) 2( N^2 n \frac{\xi}{2}-N^2 \sum_\alpha m^\alpha) \nonumber \\
    &+&\Big(\frac{J_1^4}{(N-1)^2}+3\frac{J_1^2\mu_1^2}{N-1}-2\mu_1\frac{J_1^4}{(N-1)^2}-2\mu_1^3\frac{J_1^2}{N-1}-\frac{J_1^6}{(N-1)^3}-\mu_1^2\frac{J_1^4}{(N-1)^2}\Big) (N \sum_\alpha (\frac{\xi}{2} -  m^\alpha t )) \nonumber \\
 &+&  \frac{\mu_1J_1^4}{(N-1)^2}(N^2 \sum_{\alpha \beta} \Big( \frac{\xi}{4}  (q^{\alpha \beta})^2  + t^2 m^\alpha  m^\beta  +\xi t   q^{\alpha \beta} m^{\beta}\Big))\nonumber \\
 &+&2\frac{J_1^4}{(N-1)^2}(\mu_1 -\mu_1^2)(N^2 \sum_{\alpha \beta}\left(\frac{\xi^2}{4} (q^{\alpha \beta})^2+ t^2 q^{\alpha \beta}-\frac{\xi t }{2} q^{\alpha \beta} ( m^\beta+m^\alpha)  \right))   \nonumber \\
 &+& \frac{J_1^6}{(N-1)^3}(1-2\mu_1) (N^3 \sum_{\alpha \beta \gamma}  q^{\alpha \beta} \Big( \frac{1}{8} \xi ^3  q^{ \beta\gamma }  q^{ \gamma\alpha }    +  q^{\beta \gamma} (\frac{\xi }{2}   t^2-\frac{3t}{4} m^\gamma)+\frac{t^2}{2} \xi  
   m^\alpha m^\gamma  +\frac{1}{2} \xi t^2   (m^\gamma)^2 -t^3 
   m^\gamma \Big)) \nonumber \\
&& -  \frac{J_1^6}{2(N-1)^3} 2N^3 \sum_{\alpha \beta} q^{\alpha\beta}(\frac{\xi^2 }{4}q^{\alpha\beta} +  t^2-  t \xi  m^\alpha )   \nonumber \\
   &-&\frac{J_1^6}{(N-1)^3}\Big(\frac{N^2 n^2 \xi^2}{4}-2 \frac{N^2 n\xi}{2} t \sum_{\alpha}  m^\alpha + N^2 t^2 (\sum_{\alpha }  m^\alpha)^2\Big) \nonumber \\
& &-\frac{1}{2}\frac{J_1^8}{(N-1)^4}\Big(N^4\sum_{\alpha \beta \gamma \delta} (\frac{1}{16} \xi ^4 q^{\alpha \gamma } q^{\beta \delta } q^{\gamma \delta
   } q^{\alpha \beta }-\frac{1}{2} \xi  t^3 m^{\beta } m^{\delta }
   m^{\beta } q^{\alpha \gamma }-\frac{1}{2} \xi  t^3 m^{\delta }
   m^{\delta } m^{\beta } q^{\alpha \gamma } \nonumber \\
   & & \ \ \ \ \ \ \ \ +\frac{1}{4} \xi ^2 t^2
   m^{\beta } m^{\beta } q^{\alpha \gamma } q^{\gamma \delta
   }+\frac{1}{4} \xi ^2 t^2 m^{\beta } m^{\delta } q^{\alpha \gamma }
   q^{\alpha \beta }+\frac{1}{4} \xi ^2 t^2 m^{\delta } m^{\beta }
   q^{\alpha \gamma } q^{\gamma \delta }+\frac{1}{4} \xi ^2 t^2 m^{\delta
   } m^{\delta } q^{\alpha \gamma } q^{\alpha \beta }\nonumber \\
   & & \ \ \ \ \ \ \ \ +\frac{1}{4} \xi ^2
   t^2 m^{\delta } m^{\beta } q^{\alpha \gamma } q^{\beta \delta
   }-\frac{1}{8} \xi ^3 t m^{\beta } q^{\alpha \gamma } q^{\gamma \delta }
   q^{\alpha \beta }-\frac{1}{8} \xi ^3 t m^{\delta } q^{\alpha \gamma }
   q^{\gamma \delta } q^{\alpha \beta }-\frac{1}{8} \xi ^3 t m^{\beta }
   q^{\alpha \gamma } q^{\beta \delta } q^{\gamma \delta }\nonumber \\
   & & \ \ \ \ \ \ \ \ -\frac{1}{8}
   \xi ^3 t m^{\delta } q^{\alpha \gamma } q^{\beta \delta } q^{\alpha
   \beta }+t^4 m^{\delta } m^{\beta } q^{\alpha \gamma } \nonumber \\
   & & \ \ \ \ \ \ \ \ -\frac{1}{2} \xi 
   t^3 m^{\beta } q^{\alpha \gamma } q^{\gamma \delta }-\frac{1}{2} \xi 
   t^3 m^{\delta } q^{\alpha \gamma } q^{\alpha \beta }+\frac{1}{4} \xi ^2
   t^2 q^{\alpha \gamma } q^{\gamma \delta } q^{\alpha \beta }) \Big)
\end{eqnarray}
In the replica symmetric ansatz (RSA), we will assume that $m^\alpha= m$, and $q^{\alpha \beta}=q$ for $\alpha\neq \beta$ and 0 otherwise. We define 
$z_n^k=\text{Tr}(\Omega_n-I)^k$,  $k_n^p=\vec 1_n^t (\Omega_n-I)^p \vec 1_n$ and $\Omega_n$ is the $n\times n$ matrix with ones everywhere. The matrix $\Omega_n$ has eigenvalues $n-1$ with multiplicity $1$, and $-1$ with multiplicity $n-1$. Thus, $z_n^k=(n-1)^k+(-1)^{k} (n-1)=(n-1)\left((n-1)^{k-1}+(-1)^k \right)$. It is not hard to see that instead $k_n^p=n(n-1)^p$.
This implies that $\lim_{n\rightarrow0} \frac{k_n^p}{n} =(-1)^p=e_p$, while $lim_{n\rightarrow0} \frac{z_n^p}{n} =c_p$, and we have the first coefficients $c_1=0$, $c_2=-1$, $c_3=2$ and $c_4=-3$. For $N\rightarrow \infty$ we get
\begin{eqnarray}
    \sum_{ij}\frac{F_{ij}}{\gamma N} &=&  -J_1^4(1-2\mu_1 ) 2 n m \nonumber \\
    &&-  nJ_1^2\mu_1^2(3-2\mu_1)  m t  \nonumber \\
 &+&  \mu_1 J_1^4 \Big( \frac{\xi}{4}  \frac{n(n-1)}{2}q^2  + t^2 n^2 m^2    +\xi t \frac{n(n-1)}{2}  q m\Big)\nonumber \\
 &+&2\mu_1 J_1^4(1 -\mu_1)\Big(\frac{\xi^2}{4} \frac{n(n-1)}{2}q^2+ \frac{n(n-1)}{2} t^2 q-\frac{\xi t }{2} q m n(n-1)  \Big)   \nonumber \\
 &+& J_1^6(1-2\mu_1)  
  \Big( \frac{1}{8} \xi ^3 q^3 z_n^3    +q^2 k_n^2 (\frac{\xi }{2}   t^2-\frac{3t}{4} m)+\frac{t^2}{2} \xi  
   q m^2 \frac{n(n-1)}{2}  +\frac{1}{2} \xi t^2  qm^2\frac{n(n-1)}{2} -t^3 
   q m nk_n^{2} \Big) \nonumber \\
&& -  J_1^6 (\frac{\xi^2 }{4}q^2\frac{n(n-1)}{2} + q t^2 \frac{n(n-1)}{2}-  t \xi  m n^2 )   \nonumber \\
& &-\frac{1}{2}J_1^8 (\frac{1}{16} \xi ^4 q^4 z_n^4-\frac{1}{2} \xi  t^3 m^3  q \frac{n^3(n-1)}{2}-\frac{1}{2} \xi  t^3 m^3  q \frac{n^3(n-1)}{2} \nonumber \\
   & & \ \ \ \ \ \ \ \ +\frac{1}{4} \xi ^2 t^2
   m^2 q^2 n k_n^2+\frac{1}{4} \xi ^2 t^2  m^2 q^2 n k_n^2+\frac{1}{4} \xi ^2 t^2  m^2 q^2 n k_n^2+\frac{1}{4} \xi ^2 t^2  m^2 q^2 n k_n^2\nonumber \\
   & & \ \ \ \ \ \ \ \ +\frac{1}{4} \xi ^2
   t^2 m^2 q^2 \frac{n^2(n-1)^2}{4}-\frac{1}{8} \xi ^3 t m q^3 k_n^3-\frac{1}{8} \xi ^3 t m q^3 k_n^3-\frac{1}{8} \xi ^3 t m q^3 k_n^3\nonumber \\
   & & \ \ \ \ \ \ \ \ -\frac{1}{8}
   \xi ^3 t m q^3 k_n^3+t^4 m^2 \frac{n^3(n-1)}{2} \nonumber \\
   & & \ \ \ \ \ \ \ \ -\frac{1}{2} \xi 
   t^3 mq^2 n k_n^2-\frac{1}{2} \xi 
   t^3 mq^2 n k_n^2+\frac{1}{4} \xi ^2
   t^2 q^3 k_n^3) 
\end{eqnarray}
and in the limit $n\rightarrow 0$, and adding the base Hamiltonian:
\begin{equation}
     \lim_{n\rightarrow 0}\lim_{N\rightarrow \infty} \frac{H_0}{n N}=\frac{J_1^2}{2}  (-\frac{\xi^2}{8} q^2+\xi t q m -\frac{ t^2}{2} q)+(h- \mu_1 t ) m
\end{equation}
\begin{eqnarray}
\lim_{n\rightarrow 0}\lim_{N\rightarrow \infty} \frac{H_0+\gamma \sum_{ij}F _{ij}}{n N} &=&   -\frac{J_1^2}{2}\frac{\xi^2}{8} q^2+\frac{J_1^2}{2}\xi t q m -\frac{J_1^2}{2}\frac{ t^2}{2} q+(h- \mu_1 t  m)\nonumber \\
&&\gamma\Big(-J_1^4(1-2\mu_1 ) 2  m -  J_1^2\mu_1^2(3-2\mu_1)t  m      -\mu_1 J_1^4\frac{\xi}{4}  \frac{1}{2}q^2    +\mu_1 J_1^4\xi t \frac{1}{2}  q m\nonumber \\
 &&-2\mu_1 J_1^4(1 -\mu_1)\frac{\xi^2}{4} \frac{1}{2}q^2- \frac{1}{2} t^22\mu_1 J_1^4(1 -\mu_1) q+\frac{\xi t }{2}2\mu_1 J_1^4(1 -\mu_1) q m      \nonumber \\
 &+&   
   \frac{1}{8}J_1^6(1-2\mu_1) \xi ^3c_3 q^3     + \frac{\xi }{2}   t^2J_1^6(1-2\mu_1)e_2q^2 -\frac{3t}{4}J_1^6(1-2\mu_1) e_2q^2 m \nonumber \\
&& -\frac{t^2}{4} J_1^6(1-2\mu_1)\xi 
   q m^2   -J_1^6(1-2\mu_1)\frac{1}{4} \xi t^2  qm^2  +J_1^6\frac{\xi^2 }{8}q^2 + J_1^6\frac{t^2}{2} q    \nonumber \\
& &-\frac{1}{16}\frac{1}{2}J_1^8 \xi ^4 c_4q^4   +\frac{1}{4}J_1^8 \xi ^3 t e_3 m q^3  -\frac{1}{8}J_1^8 \xi ^2t^2 e_3q^3\Big)  \nonumber \\
   &=&\left(h-\mu_1t+\gamma \big(-2J_1^4(1-2\mu_1 )-  J_1^2\mu_1^2(3-2\mu_1)t  \big)\right) m \nonumber \\
   &+&\gamma \Big(\frac{1}{4}J_1^8 \xi ^3 t e_3 \Big) q^3 m +\gamma \Big(-\frac{3t}{4}J_1^6(1-2\mu_1) e_2 \Big) q^2m \nonumber \\
   &+&\left(\frac{J_1^2}{2}\xi t+ \gamma\big(+\mu_1 J_1^4\xi t \frac{1}{2}+\xi t \mu_1 J_1^4(1 -\mu_1) \big)\right) qm \nonumber \\
   &+&\gamma \Big(-\frac{t^2}{4} J_1^6(1-2\mu_1)\xi-J_1^6(1-2\mu_1)\frac{1}{4} \xi t^2 \Big) qm^2 \nonumber \\
   &+&\left(-\frac{J_1^2}{2}\frac{ t^2}{2}+\gamma \big(- t^2\mu_1 J_1^4(1 -\mu_1)+J_1^6\frac{t^2}{2} \big) \right)q \nonumber \\
   &+&\left(-\frac{J_1^2}{2}\frac{\xi^2}{8}+\gamma \big(-\mu_1 J_1^4\frac{\xi}{4}  \frac{1}{2} -\mu_1 J_1^4(1 -\mu_1)\frac{\xi^2}{4}+\frac{\xi }{2}   t^2J_1^6(1-2\mu_1)e_2 \big)\right) q^2 \nonumber \\
   &+&\gamma \Big(J_1^6\frac{\xi^2 }{8}+\frac{1}{8}J_1^6(1-2\mu_1) \xi ^3c_3 \Big) q^3 \nonumber \\
   &+&\gamma \Big(-\frac{1}{16}\frac{1}{2}J_1^8 \xi ^4 c_4 -\frac{1}{8}J_1^8 \xi ^2
   t^2 e_3\Big) q^4 \nonumber \\
   &=&\sum_{ij=0} a_{ij} q^i m^j\equiv \tilde H_{RS}(q,m)
\end{eqnarray}

If we use $t=\xi$ and consider that $h=\frac{2}{\alpha \beta} x-8/3$, and substitute the numerical values $e_p=(-1)^p$ and $c_3=2$ and $c_4=-3$,
\begin{eqnarray}
\tilde H_{RS}(q,m)&=&\left(h-\mu_1\xi+\gamma \big(-2J_1^4(1-2\mu_1 )-  J_1^2\xi\mu_1^2(3-2\mu_1)  \big)\right) m \nonumber \\
   &-&\gamma \Big(\frac{1}{4}J_1^8 \xi ^4  \Big) q^3 m +\gamma \Big(-\frac{3\xi}{4}J_1^6(1-2\mu_1)  \Big) q^2m \nonumber \\
   &+&\left(\frac{J_1^2}{2}\xi^2+ \gamma\xi^2 \mu_1 J_1^4\big(\frac{3}{2} -\mu_1\big)\right) qm \nonumber \\
   &-&\gamma \Big(\frac{\xi^3}{2} J_1^6(1-2\mu_1) \Big) qm^2 \nonumber \\
   &+&\left(-\frac{J_1^2}{4} \xi^2+\gamma J_1^4\xi^2\big(- \mu_1 (1 -\mu_1)+\frac{J_1^2}{2} \big) \right)q \nonumber \\
   &+&\left(-\frac{J_1^2}{2}\frac{\xi^2}{8}+\gamma J_1^4\xi \Big(- \frac{\mu_1}{8}   -\mu_1 (1 -\mu_1)\frac{\xi}{4}+\frac{\xi^2 }{2}   J_1^2(1-2\mu_1) \Big)\right) q^2 \nonumber \\
   &+&\gamma J_1^6\frac{\xi^2 }{8}\Big(1+2(1-2\mu_1) \xi  \Big) q^3 \nonumber \\
   &+&\frac{ 7 \gamma J_1^8}{32} \xi ^4   q^4 \nonumber \\
\end{eqnarray}

and for $\mu_1=0$ we have a great deal of simplifications, namely
\begin{eqnarray}
\tilde H_{RS}(q,m)&=&\left(h-2\gamma J_1^4  \right) m -\gamma \Big(\frac{1}{4}J_1^8 \xi ^4  \Big) q^3 m +\gamma \Big(-\frac{3t}{4}J_1^6  \Big) q^2m +\left(\frac{J_1^2}{2}\xi^2 \big)\right) qm \nonumber \\
   &-&\gamma \Big(\frac{\xi^3}{2} J_1^6 \Big) qm^2 +\left(-\frac{J_1^2}{2}\frac{ \xi^2}{2}+\gamma \big(+J_1^6\frac{t^2}{2} \big) \right)q +\left(-\frac{J_1^2}{2}\frac{\xi^2}{8}+\gamma J_1^4\xi\big(   +\frac{\xi^2 }{2}   J_1^2 \big)\right) q^2 \nonumber \\
   &+&\gamma J_1^6\frac{\xi^2 }{8}\Big(1+2 \xi  \Big) q^3 +\frac{ 7 \gamma J_1^8}{32} \xi ^4   q^4 \nonumber \\
\end{eqnarray}

\section{Converting correction to mean field parameters}
First we define
\begin{eqnarray}
    m^\alpha&=&\frac{1}{N}\sum_{i} \sigma_i^\alpha\\
        q^{\alpha \beta}&=&\frac{1}{N}\sum_{i} \sigma_i^\alpha \sigma_i^\beta\\
    \rho^{\alpha \beta \gamma}&=&\frac{1}{N}\sum_{i} \sigma_i^\alpha \sigma_i^\beta \sigma_i^\gamma\\
    \eta^{\alpha \beta \gamma \delta}&=&\frac{1}{N}\sum_{i} \sigma_i^\alpha \sigma_i^\beta \sigma_i^\gamma \sigma_i^\delta
\end{eqnarray}

From which we have that, using the fact that $\gamma$ is small, that

\begin{equation}
        \langle e^{\sum_{ij} b_{ij} \Omega_{ij}}\rangle_{P(\Omega_{ij})}\approx \mathcal N \prod_{i,j} e^{C_{ij} b_{ij}+Q_{ij} b_{ij}^2}e^{\gamma C_{ij}} 
\end{equation}
where $C_{ij}=\delta_{ij} \mu_1$ and $Q_{ij}=\delta_{ij} \frac{\rho_1^2}{2}+(1-\delta_{ij}) \frac{J_1^2}{2 (N-1)}$.
We are now interested in the result of the calcuation of the previous terms. We now use $\pmb{b}_{ij}=\sum_{\alpha}( \frac{\xi}{2} \sigma_i^\alpha \sigma_j^\alpha-\sigma_i^\alpha t_j)$.  Also, we define $t=\frac{1}{N} \sum_j t_j$, $\sum_i \sigma_i^\alpha = N m^\alpha$, and $\sum_i \sigma_i^\alpha \sigma_i^\beta=N q^{\alpha \beta}$.
\subsection{Base Hamiltonian}
The first terms we have the $\gamma$ independent terms, which are
\begin{eqnarray}
   \sum_{ij} C_{ij} \pmb{b}_{ij}&=& \mu_1  \sum_{\alpha } \sum_{ij}  \delta_{ij} (\frac{\xi}{2} \sigma_i^\alpha \sigma_j^\alpha- \sigma_i^\alpha t_j)\nonumber \\
   &=&\mu_1\sum_{\alpha } \sum_{i} (\frac{\xi}{2} - \sigma_i^\alpha t_j)\nonumber \\
   \text{mft} &\rightarrow &-N \mu_1 t \sum_{\alpha} m^\alpha
\end{eqnarray}
And we also have 
\begin{eqnarray}
   \sum_{ij} Q_{ij} \pmb{b}_{ij}^2&=&   \sum_{\alpha \beta } \sum_{ij}  \Big(\delta_{ij} \frac{\rho_1^2}{2}+(1-\delta_{ij}) \frac{J_1^2}{2 (N-1)}\Big) (\frac{\xi}{2} \sigma_i^\alpha \sigma_j^\alpha- \sigma_i^\alpha t_j)(\frac{\xi}{2} \sigma_i^\beta \sigma_j^\beta- \sigma_i^\beta t_j)\nonumber \\
   &=&\sum_{\alpha \beta } \sum_{ij}  \Big(\delta_{ij} (\frac{\rho_1^2}{2}-\frac{J_1^2}{2 (N-1)})+ \frac{J_1^2}{2 (N-1)}\Big) (\frac{\xi^2}{4} \sigma_i^\alpha \sigma_j^\alpha \sigma_i^\beta \sigma_j^\beta-\frac{\xi}{2} \sigma_i^\alpha \sigma_i^\beta \sigma_j^\beta- \frac{\xi}{2} \sigma_i^\beta \sigma_i^\alpha \sigma_j^\alpha + t_j^2 \sigma_i^\alpha \sigma_i^\beta  )\nonumber \\
   &=&(\frac{\rho_1^2}{2}-\frac{J_1^2}{2 (N-1)})\sum_{\alpha \beta } \sum_{i}    (\frac{\xi^2}{4} -\frac{\xi t}{2} \sigma_i^\alpha - \frac{\xi t}{2} \sigma_i^\beta  + t_j^2 \sigma_i^\alpha \sigma_i^\beta  )\nonumber \\
   &+& \frac{J_1^2}{2 (N-1)}\sum_{\alpha \beta } \sum_{ij}(\frac{\xi^2}{4} \sigma_i^\alpha \sigma_j^\alpha \sigma_i^\beta \sigma_j^\beta-\frac{\xi t}{2} \sigma_i^\alpha \sigma_i^\beta \sigma_j^\beta- \frac{\xi t}{2} \sigma_i^\beta \sigma_i^\alpha \sigma_j^\alpha + t_j^2 \sigma_i^\alpha \sigma_i^\beta  )\nonumber \\
   \text{mft}&=&-(\frac{\rho_1^2}{2}-\frac{J_1^2}{2 (N-1)})N    ( -\frac{\xi t}{2}n \sum_{\alpha} m^\alpha - \frac{\xi t}{2}n \sum_{\beta} m^\beta  + t^2 \sum_{\alpha \beta} q^{\alpha \beta}  )\nonumber \\
   &+& \frac{J_1^2}{2 (N-1)}N^2 \sum_{\alpha \beta } (\frac{\xi^2}{4} (q^{\alpha \beta})^2-\frac{\xi t}{2} q^{\alpha \beta} m^\beta- \frac{\xi t}{2} q^{\alpha \beta} m^\alpha + t^2 q^{\alpha \beta}   )
\end{eqnarray}
Of course, we also have an external field term which contributes a term
\begin{equation}
    H_{h}=h\sum_{\alpha} \sum_i \sigma_i^\alpha=h N \sum_\alpha m^\alpha.
\end{equation}
In the case $\mu_1\neq0$ and $\rho_1^2=\frac{J_1^2}{N}$ we have for $N\gg 1$

\begin{eqnarray}
 \lim_{n\rightarrow 0}\frac{H_0}{N }&=&   \frac{J_1^2}{2} \sum_{\alpha \beta } (\frac{\xi^2}{4} (q^{\alpha \beta})^2-\frac{\xi t}{2} q^{\alpha \beta} m^\beta- \frac{\xi t}{2} q^{\alpha \beta} m^\alpha + t^2 q^{\alpha \beta}- \mu_1 t \sum_{\alpha } m^\alpha \nonumber \\
 \text{Replica Symmetric}&=& \frac{J_1^2}{2}  (-\frac{\xi^2}{8} q^2+\xi t q m -\frac{ t^2}{2} q)+(h- \mu_1 t)  m.
\end{eqnarray}

\subsection{Functional dependence of the corrections}
In the correction terms we will have to calculate many of these terms, which we report here.

We have
\begin{eqnarray}
    \sum_{ij} \pmb{b}_{ii}&=&\sum_{ij} \pmb{b}_{jj}= N \sum_{\alpha} \sum_i(\frac{\xi}{2} \sigma_i^\alpha \sigma_i ^\alpha -\sigma_i^\alpha t_i) =N^2 n \frac{\xi}{2}-\sum_\alpha \sum_i \sigma_i^\alpha t_i. \nonumber \\
\text{mft}    &\rightarrow& N^2 n \frac{\xi}{2}-N^2 \sum_\alpha m^\alpha
\end{eqnarray}
where we note that in order to complete the calculation we are forced to use the mean field assumption $t_k \equiv t$.
Similarly 
\begin{eqnarray}
\sum_{i} \pmb{b}_{ii} \sum_j \pmb{b}_{jj}&=&(N n \frac{\xi}{2}-\sum_\alpha \sum_i \sigma_i^\alpha t_i)^2=\frac{N^2 n^2 \xi^2}{4}-2 \frac{N n\xi}{2} \sum_{\alpha} \sum_i \sigma_i^\alpha t_i+ \sum_{\alpha \beta}  \sum_{ij} \sigma_i^\alpha \sigma_j^\beta t_i t_j \nonumber \\
\text{mft} &\rightarrow& \frac{N^2 n^2 \xi^2}{4}-2 \frac{N^2 n\xi}{2} t \sum_{\alpha}  m^\alpha + N^2 t^2 (\sum_{\alpha }  m^\alpha)^2
\end{eqnarray}

Also,
\begin{eqnarray}
    \sum_{ij} \sum_k \pmb{b}_{ik}^2&=&\sum_{ij} \sum_k \pmb{b}_{k j}^2=N \sum_{i,k}(\sum_{\alpha} (\frac{\xi}{2}\sigma_i^\alpha \sigma_k^\alpha -\sigma_i^\alpha t_k))^2 \nonumber \\
    &=& N \sum_{\alpha \beta} \sum_{i,k}(\frac{\xi^2}{4} \sigma_i^\alpha \sigma_i^\beta \sigma_k^\alpha \sigma_k^\beta-\xi \sigma_i^\alpha \sigma_i^\beta \sigma_k^\alpha t_k+ \sigma_i^\alpha \sigma_i^\beta t_k^2) \nonumber \\
    &=&N \sum_{\alpha \beta} (\frac{\xi^2}{4} ( \sum_i \sigma_i^\alpha \sigma_i^\beta)^2+ (\sum_i \sigma^\alpha_i \sigma_i^\beta) \sum_k t_k^2- \xi (\sum_i \sigma_i^\alpha \sigma_i^\beta)(\sum_k \sigma_k^\alpha t_k) )\nonumber \\
\text{mft} &\rightarrow&    N \sum_{\alpha \beta} (\frac{\xi^2 N^2}{4} ( q^{\alpha \beta})^2+ N^2(q^{\alpha\beta}) t^2- N^2 t \xi q^{\alpha \beta} m^\alpha ) \nonumber \\
&=&N^3 \sum_{\alpha \beta} q^{\alpha\beta}(\frac{\xi^2 }{4}q^{\alpha\beta} +  t^2-  t \xi  m^\alpha )
\end{eqnarray}

\begin{eqnarray}
    \sum_{ij}  \pmb{b}_{ij}&=&\sum_{\alpha}\sum_{ij} \left(\frac{\xi}{2}\sigma_i^\alpha \sigma_j^\alpha -\sigma_i^\alpha t_j \right)=\sum_\alpha (\frac{\xi}{2}  (\sum_i \sigma_i^\alpha)^2- \sum_i \sigma_i^\alpha \sum_j t_j) \nonumber \\
    \text{mft} &\rightarrow& N^2 \sum_\alpha (\frac{\xi}{2} (m^\alpha)^2 -  m^\alpha t )
\end{eqnarray}
\begin{eqnarray}
    \sum_{ij} \delta_{ij} \pmb{b}_{ij}&=&\sum_{\alpha}\sum_{ij} \delta_{ij}\left(\frac{\xi}{2}\sigma_i^\alpha \sigma_j^\alpha -\sigma_i^\alpha t_j \right)=\sum_\alpha (\frac{\xi}{2} \sum_i (\sigma_i^\alpha)^2- \sum_i \sigma_i^\alpha t_i) \nonumber \\
    \text{mft} &\rightarrow& N \sum_\alpha (\frac{\xi}{2} -  m^\alpha t )
\end{eqnarray}
Let us now look at terms of the form:
\begin{eqnarray}
    \sum_{ij} \sum_t \pmb{b}_{it}\pmb{b}_{tj}&=& \sum_{\alpha \beta} \sum_{ijk} \left( \frac{\xi}{2}\sigma_i^\alpha \sigma_k^\alpha -\sigma_i^\alpha t_k \right)\left(\frac{\xi}{2}\sigma_k^\beta \sigma_j^\beta -\sigma_k^\beta t_j \right) \nonumber \\
    &=&\sum_{\alpha \beta}\Big( \left(\frac{\xi^2}{4} \sum_i \sigma_i^\alpha \sum_j \sigma_j^\beta \sum_k \sigma_k^\alpha \sigma_k^\beta+ \sum_i \sigma_i^\alpha \sum_j t_j \sum_k \sigma_k^\beta t_k \right) \nonumber \\
    &-& \frac{\xi}{2} \left(\sum_k \sigma_k^\alpha \sigma_k^\beta \sum_i \sigma_i^\alpha \sum_j t_j+\sum_k t_k \sigma_k^\beta \sum_i \sigma_i^\alpha \sum_j \sigma_j^\beta\right) \Big)    \nonumber \\
\text{mft}    &\rightarrow& N^3\sum_{\alpha \beta}\Big(\frac{\xi^2}{4} m^\alpha q^{\alpha \beta} m^{\beta}+  t^2 m^\alpha  m^\beta  
    - \frac{\xi}{2}  t\left( q^{\alpha \beta} m^\alpha+ m^\alpha (m^\beta)^2\right) \Big)
\end{eqnarray}
while
\begin{eqnarray}
    \sum_{ij} \delta_{ij}\sum_t \pmb{b}_{it}\pmb{b}_{tj}&=&\sum_{\alpha,\beta} \sum_{ik} \pmb{b}_{ik} \pmb{b}_{ki}= \sum_{\alpha \beta} \sum_{ik} (\frac{\xi}{2}\sigma_i^\alpha \sigma_k^\alpha -\sigma_i^\alpha t_k)(\frac{\xi}{2}\sigma_k^\beta \sigma_i^\beta -\sigma_k^\beta t_i)
    \nonumber \\
    &=& \sum_{\alpha \beta} \Big( \frac{\xi}{4} \sum_i \sigma_i^\alpha \sigma_i^\beta \sum_k \sigma_k^\alpha \sigma_k^\beta+ \sum_i \sigma_i^\alpha t_i \sum_k \sigma_k^\beta t_k \nonumber \\
    &-& \frac{\xi}{2} \left(\sum_k \sigma_k^\beta \sigma_k^\alpha \sum_i \sigma_i^\alpha t_i+\sum_i \sigma_i^\alpha \sigma_i^\beta \sum_k t_k\sigma_k^\beta\right)\Big) \nonumber \\
    &=&\sum_{\alpha \beta} \Big( \frac{\xi}{4} \sum_i \sigma_i^\alpha \sigma_i^\beta \sum_k \sigma_k^\alpha \sigma_k^\beta+ \sum_i \sigma_i^\alpha t_i \sum_k \sigma_k^\beta t_k  +\xi \left(\sum_k \sigma_k^\beta \sigma_k^\alpha \sum_i \sigma_i^\alpha t_i\right)\Big)\nonumber \\
\text{mft} &\rightarrow& N^2 \sum_{\alpha \beta} \Big( \frac{\xi}{4}  (q^{\alpha \beta})^2  + t^2 m^\alpha  m^\beta  +\xi t   q^{\alpha \beta} m^{\beta}\Big)
\end{eqnarray}
\begin{eqnarray}   
     \sum_i \pmb{b}_{i i}^3  &=& \sum_{\alpha \beta \gamma}\sum_i(\frac{\xi}{2} -\sigma_i^\alpha t)(\frac{\xi}{2} -\sigma_i^\beta t)(\frac{\xi}{2} -\sigma_i^\gamma t)\nonumber \\
     &=&\sum_{\alpha \beta \gamma} N (-t^3 \rho^{\alpha \beta \gamma}+\frac{1}{2}
   \xi  t^2 q^{\alpha \beta }+\frac{1}{2}
   \xi  t^2 q^{\alpha \gamma }+\frac{1}{2}
   \xi  t^2 q^{\beta \gamma }-\frac{1}{4}
   \xi ^2 t m^{\alpha }-\frac{1}{4} \xi ^2 t
   m^{\beta }-\frac{1}{4} \xi ^2 t
   m^{\gamma }+\frac{\xi ^3}{8}) \nonumber \\
\end{eqnarray}

\begin{eqnarray}       
    \sum_{i}\pmb{b}_{i i}\sum_t \pmb{b}_{i t}\pmb{b}_{t i}  &=&  \sum_{\alpha \beta \gamma} \sum_{it}\left(\frac{\xi }{2}-t \sigma_i^{\alpha }\right)
   \left(\frac{1}{2} \xi  \sigma_i^{\beta }
   \sigma_k^{\beta }-t \sigma_i^{\beta }\right)
   \left(\frac{1}{2} \xi  \sigma_i^{\gamma }
   \sigma_k^{\gamma }-t \sigma_k^{\gamma }\right)\nonumber \\
    &=&\sum_{\alpha \beta \gamma}N^2 \Big( \frac{1}{8} \xi ^3 q^{\beta \gamma }
   q^{\beta \gamma }-t^3 m^{\gamma }
   q^{\alpha \beta } +\frac{1}{2} \xi  t^2
   q^{\alpha \beta } q^{\beta \gamma
   }+\frac{1}{2} \xi  t^2 m^{\gamma }
   \rho^{\alpha \beta \gamma }\nonumber \\
   & &\ \ \ \ \ \ +\frac{1}{2} \xi
    t^2 m^{\beta } m^{\gamma }-\frac{1}{4}
   \xi ^2 t q^{\beta \gamma } \rho^{\alpha
   \beta \gamma }-\frac{1}{4} \xi ^2 t
   m^{\beta } \rho^{\beta \gamma
   }-\frac{1}{4} \xi ^2 t m^{\gamma }
   q^{\beta \gamma }\Big) \nonumber \\ 
\end{eqnarray}

\begin{eqnarray}    
    \sum_{ij}\pmb{b}_{i j}^2(\pmb{b}_{i i}+\pmb{b}_{jj})  &=&\sum_{\alpha \beta \gamma} \sum_{ij}\left(\frac{1}{2} \xi  \sigma_i^{\alpha }
   \sigma_j^{\alpha }-t \sigma_i^{\alpha }\right)
   \left(\frac{1}{2} \xi  \sigma_i^{\beta }
   \sigma_j^{\beta }-t \sigma_i^{\beta }\right) \left(-t
   \sigma_i^{\gamma }-t \sigma_j^{\gamma }+\xi \right)\nonumber \\
   &=& \sum_{\alpha \beta \gamma}N^2(\frac{1}{4} \xi ^3 (q^{\alpha +\beta })^2 
   -t^3 m^{\gamma }
   q^{\alpha \beta }+\frac{1}{2} \xi  t^2
   m^{\alpha } \rho^{\alpha \beta \gamma
   }+\frac{1}{2} \xi  t^2 m^{\beta }
   \rho^{\alpha \beta \gamma }+\frac{1}{2} \xi
    t^2 q^{\alpha \beta } q^{\alpha
   \gamma }+\frac{1}{2} \xi  t^2 q^{\alpha
   \beta } q^{\beta +\gamma }\nonumber \\
   &-&\frac{1}{4}
   \xi ^2 t q^{\alpha \beta } \rho^{\alpha
   \beta \gamma }-\frac{1}{4} \xi ^2 t
   q^{\alpha \beta } \rho^{\alpha \beta
   \gamma }-\frac{1}{2} \xi ^2 t m^{\alpha }
   q^{\alpha \beta }-\frac{1}{2} \xi ^2 t
   m^{\beta } q^{\alpha \beta }-t^3
   \rho^{\alpha \beta \gamma }+\xi  t^2
   q^{\alpha \beta })
\end{eqnarray}

\begin{eqnarray}
    \sum_{ij} \pmb{b}_{ij}\sum_t \pmb{b}_{it}\pmb{b}_{tj}&=&\sum_{\alpha \beta \gamma} \sum_{ijk}\Big( \frac{1}{8} \xi ^3 \left(\sigma_i^{\alpha } \sigma_i^{\beta }\right) \left(\sigma_j^{\alpha }
   \sigma_j^{\gamma }\right) \left(\sigma_k^{\gamma } \sigma_k^{\beta }\right)-\frac{1}{4} \xi ^2
   \left(\sigma_i^{\alpha } \sigma_i^{\beta }\right) \sigma_j^{\alpha } \left(\sigma_k^{\beta } \sigma_k^{\gamma
   }\right) t_j-\frac{1}{4} \xi ^2 \left(\sigma_i^{\alpha } \sigma_i^{\beta }\right) \sigma_j^{\gamma }
   \left(\sigma_k^{\beta } \sigma_k^{\gamma }\right) t_j \nonumber \\
   & &\ \ \ \ +\frac{1}{2} \xi  \left(\sigma_i^{\beta }
   \sigma_i^{\alpha }\right) \left(\sigma_k^{\beta } \sigma_k^{\gamma }\right) t_j^2-\frac{1}{4} \xi ^2
   \left(\sigma_i^{\alpha } \sigma_i^{\beta }\right) \left(\sigma_j^{\alpha } \sigma_j^{\gamma }\right)
   \sigma_k^{\gamma } t_k+\frac{1}{2} \xi  \left(\sigma_i^{\alpha } \sigma_i^{\beta }\right)
   \sigma_j^{\alpha } \sigma_k^{\gamma } t_j t_k \nonumber \\
   & &\ \ \ \ \ +\frac{1}{2} \xi  \left(\sigma_i^{\beta } \sigma_i^{\alpha
   }\right) \sigma_j^{\gamma } \sigma_k^{\gamma } t_j t_k-\left(\sigma_i^{\alpha } \sigma_i^{\beta }\right)
   \sigma_k^{\gamma } t_j^2 t_k\Big) \nonumber \\
\text{mft}   &\rightarrow& \sum_{\alpha \beta \gamma} \Big( \frac{1}{8} \xi ^3 N^3 q^{\alpha \beta} q^{\alpha \gamma} q^{\gamma \beta}-\frac{t}{4} \xi ^2
   N^3 q^{\alpha \beta}m^\alpha q^{\beta \gamma} -\frac{t}{4} \xi ^2 N^3 q^{\alpha \beta} m^\gamma
   q^{\beta \gamma}  \nonumber \\
   & &\ \ \ \ +\frac{\xi }{2}   N^3 t^2 q^{\beta \alpha} q^{\beta \gamma}-\frac{t}{4} N^3 \xi ^2
  q^{\alpha \beta} q^{\alpha \gamma}
   m^\gamma +\frac{t^2}{2} \xi   N^3 q^{\alpha \beta}
   m^\alpha m^\gamma 
   +\frac{1}{2} \xi t^2 N^3 q^{\alpha\beta} (m^\gamma)^2 -t^3 N^3 q^{\alpha \beta}
   m^\gamma \Big) \nonumber \\
 &=& N^3 \sum_{\alpha \beta \gamma}  q^{\alpha \beta} \Big( \frac{1}{8} \xi ^3  q^{ \beta\gamma }  q^{ \gamma\alpha }    +  q^{\beta \gamma} (\frac{\xi }{2}   t^2-\frac{3t}{4} m^\gamma)+\frac{t^2}{2} \xi  
   m^\alpha m^\gamma  +\frac{1}{2} \xi t^2   (m^\gamma)^2 -t^3 
   m^\gamma \Big) \nonumber \\
\end{eqnarray}

\begin{eqnarray}
    \sum_{ij} (\sum_t \pmb{b}_{it}\pmb{b}_{tj})^2
    &=&\sum_{\alpha \beta \gamma \delta} \sum_{ijkt} \Big(\frac{1}{16} \xi ^4 \left(\sigma_i^{\alpha } \sigma_i^{\gamma }\right) \left(\sigma_j^{\beta }
   \sigma_j^{\delta }\right) \left(\sigma_k^{\alpha } \sigma_k^{\beta }\right) \left(s_m^{\gamma }
   s_m^{\delta }\right) \nonumber \\
   & & \ \ \ \ \ \ - \frac{1}{8} \xi ^3 \left(\sigma_i^{\alpha } \sigma_i^{\gamma }\right)
   \sigma_j^{\beta } \left(\sigma_k^{\alpha } \sigma_k^{\beta }\right) \left(s_m^{\gamma } s_m^{\delta
   }\right) t_j-\frac{1}{8} \xi ^3 \left(\sigma_i^{\alpha } \sigma_i^{\gamma }\right) \sigma_j^{\delta
   } \left(\sigma_k^{\alpha } \sigma_k^{\beta }\right) \left(s_m^{\gamma } s_m^{\delta }\right)
   t_j \nonumber \\
   & & \ \ \ \ \ \ +\frac{1}{4} \xi ^2 \left(\sigma_i^{\alpha } \sigma_i^{\gamma }\right) \left(\sigma_k^{\alpha }
   \sigma_k^{\beta }\right) \left(s_m^{\gamma } s_m^{\delta }\right) t_j^2-\frac{1}{8} \xi ^3
   \left(\sigma_i^{\alpha } \sigma_i^{\gamma }\right) \left(\sigma_j^{\beta  } \sigma_j^{\delta
   }\right) \sigma_k^{\beta } \left(s_m^{\gamma } s_m^{\delta }\right) t_k \nonumber \\
   & & \ \ \ \ \ \ +\frac{1}{4} \xi ^2
   \left(\sigma_i^{\alpha } \sigma_i^{\gamma }\right) \sigma_j^{\beta } \sigma_k^{\beta } \left(s_m^{\gamma
   } s_m^{\delta }\right) t_j t_k+\frac{1}{4} \xi ^2 \left(\sigma_i^{\alpha } \sigma_i^{\gamma
   }\right) \sigma_j^{\delta } \sigma_k^{\beta } \left(s_m^{\gamma } s_m^{\delta }\right) t_j
   t_k-\frac{1}{2} \xi  \left(\sigma_i^{\alpha } \sigma_i^{\gamma }\right) \sigma_k^{\beta }
   \left(s_m^{\gamma } s_m^{\delta }\right) t_j^2 t_k \nonumber \\
   & & \ \ \ \ \ \ -\frac{1}{8} \xi ^3
   \left(\sigma_i^{\alpha } \sigma_i^{\gamma }\right) \left(\sigma_j^{\beta } \sigma_j^{\delta }\right)
   \left(\sigma_k^{\alpha } \sigma_k^{\beta }\right) s_m^{\delta } t_m +\frac{1}{4} \xi ^2
   \left(\sigma_i^{\alpha } \sigma_i^{\gamma }\right) \sigma_j^{\beta } \left(\sigma_k^{\alpha } \sigma_k^{\beta
   }\right) s_m^{\delta } t_j t_m \nonumber \\
   & & \ \ \ \ \ \ +\frac{1}{4} \xi ^2 \left(\sigma_i^{\alpha } \sigma_i^{\gamma
   }\right) \sigma_j^{\delta } \left(\sigma_k^{\alpha } \sigma_k^{\beta }\right) s_m^{\delta } t_j
   t_m-\frac{1}{2} \xi  \left(\sigma_i^{\alpha } \sigma_i^{\gamma }\right) \left(\sigma_k^{\alpha }
   \sigma_k^{\beta }\right) s_m^{\delta } t_j^2 t_m \nonumber \\
   & & \ \ \ \ \ \ +\frac{1}{4} \xi ^2 \left(\sigma_i^{\alpha }
   \sigma_i^{\gamma }\right) \left(\sigma_j^{\beta } \sigma_j^{\delta }\right) \sigma_k^{\beta } s_m^{\delta
   } t_k t_m-\frac{1}{2} \xi  \left(\sigma_i^{\alpha } \sigma_i^{\gamma }\right) \sigma_j^{\beta }
   \sigma_k^{\beta } s_m^{\delta } t_j t_k t_m \nonumber \\
   & & \ \ \ \ \ \ -\frac{1}{2} \xi  \left(\sigma_i^{\alpha }
   \sigma_i^{\gamma }\right) \sigma_j^{\delta } \sigma_k^{\beta } s_m^{\delta } t_j t_k
   t_m+\left(\sigma_i^{\alpha } \sigma_i^{\gamma }\right) \sigma_k^{\beta } s_m^{\delta } t_j^2 t_k t_m\Big) \nonumber \\
\text{mft} & \rightarrow &\sum_{\alpha \beta \gamma \delta}  \Big(\frac{N^4}{16} \xi ^4  q^{\alpha \gamma}  q^{\beta \delta}  q^{\alpha \beta}  q^{\gamma \delta} - \frac{t}{8} \xi ^3 N^4 q^{\alpha \gamma}
    m^\beta   q^{\alpha \beta}  q^{\gamma \delta} -\frac{t}{8} \xi ^3 N^4 q^{\alpha \gamma} m^\delta q^{\alpha \beta}  q^{\gamma \delta}
    \nonumber \\
   & & \ \ \ \ \ \ +\frac{t^2}{4} \xi ^2 N^4 q^{\alpha \gamma}  q^{\alpha \beta}   q^{\gamma \delta} -\frac{t}{8} N^4 \xi ^3
    q^{\alpha \gamma}  q^{\beta \delta}  m^\beta q^{\gamma \delta}  \nonumber \\
   & & \ \ \ \ \ \ +\frac{t^2}{4} N^4 \xi ^2
     q^{\alpha \gamma} (m^\beta)^2  q^{\gamma \delta} +\frac{t^2}{4} N^4 \xi ^2  q^{\alpha \gamma}m^\delta m^\beta q^{\gamma \delta}-\frac{N^4}{2} t^3 \xi  q^{\alpha \gamma} m^\beta
   q^{\gamma \delta}  \nonumber \\
   & & \ \ \ \ \ \ -\frac{N^4}{8}t  \xi ^3
   q^{\alpha \gamma} q^{\beta \delta}
   q^{\alpha \beta} m^\delta +\frac{N^4}{4} t^2 \xi ^2
   q^{\alpha \gamma} q^{\alpha \beta}  m^\delta m^\beta  \nonumber \\
   & & \ \ \ \ \ \ +\frac{N^4}{4} t^2 \xi ^2  q^{\alpha \gamma}  q^{\alpha \beta} (m^\delta)^2 -\frac{N^4}{2} t^3\xi   q^{\alpha \gamma} q^{\alpha \beta} m^\delta  +\frac{N^4}{4} t^2 \xi ^2 q^{\alpha \gamma} q^{\beta \delta} m^\beta m^\delta -\frac{N^4}{2} t^3 \xi  q^{\alpha \gamma} (m^\beta)^2 m^\delta  \nonumber \\
   & & \ \ \ \ \ \ -\frac{N^4}{2}  t^3 \xi  q^{\alpha\gamma} (m^\delta)^2 m^\beta + t^4 N^4 q^{\alpha \gamma} m^\beta m^\delta \Big) \nonumber \\
   &=&N^4 \sum_{\alpha \beta \gamma \delta} \Big(\frac{\xi^4}{16} q^{\alpha \gamma} q^{\beta \delta} q^{\alpha \beta} q^{\gamma \delta} -\frac{t \xi^3}{2} q^{\alpha \beta} q^{\beta 
   \gamma } q^{\gamma \delta} m^\delta \nonumber \\
   & &\ \ \ \ \ \ \ \ +  \frac{\xi^2 t^2}{4}(q^{\alpha \gamma} q^{\gamma \delta} ((m^\beta)^2+m^\delta)+q^{\alpha \gamma} q^{\beta \delta} m^\beta m^\delta+q^{\alpha \gamma} q^{\gamma \delta}m^\alpha m^\beta)\nonumber \\
   & &\ \ \ \ \ \ \ \ -t^3 \xi ( q^{\alpha \gamma} q^{\gamma \delta} m^\beta +  q^{\alpha \gamma} (m^\beta)^2 m^\delta)+ t^4 q^{\alpha \gamma} m^\beta m^\delta)
   \Big)
\end{eqnarray}

Also, for the next asymmetrical calculation we need
\begin{eqnarray}
    \sum_{ij} ( \pmb{b}_{ii}+ \pmb{b}_{jj})  \pmb{b}_{ij}&=&\sum_{\alpha \beta}
     \sum_{ij} (\frac{\xi}{2} \sigma_i^\alpha \sigma_j^\alpha- \sigma_i^\alpha t_j)(\xi - \sigma_i^\alpha t_i- \sigma_j^\alpha t_j)\nonumber \\
     &=& \sum_{\alpha \beta} \sum_{ij}(\frac{1}{2} \xi ^2 \sigma_i^{\alpha } \sigma_j^{\alpha }+t^2 \sigma_i^{\alpha } \sigma_j^{\beta
   }-\frac{1}{2} \xi  t \sigma_i^{\alpha } \sigma_j^{\alpha} \sigma_j^{\beta }-\frac{1}{2} \xi  t
   \sigma_j^{\alpha } \sigma_i^{\alpha}\sigma_i^{\beta }+t^2 \sigma_i^{\alpha}\sigma_i^{\beta }-\xi  t \sigma_i^{\alpha
   }) \nonumber \\
 \text{mft}    &\rightarrow&\sum_{\alpha \beta} N^2(\frac{1}{2} \xi ^2 m^{\alpha } m^{\alpha }+t^2 m^{\alpha } m^{\beta
   }-\frac{1}{2} \xi  t m^{\alpha } m^{\alpha +\beta }-\frac{1}{2} \xi  t
   m^{\alpha } q^{\alpha \beta }+t^2 q^{\alpha \beta }-\xi  t m^{\alpha
   })
\end{eqnarray}

And for the last calculation we need the term
\begin{eqnarray}
    \sum_{ij} ( \pmb{b}_{ij})^2&=&\sum_{\alpha \beta} \sum_{ij}( \frac{\xi}{2} \sigma_i^\alpha \sigma_j^\alpha -\sigma_i^\alpha t_j)( \frac{\xi}{2} \sigma_i^\beta \sigma_j^\beta -\sigma_i^\beta t_j) \nonumber \\
    &=&\sum_{\alpha \beta}\sum_{ij}\left(\frac{\xi^2}{4} \sigma_i^\alpha \sigma_j^\alpha \sigma_i^\beta \sigma_j^\beta +\sigma_i^\alpha \sigma_i^\beta t_j^2- \frac{\xi}{2} (\sigma_i^\alpha \sigma_j^\alpha \sigma_i^\beta t_j+\sigma_i^\beta \sigma_j^\beta \sigma_i^\alpha t_j) \right) \nonumber \\
    \text{mft} &\rightarrow&  N^2 \sum_{\alpha \beta}\left(\frac{\xi^2}{4} (q^{\alpha \beta})^2+ t^2 q^{\alpha \beta}-\frac{\xi t }{2} q^{\alpha \beta} ( m^\beta+m^\alpha)  \right)
\end{eqnarray}

\begin{eqnarray}
\sum_{ij}\delta_{ij}\pmb{b}_{i j}^4 &=& \sum_{\alpha \beta \gamma \delta} \sum_i(\frac{\xi}{2}-\sigma_i^\alpha t_i)(\frac{\xi}{2}-\sigma_i^\beta t_i)(\frac{\xi}{2}-\sigma_i^\gamma t_i)(\frac{\xi}{2}-\sigma_i^\delta t_i)\nonumber \\
&=&N\sum_{\alpha \beta \gamma \delta} (t^4 \eta^{\alpha \beta \gamma \delta }-\frac{1}{2} \xi  t^3 \rho^{\alpha \beta
   \gamma }-\frac{1}{2} \xi  t^3 \rho^{\alpha \beta \delta }-\frac{1}{2} \xi 
   t^3 \rho^{\alpha \gamma \delta }-\frac{1}{2} \xi  t^3 \rho^{\beta \gamma
   \delta }+\frac{1}{4} \xi ^2 t^2 q^{\alpha \beta } \nonumber \\
   & &\ \ \ \ \ \ \ \ \ \ +\frac{1}{4} \xi ^2 t^2
   q^{\alpha +\gamma }+\frac{1}{4} \xi ^2 t^2 q^{\alpha \delta }+\frac{1}{4}
   \xi ^2 t^2 q^{\beta \gamma }+\frac{1}{4} \xi ^2 t^2 q^{\beta \delta
   }+\frac{1}{4} \xi ^2 t^2 q^{\gamma \delta }-\frac{1}{8} \xi ^3 t m^{\alpha} \nonumber \\ 
   & &\ \ \ \ \ \ \ \ \ \ -\frac{1}{8} \xi ^3 t m^{\beta }-\frac{1}{8} \xi ^3 t m^{\gamma
   }-\frac{1}{8} \xi ^3 t m^{\delta }+\frac{\xi ^4}{16})
\end{eqnarray}
\begin{eqnarray}
\sum_{ij}\pmb{b}_{i j}^2(\pmb{b}_{j j}+\pmb{b}_{ii})^2&=&\sum_{\alpha \beta \gamma \delta}\sum_{ij}(\frac{\xi}{2}\sigma_i^\alpha \sigma_j^\alpha- \sigma_i^\alpha t)(\frac{\xi}{2}\sigma_i^\beta \sigma_j^\beta- \sigma_i^\beta t)(\xi- (\sigma_i^\gamma -\sigma_j^\gamma)t)(\xi- (\sigma_i^\delta -\sigma_j^\delta)t) \nonumber \\
&=&N^2\sum_{\alpha \beta \gamma \delta}
(\frac{1}{4} \xi ^4 q^{\alpha \beta } q^{\alpha \beta }+t^4 m^{\gamma }
   \rho^{\alpha \beta \delta }+t^4 m^{\delta } \rho^{\alpha \beta \gamma }+t^4
   q^{\alpha \beta } q^{\gamma \delta }-\frac{1}{2} \xi  t^3 m^{\alpha }
   \eta^{\alpha \beta \gamma \delta } \nonumber \\
   && \ \ \ \ \ \ \ -\frac{1}{2} \xi  t^3 m^{\beta }
   \eta^{\alpha \beta \gamma \delta }-\frac{1}{2} \xi  t^3 q^{\alpha \gamma }
   \rho^{\alpha \beta \delta }-\frac{1}{2} \xi  t^3 q^{\beta \gamma }
   \rho^{\alpha \beta \delta }-\frac{1}{2} \xi  t^3 q^{\alpha \delta }
   \rho^{\alpha \beta \gamma }\nonumber \\
   && \ \ \ \ \ \ \ -\frac{1}{2} \xi  t^3 q^{\beta \delta }
   \rho^{\alpha \beta \gamma }-\frac{1}{2} \xi  t^3 q^{\alpha \beta }
   \rho^{\alpha \gamma \delta }-\frac{1}{2} \xi  t^3 q^{\alpha \beta }
   \rho^{\beta \gamma \delta }-\xi  t^3 m^{\gamma } q^{\alpha \beta }-\xi 
   t^3 m^{\delta } q^{\alpha \beta } \nonumber \\
   && \ \ \ \ \ \ \ +\frac{1}{4} \xi ^2 t^2 q^{\alpha
   \beta } \eta^{\alpha \beta \gamma \delta }+\frac{1}{4} \xi ^2 t^2
   \rho^{\alpha \beta \delta } \rho^{\alpha \beta \gamma }+\frac{1}{4} \xi ^2
   t^2 \rho^{\alpha \beta \gamma } \rho^{\alpha \beta \delta }\nonumber \\
   && \ \ \ \ \ \ \ +\frac{1}{4} \xi
   ^2 t^2 q^{\alpha \beta } \eta^{\alpha \beta \gamma \delta }+\frac{1}{2}
   \xi ^2 t^2 m^{\alpha } \rho^{\alpha \beta \gamma }+\frac{1}{2} \xi ^2 t^2
   m^{\beta } \rho^{\alpha \beta \gamma }+\frac{1}{2} \xi ^2 t^2 q^{\alpha
   \beta } q^{\alpha \gamma }\nonumber \\
   && \ \ \ \ \ \ \ +\frac{1}{2} \xi ^2 t^2 q^{\alpha \beta }
   q^{\beta \gamma } +\frac{1}{2} \xi ^2 t^2 m^{\alpha } \rho^{\alpha \beta
   \delta }+\frac{1}{2} \xi ^2 t^2 m^{\beta } \rho^{\alpha \beta \delta
   }+\frac{1}{2} \xi ^2 t^2 q^{\alpha \beta } q^{\alpha \delta }+\frac{1}{2}
   \xi ^2 t^2 q^{\alpha \beta } q^{\beta \delta }\nonumber \\
   && \ \ \ \ \ \ \ -\frac{1}{4} \xi ^3 t
   q^{\alpha \beta } \rho^{\alpha \beta \gamma } -\frac{1}{4} \xi ^3 t
   q^{\alpha \beta } \rho^{\alpha \beta \gamma }-\frac{1}{4} \xi ^3 t
   q^{\alpha \beta } \rho^{\alpha \beta \delta } \nonumber \\
   && \ \ \ \ \ \ \ -\frac{1}{4} \xi ^3 t
   q^{\alpha +\beta } \rho^{\alpha \beta \delta }-\frac{1}{2} \xi ^3 t
   m^{\alpha } q^{\alpha \beta }-\frac{1}{2} \xi ^3 t m^{\beta }
   q^{\alpha \beta }\nonumber \\
   && \ \ \ \ \ \ \ +t^4 \eta^{\alpha \beta \gamma \delta }-\xi  t^3
   \rho^{\alpha \beta \gamma }-\xi  t^3 \rho^{\alpha \beta \delta }+\xi ^2 t^2
   q^{\alpha \beta })
\end{eqnarray}
\begin{eqnarray}
 \sum_{ij}(\sum_t \pmb{b}_{i t}\pmb{b}_{t j})^2&=&\sum_{ij lk} \pmb{b}_{i l}\pmb{b}_{l j}\pmb{b}_{i k}\pmb{b}_{k j} =\sum_{\alpha \beta \gamma \delta}\sum_{ijlt} \left(\frac{\xi}{2}   \sigma_i^{\gamma } \sigma_k^{\gamma }-t \sigma_i^{\gamma }\right)
   \left(\frac{\xi}{2}   \sigma_i^{\alpha } \sigma_l^{\alpha }-t \sigma_i^{\alpha }\right)
   \left(\frac{\xi}{2}   \sigma_j^{\delta } \sigma_k^{\delta }-t \sigma_k^{\delta }\right)
   \left(\frac{\xi}{2}   \sigma_j^{\beta } \sigma_l^{\beta }-t \sigma_l^{\beta }\right) \nonumber \\
 &=&N^4\sum_{\alpha \beta \gamma \delta} (\frac{1}{16} \xi ^4 q^{\alpha \gamma } q^{\beta \delta } q^{\gamma \delta
   } q^{\alpha \beta }-\frac{1}{2} \xi  t^3 m^{\beta } m^{\delta }
   m^{\beta } q^{\alpha \gamma }-\frac{1}{2} \xi  t^3 m^{\delta }
   m^{\delta } m^{\beta } q^{\alpha \gamma } \nonumber \\
   & & \ \ \ \ \ \ \ \ +\frac{1}{4} \xi ^2 t^2
   m^{\beta } m^{\beta } q^{\alpha \gamma } q^{\gamma \delta
   }+\frac{1}{4} \xi ^2 t^2 m^{\beta } m^{\delta } q^{\alpha \gamma }
   q^{\alpha \beta }+\frac{1}{4} \xi ^2 t^2 m^{\delta } m^{\beta }
   q^{\alpha \gamma } q^{\gamma \delta }+\frac{1}{4} \xi ^2 t^2 m^{\delta
   } m^{\delta } q^{\alpha \gamma } q^{\alpha \beta }\nonumber \\
   & & \ \ \ \ \ \ \ \ +\frac{1}{4} \xi ^2
   t^2 m^{\delta } m^{\beta } q^{\alpha \gamma } q^{\beta \delta
   }-\frac{1}{8} \xi ^3 t m^{\beta } q^{\alpha \gamma } q^{\gamma \delta }
   q^{\alpha \beta }-\frac{1}{8} \xi ^3 t m^{\delta } q^{\alpha \gamma }
   q^{\gamma \delta } q^{\alpha \beta }-\frac{1}{8} \xi ^3 t m^{\beta }
   q^{\alpha \gamma } q^{\beta \delta } q^{\gamma \delta }\nonumber \\
   & & \ \ \ \ \ \ \ \ -\frac{1}{8}
   \xi ^3 t m^{\delta } q^{\alpha \gamma } q^{\beta \delta } q^{\alpha
   \beta }+t^4 m^{\delta } m^{\beta } q^{\alpha \gamma } \nonumber \\
   & & \ \ \ \ \ \ \ \ -\frac{1}{2} \xi 
   t^3 m^{\beta } q^{\alpha \gamma } q^{\gamma \delta }-\frac{1}{2} \xi 
   t^3 m^{\delta } q^{\alpha \gamma } q^{\alpha \beta }+\frac{1}{4} \xi ^2
   t^2 q^{\alpha \gamma } q^{\gamma \delta } q^{\alpha \beta })
\end{eqnarray}

\begin{eqnarray}
\sum_{ij}\delta_{ij}\pmb{b}_{i j}^2\sum_t \pmb{b}_{i t}\pmb{b}_{t j} &=& \sum_{i}\pmb{b}_{i i}^2\sum_t \pmb{b}_{i t}\pmb{b}_{t i}\nonumber \\
&=&N^2\sum_{\alpha \beta \gamma \delta}(\frac{1}{16} \xi ^4 q^{\gamma \delta } q^{\gamma \delta }+t^4 \sigma_k^{\delta }
   \rho^{\alpha \beta \gamma }-\frac{1}{2} \xi  t^3 q^{\gamma \delta }
   \rho^{\alpha \beta \gamma }-\frac{1}{2} \xi  t^3 m^{\delta } \eta^{\alpha
   \beta \gamma \delta }-\frac{1}{2} \xi  t^3 m^{\delta } q^{\alpha +\gamma} \nonumber \\
   & &\ \ \ \ \ \ \ -\frac{1}{2} \xi  t^3 m^{\delta } q^{\beta \gamma }+\frac{1}{4} \xi ^2
   t^2 q^{\gamma \delta } \eta^{\alpha \beta \gamma \delta }+\frac{1}{4} \xi
   ^2 t^2 q^{\alpha \gamma } q^{\gamma \delta }+\frac{1}{4} \xi ^2 t^2
   m^{\delta } \rho^{\alpha \gamma \delta }+\frac{1}{4} \xi ^2 t^2 q^{\beta
   \gamma } q^{\gamma \delta } \nonumber \\
   & &\ \ \ \ \ \ \ +\frac{1}{4} \xi ^2 t^2 m^{\delta } \rho^{\beta \gamma \delta }+\frac{1}{4} \xi ^2 t^2 m^{\gamma } m^{\delta
   }-\frac{1}{8} \xi ^3 t q^{\gamma \delta } \rho^{\alpha \gamma \delta} \nonumber \\
   & &\ \ \ \ \ \ \ -\frac{1}{8} \xi ^3 t q^{\gamma \delta } \rho^{\beta \gamma \delta
   }-\frac{1}{8} \xi ^3 t m^{\gamma } q^{\gamma \delta }-\frac{1}{8} \xi ^3 t
   m^{\delta } q^{\gamma \delta })
\end{eqnarray}
\begin{eqnarray}
\sum_{ij}\pmb{b}_{i j}(\pmb{b}_{j j}+\pmb{b}_{ii}) \sum_t \pmb{b}_{i t}\pmb{b}_{t j}&=&\sum_{\alpha \beta \gamma \delta}\sum_{ij} \left(\frac{1}{2} \xi  \sigma_i^{\alpha } \sigma_j^{\alpha }-t \sigma_i^{\alpha }\right) \left(-t
   \sigma_i^{\beta }-t \sigma_j^{\beta }+\xi \right) \left(\frac{1}{2} \xi  \sigma_i^{\gamma }
   \sigma_k^{\gamma }-t \sigma_i^{\gamma }\right) \left(\frac{1}{2} \xi  \sigma_j^{\delta }
   \sigma_k^{\delta }-t \sigma_k^{\delta }\right)\nonumber \\
   &=& N^2 \sum_{\alpha \beta \gamma \delta}(\frac{1}{8} \xi ^4 q^{\alpha \gamma } q^{\alpha \delta } q^{\gamma \delta
   }+t^4 m^{\beta } m^{\delta } q^{\alpha \gamma }-\frac{1}{2} \xi  t^3
   m^{\alpha } m^{\delta } \rho^{\alpha \beta \gamma }-\frac{1}{2} \xi  t^3
   m^{\beta } q^{\alpha \gamma } q^{\gamma \delta }\nonumber \\
   & & \ \ \ \ \ \ \ \ -\frac{1}{2} \xi  t^3
   m^{\delta } q^{\alpha \gamma } q^{\alpha \beta }-\frac{1}{2} \xi  t^3
   m^{\delta } m^{\delta } q^{\alpha \beta \gamma }-\frac{1}{2} \xi  t^3
   m^{\delta } q^{\alpha \gamma } q^{\beta \delta }+\frac{1}{4} \xi ^2 t^2
   m^{\alpha } q^{\gamma \delta } \rho^{\alpha +\beta +\gamma }\nonumber \\
   & & \ \ \ \ \ \ \ \ +\frac{1}{4}
   \xi ^2 t^2 q^{\alpha \gamma } q^{\alpha \beta } m^{\gamma +\delta
   }+\frac{1}{4} \xi ^2 t^2 m^{\delta } q^{\gamma \delta } \rho^{\alpha \beta
   \gamma }+\frac{1}{4} \xi ^2 t^2 m^{\delta } q^{\alpha \delta }
   \rho^{\alpha \beta \gamma }\nonumber \\
   & & \ \ \ \ \ \ \ \ +\frac{1}{4} \xi ^2 t^2 q^{\alpha \gamma }
   q^{\beta \delta } q^{\gamma \delta } +\frac{1}{4} \xi ^2 t^2 m^{\delta }
   q^{\alpha \gamma } \rho^{\alpha \beta \delta }+\frac{1}{2} \xi ^2 t^2
   m^{\alpha } m^{\delta } q^{\alpha \gamma }+\frac{1}{2} \xi ^2 t^2
   m^{\delta } m^{\delta } q^{\alpha \gamma } \nonumber \\
   & & \ \ \ \ \ \ \ \ -\frac{1}{8} \xi ^3 t
   q^{\alpha \delta } q^{\gamma \delta } \rho^{\alpha \beta \gamma
   }-\frac{1}{8} \xi ^3 t q^{\alpha \gamma } q^{\gamma \delta } \rho^{\alpha
   \beta \delta } -\frac{1}{4} \xi ^3 t m^{\alpha } q^{\alpha \gamma }
   q^{\gamma \delta }-\frac{1}{4} \xi ^3 t m^{\delta } q^{\alpha \gamma }
   q^{\gamma \delta }
   \nonumber \\
   & & \ \ \ \ \ \ \ \ -\frac{1}{4} \xi ^3 t m^{\delta } q^{\alpha \gamma }
   q^{\alpha \delta }+t^4 m^{\delta } \rho^{\alpha \beta \gamma
   }-\frac{1}{2} \xi  t^3 q^{\gamma q\delta } \rho^{\alpha \beta \gamma }-\xi 
   t^3 m^{\delta } q^{\alpha \gamma }+\frac{1}{2} \xi ^2 t^2 q^{\alpha
   \gamma } q^{\gamma q\delta })\nonumber \\
\end{eqnarray}
 
\begin{eqnarray}
\sum_{ij} \delta_{ij}\pmb{b}_{i j}^3(\pmb{b}_{j j}+\pmb{b}_{ii})&=&2\sum_{i} \pmb{b}_{i i}^4=2\sum_{\alpha \beta \gamma \delta}\sum_i \left(\frac{\xi }{2}-t \sigma_i^{\alpha }\right) \left(\frac{\xi }{2}-t \sigma_i^{\beta
   }\right) \left(\frac{\xi }{2}-t \sigma_i^{\gamma }\right) \left(\frac{\xi }{2}-t
   \sigma_i^{\delta }\right)  \nonumber \\
&=&2 N \sum_{\alpha \beta \gamma \delta} (
t^4 \eta^{\alpha \beta \gamma \delta }-\frac{1}{2} \xi  t^3 \rho^{\alpha \beta
   \gamma }-\frac{1}{2} \xi  t^3 \rho^{\alpha +\beta +\delta }-\frac{1}{2} \xi 
   t^3 \rho^{\alpha \gamma \delta }-\frac{1}{2} \xi  t^3 \rho^{\beta \gamma
   \delta }
   \nonumber \\
   & &\ \ \ \ \ \ \ \ +\frac{1}{4} \xi ^2 t^2 q^{\alpha \beta }+\frac{1}{4} \xi ^2 t^2
   q^{\alpha \gamma }+\frac{1}{4} \xi ^2 t^2 q^{\alpha \delta }+\frac{1}{4}
   \xi ^2 t^2 q^{\beta \gamma }+\frac{1}{4} \xi ^2 t^2 q^{\beta \delta
   }+\frac{1}{4} \xi ^2 t^2 q^{\gamma \delta }
   \nonumber \\
   & &\ \ \ \ \ \ \ \ -\frac{1}{8} \xi ^3 t m^{\alpha
   }-\frac{1}{8} \xi ^3 t m^{\beta }-\frac{1}{8} \xi ^3 t m^{\gamma
   }-\frac{1}{8} \xi ^3 t m^{\delta }+\frac{\xi ^4}{16})
\end{eqnarray}

\subsection{Corrections}
\begin{eqnarray}
 \frac{F^3_{ij}}{\gamma N}  &=&   \underbrace{\rho_1^2 \big(\rho_1^2 \delta_{ij}+(1-\delta_{ij}) \frac{J_1^2}{N-1}\big) (\pmb{b}_{jj}+\pmb{b}_{ii})}_{F3_A} \nonumber \\
 &+& \underbrace{\rho_1^2(3\mu_1^2+\rho_1^2)\delta_{ij} \pmb{b}_{ij}}_{F3_B}\nonumber \\
 &+&\underbrace{3 \mu_1 \rho_1^2 \delta_{ij} \pmb{b}_{ij}^2}_{F3_C}  \nonumber \\
&+&   \underbrace{\mu_1  \frac{J_1^4}{(N-1)^2}\sum_{t}\big(\delta_{ij}(1-\delta_{it})(1-\delta_{jt}) \pmb{b}_{it}\pmb{b}_{tj}\big)}_{F3_D}
  \nonumber \\
  &+&   \underbrace{\mu_1  \frac{J_1^4}{(N-1)^2}\sum_{t}\big(\delta_{it}(1-\delta_{tj})   \pmb{b}_{tj}^2+
 (1-\delta_{it}) \delta_{tj} \pmb{b}_{it}^2\big)}_{F3_E}
  \nonumber \\
&+& \underbrace{\big(\rho_1^2 \delta_{ij}+(1-\delta_{ij}) \frac{J_1^2}{N-1}\big) \pmb{b}_{ij} \sigma_1^4 \delta_{ij} \pmb{b}_{ii}\pmb{b}_{jj}}_{F3_F}\\
&+& \underbrace{\big(\rho_1^2 \delta_{ij}+(1-\delta_{ij}) \frac{J_1^2}{N-1}\big) \pmb{b}_{ij}  (1-\delta_{ij}) \frac{\rho_1^2J_1^2}{N-1}(\pmb{b}_{ij}\pmb{b}_{jj}+\pmb{b}_{ii}\pmb{b}_{ij})}_{F3_G}\\
&+& \underbrace{\big(\rho_1^2 \delta_{ij}+(1-\delta_{ij}) \frac{J_1^2}{N-1}\big) \pmb{b}_{ij} \sum_t \frac{J_1^4}{(N-1)^2}(1-\delta_{it})(1-\delta_{tj})\pmb{b}_{it}\pmb{b}_{tj} }_{F3_H}
\nonumber \\  
\end{eqnarray}
Asymmetric F3:
\begin{eqnarray}
    F3_A&=&\sum_{ij}\big(\rho_1^2 \delta_{ij}+(1-\delta_{ij}) \frac{J_1^2}{N-1}\big) \rho_1^2(\pmb{b}_{ii}+\pmb{b}_{jj})\nonumber \\
    &=& \sum_{ij}\big(\rho_1^2 \delta_{ij}+(1-\delta_{ij}) \frac{J_1^2}{N-1}\big) \rho_1^2(\sum_{\alpha}( \frac{\xi}{2} \sigma_i^\alpha \sigma_i^\alpha-\sigma_i^\alpha t_i)+\sum_{\beta}( \frac{\xi}{2} \sigma_j^\beta \sigma_j^\beta-\sigma_j^\beta t_j))\nonumber \\
&\approx&-\rho_1^2\sum_{\alpha}\sum_{ij}\big(\rho_1^2 \delta_{ij}(  \sigma_i^\alpha t_i+\sigma_j^\alpha t_j)+(1-\delta_{ij}) \frac{J_1^2}{N-1}(  \sigma_i^\alpha t_i+\sigma_j^\alpha t_j)\big)  \nonumber \\
&=&-\rho_1^2 t \sum_{\alpha}\big(2\rho_1^2 \sum_{i}  \sigma_i^\alpha + \frac{J_1^2}{N-1} \sum_{ij}(1-\delta_{ij})   (  \sigma_i^\alpha +\sigma_j^\alpha )\big)\nonumber \\
&=&-\rho_1^2 t \sum_{\alpha}\big(2\rho_1^2 \sum_{i}  \sigma_i^\alpha + 2 \frac{ J_1^2}{N-1} \sum_{i<j}  (\sigma_i^\alpha +\sigma_j^\alpha )\big)\nonumber \\
&=&-\rho_1^2 t \sum_{\alpha}\big(2\rho_1^2 \sum_{i}  \sigma_i^\alpha + 4  J_1^2 \sum_j \sigma_j^\alpha \big)\nonumber \\
&=&-\rho_1^2 t \sum_{\alpha}\big(2\rho_1^2 + 4  J_1^2 \big)\sum_{i}  \sigma_i^\alpha \nonumber \\
&\rightarrow&-\rho_1^2 t \big(2\rho_1^2 + 4  J_1^2 \big) N \sum_{\alpha} m^\alpha
\end{eqnarray}

\begin{eqnarray}
F3_B&=&\rho_1^2(\rho_1^2+3 \mu_1^2) \sum_{ij} \delta_{ij} \pmb{b}_{ij}\nonumber \\
\text{mft} &\approx&\rho_1^2(\rho_1^2+3 \mu_1^2)t N\sum_{\alpha} m^\alpha
\end{eqnarray}

\begin{eqnarray}
    F3_C&=&3 \mu_1 \rho_1^2 \sum_{ij} \delta_{ij} \pmb{b}_{ij}^2\nonumber \\
    \text{mean field}&=&3 \mu_1 \rho_1^2 \sum_{\alpha,\beta}\sum_{i} ( -\frac{\xi}{2}(\sigma_i^\alpha+\sigma_i^\beta )t +\sigma_i^\alpha \sigma_i^\beta t^2)\nonumber \\   
   &\rightarrow&3 N \mu_1 \rho_1^2 \sum_{\alpha,\beta} (q^{\alpha \beta}t^2 -\frac{\xi}{2}(m^\alpha+m^\beta )t )\nonumber \\   
\end{eqnarray}

\begin{eqnarray}
   F3_{D}&=& \mu_1 \frac{J_1^4}{(N-1)^2} \sum_{ij}\sum_{t}\big(\delta_{ij}(1-\delta_{it})(1-\delta_{jt}) \pmb{b}_{it}\pmb{b}_{tj}\big) \nonumber \\
   \text{mean field}&=& \mu_1 \frac{J_1^4}{(N-1)^2} \sum_{\alpha \beta}\big(\sum_{j}\sum_{r}(\frac{\xi}{2}\sigma_j^\alpha \sigma_r^\alpha-\sigma_j^\alpha  t)(\frac{\xi}{2}\sigma_r^\beta \sigma_j^\beta-\sigma_r^\beta t)-   N(q^{\alpha \beta}t^2 -\frac{\xi}{2}(m^\alpha+m^\beta )t ))\nonumber \\    
&=& \mu_1 \frac{J_1^4}{(N-1)^2} \sum_{\alpha \beta}\big(\sum_{j}\sum_{r}(\frac{\xi^2}{4}\sigma_j^\alpha \sigma_r^\alpha \sigma_r^\beta \sigma_j^\beta-\frac{\xi}{2}(\sigma_j^\alpha \sigma_r^\alpha \sigma_r^\beta+\sigma_r^\beta \sigma_j^\beta \sigma_j^\alpha)t+t^2 \sigma_r^\beta \sigma_j^\alpha  )-   N(q^{\alpha \beta}t^2 -\frac{\xi}{2}(m^\alpha+m^\beta )t ))\nonumber \\  
&\rightarrow & \mu_1 \frac{J_1^4}{(N-1)^2} \sum_{\alpha \beta}\big((\frac{\xi^2}{4}N^2 (q^{\alpha \beta})^2-\frac{\xi N^2}{2}(q^{\alpha \beta} m^\beta+q^{\alpha \beta} m^\alpha)t+N^2 t^2 m^\alpha m^\beta )-   N(q^{\alpha \beta}t^2 -\frac{\xi}{2}(m^\alpha+m^\beta )t ))\nonumber \\  
\end{eqnarray}

\begin{eqnarray}
  F3_E&=&\mu_1  \frac{J_1^4}{(N-1)^2} \sum_{ij}\sum_{t}\big((1-\delta_{tj})\delta_{it}   \pmb{b}_{tj}^2+
 (1-\delta_{it})\delta_{tj}  \pmb{b}_{it}^2\big) \nonumber \\
&=&2 \mu_1  \frac{J_1^4}{(N-1)^2} \sum_{\alpha \beta}\big((\frac{\xi^2}{4}N^2 (q^{\alpha \beta})^2-\frac{\xi N^2}{2}(q^{\alpha \beta} m^\beta+q^{\alpha \beta} m^\alpha)t+N^2 t^2 m^\alpha m^\beta )-   N(q^{\alpha \beta}t^2 -\frac{\xi}{2}(m^\alpha+m^\beta )t )) \nonumber \\ 
\end{eqnarray}

\begin{eqnarray}
    F3_F&=&\sum_{ij}\big(\rho_1^2 \delta_{ij}+(1-\delta_{ij}) \frac{J_1^2}{N-1}\big) \pmb{b}_{ij} \sigma_1^4 \delta_{ij} \pmb{b}_{ii}\pmb{b}_{jj} \nonumber \\
    &\approx&\sigma_1^6\sum_{\alpha \beta \gamma}\sum_i (-t^3 \sigma_i^{\alpha} \sigma_i^{\beta} \sigma_i^{\gamma }+\frac{1}{2}
   \xi  t^2 \sigma_i^{\alpha}\sigma_i^{\beta }+\frac{1}{2}
   \xi  t^2 \sigma_i^{\alpha}\sigma_i^{\gamma }+\frac{1}{2}
   \xi  t^2 \sigma_i^{\beta}\sigma_i^{\gamma }-\frac{1}{4}
   \xi ^2 t \sigma_i^{\alpha }-\frac{1}{4} \xi ^2 t
   \sigma_i^{\beta }-\frac{1}{4} \xi ^2 t
   \sigma_i^{\gamma }+\frac{\xi ^3}{8} )\nonumber \\
   &=&N \sigma_1^6\sum_{\alpha \beta \gamma} (-t^3  \rho^{\alpha \beta \gamma}+\frac{1}{2}
   \xi  t^2  q^{\alpha \beta}+\frac{1}{2}
   \xi  t^2  q^{\alpha \gamma}+\frac{1}{2}
   \xi  t^2  q^{\beta \gamma}-\frac{1}{4}
   \xi ^2 t  m^\alpha-\frac{1}{4} \xi ^2 t
    m^\beta-\frac{1}{4} \xi ^2 t
    m^\gamma )\nonumber \\
\end{eqnarray}

\begin{eqnarray}
    F3_G&=&\sum_{ij} \big(\rho_1^2 \delta_{ij}+(1-\delta_{ij}) \frac{J_1^2}{N-1}\big) \pmb{b}_{ij}  (1-\delta_{ij}) \frac{\rho_1^2J_1^2}{N-1}(\pmb{b}_{ij}\pmb{b}_{jj}+\pmb{b}_{ii}\pmb{b}_{ij})\nonumber \\
    &=&\frac{2\rho_1^2J_1^4}{(N-1)^2} N^2\sum_{\alpha \beta \gamma}((\frac{1}{8} \xi ^3 )(q^{\alpha \beta})^2
   +\frac{1}{2} \xi  t^2
   m^\alpha \rho^{\alpha \beta \gamma}+\frac{1}{2} \xi  t^2 m^\beta 
   \rho^{\alpha \beta \gamma}-\frac{1}{4} \xi
   ^2 t q^{\alpha \beta} \rho^{\alpha \beta \gamma} \nonumber \\
   & &\ \ \ \ \ \ \ \ \ \ \ \ \ \ \ \ \ \ \ \-\frac{1}{4} \xi ^2 t m^\alpha
   q^{\alpha \beta }-\frac{1}{4} \xi ^2 t
   m^{\beta } \rho^{\alpha \beta }-t^3
   \rho^{\alpha \beta \gamma }+\frac{1}{2} \xi
    t^2 q^{\alpha \beta }\nonumber \\
    & &\ \ \ \ \ \ \ \ \ \ \ \ \ \ \ \ \ \ \ \  +\frac{1}{8} \xi ^3 (q^{\alpha \beta })^2
   -t^3 m^\gamma
   q^{\alpha \beta }+\frac{1}{2} \xi  t^2
   q^{\alpha \beta } q^{\alpha \gamma
   }+\frac{1}{2} \xi  t^2 q^{\alpha \beta }
   q^{\beta \gamma } \nonumber \\
   & &\ \ \ \ \ \ \ \ \ \ \ \ \ \ \ \ \ \ \ \-\frac{1}{4} \xi ^2 t
   q^{\alpha \beta } \rho^{\alpha \beta
   \gamma }-\frac{1}{4} \xi ^2 t m^{\alpha }
   q^{\alpha \beta }-\frac{1}{4} \xi ^2 t
   m^{\beta } q^{\alpha \beta
   }+\frac{1}{2} \xi  t^2 q^{\alpha \beta }\nonumber \\
    & &\ \ \ \ \ \ \ \ \ \ \ \ \ \ \ \ \ \ \ \ -\frac{2}{N}  (-t^3  \rho^{\alpha \beta \gamma}+\frac{1}{2}
   \xi  t^2  q^{\alpha \beta}+\frac{1}{2}
   \xi  t^2  q^{\alpha \gamma}+\frac{1}{2}
   \xi  t^2  q^{\beta \gamma}-\frac{1}{4}
   \xi ^2 t  m^\alpha-\frac{1}{4} \xi ^2 t
    m^\beta-\frac{1}{4} \xi ^2 t
    m^\gamma ))   \nonumber \\    
\end{eqnarray}

\begin{eqnarray}
   F3_H&=&  \sum_{ij} \big(\rho_1^2 \delta_{ij}+(1-\delta_{ij}) \frac{J_1^2}{N-1}\big) \pmb{b}_{ij} \sum_t \frac{J_1^4}{(N-1)^2}(1-\delta_{it})(1-\delta_{tj})\pmb{b}_{it}\pmb{b}_{tj} \nonumber \\
        \text{mft}   &=& \frac{J_1^4}{(N-1)^2}\sum_{\alpha \beta \gamma}\Big( (\rho_1^2-\frac{J_1^2}{N-1})\sum_{i}    \Big(\sum_k(\frac{\xi}{2}\sigma_i^\alpha \sigma_k^\alpha- \sigma_i^\alpha t)(\frac{\xi}{2}\sigma_k^\beta \sigma_i^\beta -\sigma_k^\beta t)-(\frac{\xi}{2}- \sigma_i^\alpha t)(\frac{\xi}{2} -\sigma_i^\beta t)\Big)(\frac{\xi}{2}- \sigma_i^\gamma t)\nonumber \\
   &+&\frac{J_1^2}{N-1}\sum_{ij}   \Big(\sum_k (\frac{\xi}{2}\sigma_i^\alpha \sigma_k^\alpha- \sigma_i^\alpha t)(\frac{\xi}{2}\sigma_k^\beta \sigma_j^\beta -\sigma_k^\beta t)-((\frac{\xi}{2}- \sigma_i^\alpha t)+(\frac{\xi}{2}- \sigma_j^\alpha t))(\frac{\xi}{2}\sigma_i^\beta \sigma_j^\beta- \sigma_i^\beta t) \nonumber \\
   & &\ \ \ \ \ \ \ \ \ \ \ \ \ \ + \delta_{ij}(\frac{\xi}{2}\sigma_i^\alpha \sigma_j^\alpha- \sigma_i^\alpha t)(\frac{\xi}{2}\sigma_i^\beta \sigma_j^\beta- \sigma_i^\beta t)\Big) (\frac{\xi}{2}\sigma_i^\gamma \sigma_j^\gamma- \sigma_i^\gamma t)\Big)   \nonumber \\
&=& \frac{J_1^4}{(N-1)^2}\sum_{\alpha \beta \gamma}\Big( (\rho_1^2-\frac{J_1^2}{N-1})   \big(\frac{1}{8} \xi ^3 m^\alpha m^{\beta }
   q^{\alpha \beta }+\frac{1}{2} \xi  t^2
   (m^{\beta })^2 q^{\alpha
   \gamma }-\frac{1}{4} \xi ^2 t m^{\beta }
   q^{\alpha \gamma } q^{\alpha \beta
   }-\frac{1}{4} \xi ^2 t m^{\alpha }
   (m^{\beta })^2-t^3 m^{\beta }
   q^{\alpha \gamma }\nonumber \\
   & & \ \ \ \ \ \ \ \ \  \ \ \ \ \ \ \ \ \ +\frac{1}{2} \xi  t^2
   q^{\alpha \gamma } q^{\alpha \beta
   }+\frac{1}{2} \xi  t^2 m^{\alpha }
   m^{\beta }-\frac{1}{4} \xi ^2 t
   m^{\alpha } q^{\alpha \beta })\nonumber \\
&-&\sum_{i}(-t^3 m^{\beta } q^{\alpha \gamma
   }+\frac{1}{2} \xi  t^2 m^{\alpha }
   m^{\beta }+\frac{1}{2} \xi  t^2
   m^{\gamma } m^{\beta }+\frac{1}{2} \xi 
   t^2 q^{\alpha \gamma }-\frac{1}{4} \xi ^2
   t m^{\alpha }-\frac{1}{4} \xi ^2 t
   m^{\gamma }-\frac{1}{4} \xi ^2 t
   m^{\beta }+\frac{\xi ^3}{8})\big)\nonumber \\
&+&\frac{J_1^2}{N-1}\big(
\sum_{ijk}(\frac{1}{8} \xi ^3 q^{\alpha \gamma }
   q^{\beta \gamma } q^{\alpha \beta
   }+\frac{1}{2} \xi  t^2 m^{\beta }
   m^{\beta } q^{\alpha \gamma
   }+\frac{1}{2} \xi  t^2 m^{\gamma }
   m^{\beta } q^{\alpha \gamma
   }-\frac{1}{4} \xi ^2 t m^{\beta }
   q^{\alpha \gamma } q^{\alpha \beta
   }\nonumber \\
   & & \ \ \ \ \ \ \ \ \  \ \ \ \ \ \ \ \ \-\frac{1}{4} \xi ^2 t m^{\gamma }
   m^{\alpha +\gamma } m^{\alpha +\beta
   }-\frac{1}{4} \xi ^2 t m^{\beta }
   q^{\alpha \gamma } q^{\beta \gamma
   }-t^3 m^{\beta } q^{\alpha \gamma
   }+\frac{1}{2} \xi  t^2 q^{\alpha \gamma }
   q^{\alpha \beta })\nonumber \\
&-&\sum_{i}(-t^3 \rho^{\alpha \beta \gamma }+\frac{1}{2}
   \xi  t^2 q^{\alpha \beta }+\frac{1}{2}
   \xi  t^2 q^{\alpha \gamma }+\frac{1}{2}
   \xi  t^2 q^{\beta \gamma }-\frac{1}{4}
   \xi ^2 t m^{\alpha }-\frac{1}{4} \xi ^2 t
   m^{\beta }-\frac{1}{4} \xi ^2 t
   m^{\gamma }+\frac{\xi ^3}{8})\big)\Big)\nonumber \\
  \nonumber \\
\end{eqnarray}

\begin{eqnarray}
-\frac{2F_{ij}^4}{\gamma N}&=&\Big(\ \ \ 
  \underbrace{4\mu_1 \rho_1^2   \big(\rho_1^2 \delta_{ij}+(1-\delta_{ij}) \frac{J_1^2}{N-1}\big)  (\pmb{b}_{ii}+\pmb{b}_{jj})}_{F4_A} \nonumber \\
   & &\ \ \ \ \ +\underbrace{2\sigma_1^4  \big(\rho_1^2 \delta_{ij}+(1-\delta_{ij}) \frac{J_1^2}{N-1}\big)  \pmb{b}_{ii}\pmb{b}_{jj}}_{F4_B} \nonumber \\
& &\ \ \ \ \ +\underbrace{4(\mu_1 \sigma_1^4+\mu_1^3 \rho_1^2)\delta_{ij}   \pmb{b}_{ij}}_{F4_C}\nonumber \\
& &\ \ \ \ \ +\underbrace{\sum_t \big(\rho_1^2 \delta_{it}+(1-\delta_{it}) \frac{J_1^2}{N-1}\big)  \big(\sigma_1^4 \delta_{tj}+(1-\delta_{tj}) \frac{J_1^4}{(N-1)^2}\big)(\pmb{b}_{it}^2+\pmb{b}_{t j}^2)}_{F4_D} \nonumber \\
& &\ \ \ \ \ + \underbrace{\sum_t (2\mu_1^2+2\rho_1^2) \delta_{ij}\big(\sigma_1^4 \delta_{t j}+(1-\delta_{t j}) \frac{J_1^4}{(N-1)^2}\big)\pmb{b}_{i t}\pmb{b}_{t j}}_{F4_E} \nonumber \\
& &\ \ \ \ \ +\underbrace{4\mu_1^2\big(\sigma_1^4 \delta_{i j}+(1-\delta_{i j}) \frac{J_1^4}{(N-1)^2}\big) \pmb{b}_{i j}^2}_{F4_F}\nonumber \\
& &\ \ \ \ \ +\underbrace{\mu_1\big(\rho_1^2 \delta_{it^\prime}+(1-\delta_{it^\prime}) \frac{J_1^2}{N-1}\big)\big(\rho_1^2 \delta_{t^\prime j}+(1-\delta_{t^\prime j}) \frac{J_1^2}{N-1}\big)\pmb{b}_{i t^\prime}\pmb{b}_{t^\prime j}}_{F4_G} \nonumber \\
& &\ \ \ \ \ +\underbrace{3\mu_1 \big(\rho_1^2 \delta_{ij}+(1-\delta_{ij}) \frac{J_1^2}{N-1}\big)\pmb{b}_{i j}\big(\rho_1^2 \delta_{it}+(1-\delta_{it}) \frac{J_1^2}{N-1}\big)\big(\rho_1^2 \delta_{t j}+(1-\delta_{t j}) \frac{J_1^2}{N-1}\big)\pmb{b}_{i t}\pmb{b}_{t j}}_{F4_H}\nonumber \\
& &\ \ \ \ \  +\big(\rho_1^2 \delta_{it^\prime}+(1-\delta_{it^\prime}) \frac{J_1^2}{N-1}\big)\big(\rho_1^2 \delta_{t^\prime j}+(1-\delta_{t^\prime j}) \frac{J_1^2}{N-1}\big)\pmb{b}_{i t^\prime}\pmb{b}_{t^\prime j} \nonumber \\
& &\ \ \ \ \ \ \ \ \ \ \cdot \underbrace{\big(\rho_1^2 \delta_{tj}+(1-\delta_{tj})\big(\rho_1^2 \delta_{it}+(1-\delta_{it}) \frac{J_1^2}{N-1}\big) \pmb{b}_{i t}\pmb{b}_{t j}\Big)}_{F4_I}
\end{eqnarray}

\begin{eqnarray}
F4_A&=&\sum_{ij}4\mu_1 \rho_1^2   \big(\rho_1^2 \delta_{ij}+(1-\delta_{ij}) \frac{J_1^2}{N-1}\big)  (\pmb{b}_{ii}+\pmb{b}_{jj}) \nonumber \\
\text{mft} &=&4\mu_1 \rho_1^2   \big(2 (\rho_1^2 -\frac{J_1^2}{N-1}) +2\frac{J_1^2 N}{N-1}\big)  \sum_{\alpha} \sum_{i}(\frac{\xi}{2}-\sigma_i^\alpha t) \nonumber \\
&\approx &-8\mu_1 \rho_1^2  t \big( \rho_1^2 -\frac{J_1^2}{N-1} +\frac{J_1^2 N}{N-1}\big) N \sum_{\alpha} m^\alpha  \nonumber \\
\end{eqnarray}

\begin{eqnarray}
F4_B&=&2\sigma_1^4  \sum_{ij} \big(\rho_1^2 \delta_{ij}+(1-\delta_{ij}) \frac{J_1^2}{N-1}\big)  \pmb{b}_{ii}\pmb{b}_{jj}\nonumber \\
&\approx&  \sigma_1^4(\rho_1^2-\frac{J_1^2 }{N-1}) \sum_{\alpha\beta}\sum_{i}(-\frac{\xi}{2}t (\sigma_i^\alpha+\sigma_i^\beta) +\sigma_i^\alpha \sigma_i^\beta t^2)+\frac{J_1^2 \sigma_1^4}{N-1}\sum_{\alpha \beta}\sum_{ij} (-\frac{\xi}{2}t (\sigma_i^\alpha+\sigma_j^\beta) +\sigma_i^\alpha \sigma_j^\beta t^2) \nonumber \\ 
&=&  \sigma_1^4(\rho_1^2-\frac{J_1^2 }{N-1}) N\sum_{\alpha\beta}(-\frac{\xi}{2}t  (m^\alpha+m^\beta) +q^{\alpha \beta} t^2)+\frac{J_1^2 \sigma_1^4}{N-1}N^2\sum_{\alpha \beta}\sum_{ij} (-\frac{\xi}{2}t (m^\alpha+m^\beta) +m^\alpha m^\beta t^2) \nonumber \\ 
\end{eqnarray}
\begin{eqnarray}
F4_C&=&\sum_{ij} 4(\mu_1 \sigma_1^4+\mu_1^3 \rho_1^2)\delta_{ij}   \pmb{b}_{ij} \nonumber \\
&=& 4(\mu_1 \sigma_1^4+\mu_1^3 \rho_1^2)\sum_{i}   \pmb{b}_{ii} \nonumber \\
\text{mft}&\approx & - 4(\mu_1 \sigma_1^4+\mu_1^3 \rho_1^2)Nt\sum_{\alpha}m^\alpha   \nonumber \\
\end{eqnarray}

\begin{eqnarray}
F4_D&=& \sum_{ij} \sum_t \big(\rho_1^2 \delta_{it}+(1-\delta_{it}) \frac{J_1^2}{N-1}\big)  \big(\sigma_1^4 \delta_{tj}+(1-\delta_{tj}) \frac{J_1^4}{(N-1)^2}\big)(\pmb{b}_{it}^2+\pmb{b}_{t j}^2) \nonumber \\
&=& \sum_{ij} \sum_t \Big(\sigma_1^4 \delta_{it}\delta_{tj}+\big((1-\delta_{it}) \delta_{tj}+\delta_{it}(1-\delta_{tj})\big)  \frac{J_1^2\rho_1^2}{N-1}+(1-\delta_{it})(1-\delta_{tj})\frac{J_1^4}{(N-1)^2}\Big)  (\pmb{b}_{it}^2+\pmb{b}_{t j}^2) \nonumber \\
&
\text{mft}
\nonumber \\
&=&  2\Big( \sigma_1^4+ \frac{J_1^2 \rho_1^2}{N-1}(N-2)+ \frac{J_1^4}{(N-1)^2}(1-N) \Big)N \sum_{\alpha \beta}  (\frac{\xi^2}{4}- \frac{\xi}{2}(m^\alpha+m^\beta) t+ q^{\alpha \beta} t^2) \nonumber \\
&+&  2\Big(\frac{J_1^2 \rho_1^2}{N-1}+\frac{J_1^4}{(N-1)^2}(N-1)\Big)N^2\sum_{\alpha \beta} (\frac{\xi^2}{4}(q^{\alpha \beta})^2 \frac{t\xi}{2}(m^\alpha q^{\alpha \beta}+q^{\alpha \beta} m^\beta)+q^{\alpha \beta}t^2)
\nonumber \\
\end{eqnarray}
\begin{eqnarray}
F4_E&=&(2\mu_1^2+2\rho_1^2)\sum_{ij} \sum_t  \delta_{ij}\big(\sigma_1^4 \delta_{t j}+(1-\delta_{t j}) \frac{J_1^4}{(N-1)^2}\big)\pmb{b}_{i t}\pmb{b}_{t j} \nonumber \\
\text{mft}&=&(2\mu_1^2+2\rho_1^2)\sum_{\alpha \beta}   \Big((\sigma_1^4-\frac{J_1^4}{(N-1)^2}) \sum_{i}(\frac{\xi}{2}- \sigma^\alpha_i t)(\frac{\xi}{2}- \sigma^\alpha_i t)\nonumber \\
& &\ \ \ \ \ \ \ \ \ \ \ \ \ \ \ \ \ \ \ \ \ \ \ \ \ \ \ \ \ \  +\frac{J_1^4}{(N-1)^2}\sum_{i,j} (\frac{\xi}{2}\sigma^\alpha_i \sigma^\alpha_j- \sigma^\alpha_i t)(\frac{\xi}{2}\sigma^\beta_j \sigma^\beta_i- \sigma^\beta_j t)\Big) \nonumber \\
&=&(2\mu_1^2+2\rho_1^2)\sum_{\alpha \beta}   \Big((\sigma_1^4-\frac{J_1^4}{(N-1)^2}) N(\frac{\xi^2}{4}- \frac{\xi}{2}(m^\alpha+m^\beta) t+ q^{\alpha \beta} t^2)\nonumber \\
& &\ \ \ \ \ \ \ \ \ \ \ \ \ \ \ \ \ \ \ \ \ \ \ \ \ \ \ \ \ \  +\frac{J_1^4}{(N-1)^2}N^2 (\frac{\xi^2}{4}(q^{\alpha \beta})^2- \frac{\xi t}{2}(m^\alpha q^{\alpha \beta}+q^{\alpha \beta} m^\beta )+t^2 m^\alpha m^\beta \Big) \nonumber \\
\end{eqnarray}

\begin{eqnarray}
    F4_F&=&4\mu_1^2\sum_{ij}\big(\sigma_1^4 \delta_{i j}+(1-\delta_{i j}) \frac{J_1^4}{(N-1)^2}\big) \pmb{b}_{i j}^2 \nonumber \\
    &=&4\mu_1^2\Big((\sigma_1^4-\frac{J_1^4}{(N-1)^2})\sum_{i} \pmb{b}_{i i}^2 +\frac{J_1^4}{(N-1)^2}\sum_{ij}\pmb{b}_{i j}^2 \Big)  \nonumber \\
   \text{mft}  &=&4\mu_1^2\sum_{\alpha \beta}\Big((\sigma_1^4-\frac{J_1^4}{(N-1)^2}) N(\frac{\xi^2}{4}- \frac{\xi}{2}(m^\alpha+m^\beta) t+ q^{\alpha \beta} t^2) \nonumber \\
    & &\ \ \ \ \ \ \ \ \ \ \ \ \ \ \ \ \ \ \ \ \ \ \ \ \ \ \ \ \ \  +\frac{J_1^4}{(N-1)^2}\sum_{ij}(\frac{\xi}{2} \sigma_i^\alpha \sigma_j^\alpha -\sigma_i^\alpha t)(\frac{\xi}{2} \sigma_i^\beta \sigma_j^\beta -\sigma_i^\beta t) \Big)  \nonumber \\ 
&=&4\mu_1^2\sum_{\alpha \beta}\Big((\sigma_1^4-\frac{J_1^4}{(N-1)^2}) N(\frac{\xi^2}{4}- \frac{\xi}{2}(m^\alpha+m^\beta) t+ q^{\alpha \beta} t^2) \nonumber \\
    & &\ \ \ \ \ \ \ \ \ \ \ \ \ \ \ \ \ \ \ \ \ \ \ \ \ \ \ \ \ \  +\frac{J_1^4 N^2}{(N-1)^2}(\frac{\xi^2}{4} (q^{\alpha \beta})^2 - \frac{\xi t}{2}(q^{\alpha \beta}m^\beta+m^\alpha q^{\alpha \beta}) -q^{\alpha \beta} t^2) \Big)  \nonumber \\     
\end{eqnarray}

\begin{eqnarray}
    F4_G&=&\mu_1\sum_{ij}\sum_{t^\prime}\big(\rho_1^2 \delta_{it^\prime}+(1-\delta_{it^\prime}) \frac{J_1^2}{N-1}\big)\big(\rho_1^2 \delta_{t^\prime j}+(1-\delta_{t^\prime j}) \frac{J_1^2}{N-1}\big)\pmb{b}_{i t^\prime}\pmb{b}_{t^\prime j} \nonumber \\
 &=& 
\mu_1\big((\rho_1^2-\frac{J_1^2}{N-1})^2 \sum_{t^\prime}\pmb{b}_{t^\prime t^\prime}^2+(\frac{J_1^2\rho_1^2}{N-1}-\frac{J_{1}^4}{(N-1)^2})\sum_{ij} \pmb{b}_{i j}(\pmb{b}_{i i}+\pmb{b}_{j j})+
 \frac{J_1^4}{(N-1)^2} \sum_{it^\prime j}\pmb{b}_{i t^\prime}\pmb{b}_{t^\prime j}\big)     \nonumber \\ 
 \text{mft} &=& 
\mu_1\sum_{\alpha \beta}\Big((\rho_1^2-\frac{J_1^2}{N-1})^2 N(\frac{\xi^2}{4}- \frac{\xi}{2}(m^\alpha+m^\beta) t+ q^{\alpha \beta} t^2)\nonumber \\
&+&N^2(\frac{J_1^2\rho_1^2}{N-1}-\frac{J_{1}^4}{(N-1)^2})(\frac{1}{2} \xi ^2 (m^\alpha)^2+t^2 m^{\alpha } m^{\beta }-\frac{1}{2}
   \xi  t m^{\alpha } q^{\alpha \beta
   }-\frac{1}{2} \xi  t m^{\alpha }
   q^{\alpha \beta }+t^2 q^{\alpha +\beta
   }-\xi  t  m^{\alpha })\nonumber \\
&+& \frac{J_1^4}{(N-1)^2}N^3 (\frac{1}{4} \xi ^2 m^{\alpha } m^{\beta }
   q^{\alpha \beta }-\frac{1}{2} \xi  t
   m^{\alpha } m^{\beta } m^{\beta }+t^2
   m^{\alpha } m^{\beta }-\frac{1}{2} \xi 
   t m^{\alpha } q^{\alpha \beta })\Big)     \nonumber \\ 
\end{eqnarray}

\begin{eqnarray}
    F4_H&=&3\mu_1 \sum_{ijt} \big(\rho_1^2 \delta_{ij}+(1-\delta_{ij}) \frac{J_1^2}{N-1}\big)\pmb{b}_{i j}\big(\rho_1^2 \delta_{it}+(1-\delta_{it}) \frac{J_1^2}{N-1}\big)\big(\rho_1^2 \delta_{t j}+(1-\delta_{t j}) \frac{J_1^2}{N-1}\big)\pmb{b}_{i t}\pmb{b}_{t j} \nonumber \\
    &=&   3\mu_1(\rho_1^2- \frac{J_1^2}{N-1})\Big((\rho_1^2- \frac{J_1^2}{N-1})^2 +2(\frac{J_1^2 \rho_1^2}{N-1}-\frac{J_1^4}{(N-1)
    ^2})+ \frac{J_1^2}{N-1}(\rho_1^2- \frac{J_1^2}{N-1})\Big)\nonumber \\
    & &\ \ \ \ \cdot \sum_{\alpha \beta \gamma} N (-t^3 \rho^{\alpha \beta \gamma}+\frac{1}{2}
   \xi  t^2 q^{\alpha \beta }+\frac{1}{2}
   \xi  t^2 q^{\alpha \gamma }+\frac{1}{2}
   \xi  t^2 q^{\beta \gamma }-\frac{1}{4}
   \xi ^2 t m^{\alpha }-\frac{1}{4} \xi ^2 t
   m^{\beta }-\frac{1}{4} \xi ^2 t
   m^{\gamma }+\frac{\xi ^3}{8}) \nonumber \\
    & & \ + \frac{3\mu_1(\rho_1^2-\frac{J_1^2}{N-1})J_1^4}{(N-1)^2}\sum_{\alpha \beta \gamma}N^2 \Big( \frac{1}{8} \xi ^3 q^{\beta \gamma }
   q^{\beta \gamma }-t^3 m^{\gamma }
   q^{\alpha \beta } +\frac{1}{2} \xi  t^2
   q^{\alpha \beta } q^{\beta \gamma
   }+\frac{1}{2} \xi  t^2 m^{\gamma }
   \rho^{\alpha \beta \gamma }\nonumber \\
   & &\ \ \ \ \ \ +\frac{1}{2} \xi
    t^2 m^{\beta } m^{\gamma }-\frac{1}{4}
   \xi ^2 t q^{\beta \gamma } \rho^{\alpha
   \beta \gamma }-\frac{1}{4} \xi ^2 t
   m^{\beta } \rho^{\beta \gamma
   }-\frac{1}{4} \xi ^2 t m^{\gamma }
   q^{\beta \gamma }\Big)  \nonumber \\  
    & & \  + 3\mu_1 \frac{J_1^2}{N-1}(\frac{J_1^2 \rho_1^2}{N-1}-\frac{J_1^4}{(N-1)^2}
    ) \sum_{\alpha \beta \gamma}N^2(\frac{1}{4} \xi ^3 (q^{\alpha +\beta })^2 
   -t^3 m^{\gamma }
   q^{\alpha \beta }+\frac{1}{2} \xi  t^2
   m^{\alpha } \rho^{\alpha \beta \gamma
   }+\frac{1}{2} \xi  t^2 m^{\beta }
   \rho^{\alpha \beta \gamma }\nonumber \\
   & &\ \ \ \ \ \ \ \ +\frac{1}{2} \xi
    t^2 q^{\alpha \beta } q^{\alpha
   \gamma }+\frac{1}{2} \xi  t^2 q^{\alpha
   \beta } q^{\beta +\gamma }-\frac{1}{4}
   \xi ^2 t q^{\alpha \beta } \rho^{\alpha
   \beta \gamma }-\frac{1}{4} \xi ^2 t
   q^{\alpha \beta } \rho^{\alpha \beta
   \gamma }\nonumber \\
   & &\ \ \ \ \ \ \ \-\frac{1}{2} \xi ^2 t m^{\alpha }
   q^{\alpha \beta }-\frac{1}{2} \xi ^2 t
   m^{\beta } q^{\alpha \beta }-t^3
   \rho^{\alpha \beta \gamma }+\xi  t^2
   q^{\alpha \beta })\nonumber \\
    & & \ + 3\mu_1 \frac{J_1^6}{(N-1)^3}N^3 \sum_{\alpha \beta \gamma}  q^{\alpha \beta} \Big( \frac{1}{8} \xi ^3  q^{ \beta\gamma }  q^{ \gamma\alpha }    +  q^{\beta \gamma} (\frac{\xi }{2}   t^2-\frac{3t}{4} m^\gamma)+\frac{t^2}{2} \xi  
   m^\alpha m^\gamma  +\frac{1}{2} \xi t^2   (m^\gamma)^2 -t^3 
   m^\gamma \Big) \nonumber \\     
\end{eqnarray}

\begin{eqnarray}
F4_I&=&\sum_{ij} \big(\rho_1^2 \delta_{it^\prime}+(1-\delta_{it^\prime}) \frac{J_1^2}{N-1}\big)\big(\rho_1^2 \delta_{t^\prime j}+(1-\delta_{t^\prime j}) \frac{J_1^2}{N-1}\big)\pmb{b}_{i t^\prime}\pmb{b}_{t^\prime j} \nonumber \\
& &\ \ \ \ \ \ \ \ \ \ \cdot \big(\rho_1^2 \delta_{tj}+(1-\delta_{tj})\big(\rho_1^2 \delta_{it}+(1-\delta_{it}) \frac{J_1^2}{N-1}\big) \pmb{b}_{i t}\pmb{b}_{t j}\nonumber \\
&=& \sum_{ij}\Big(\sum_t  
\big((\frac{J_1^4}{(N-1)^2}+\sigma_1^4) \delta_{it}\delta_{tj}+(\frac{J_1^2\rho_1^2}{N-1}-\frac{J_1^4}{(N-1)^2})((1-\delta_{it})\delta_{tj}+\delta_{it}(1-\delta_{tj}))+ \frac{J_1^4}{(N-1)^2}\big) \pmb{b}_{i t}\pmb{b}_{t j}
\Big)^2 \nonumber \\
&=& \sum_{ij}\Big(\sum_t  
\big((\frac{J_1^4}{(N-1)^2}+\sigma_1^4) \delta_{it}\delta_{tj}+(\frac{J_1^2\rho_1^2}{N-1}-\frac{J_1^4}{(N-1)^2})(\delta_{tj}+\delta_{it}-2\delta_{it}\delta_{tj})+ \frac{J_1^4}{(N-1)^2}\big) \pmb{b}_{i t}\pmb{b}_{t j}
\Big)^2 \nonumber \\
&=& \sum_{ij}\Big(  
(\frac{J_1^2}{N-1}-\rho_1^2)^2 \delta_{ij}\pmb{b}_{i j}^2 +(\frac{J_1^2\rho_1^2}{N-1}-\frac{J_1^4}{(N-1)^2})\pmb{b}_{i j}(\pmb{b}_{j j}+\pmb{b}_{ii}) + \frac{J_1^4}{(N-1)^2}\sum_t \pmb{b}_{i t}\pmb{b}_{t j} \Big)^2 \nonumber \\
&=& \Big(  
(\frac{J_1^2}{N-1}-\rho_1^2)^4 \sum_{ij}\delta_{ij}\pmb{b}_{i j}^4 +(\frac{J_1^2\rho_1^2}{N-1}-\frac{J_1^4}{(N-1)^2})^2\sum_{ij}\pmb{b}_{i j}^2(\pmb{b}_{j j}+\pmb{b}_{ii})^2 \nonumber \\
&+& \frac{J_1^8}{(N-1)^4}\sum_{ij}(\sum_t \pmb{b}_{i t}\pmb{b}_{t j})^2 +
2(\frac{J_1^2}{N-1}-\rho_1^2)^2\frac{J_1^4}{(N-1)^2} \sum_{ij}\delta_{ij}\pmb{b}_{i j}^2\sum_t \pmb{b}_{i t}\pmb{b}_{t j} \nonumber \\
&+&2(\frac{J_1^2\rho_1^2}{N-1}-\frac{J_1^4}{(N-1)^2})\frac{J_1^4}{(N-1)^2}\sum_{ij}\pmb{b}_{i j}(\pmb{b}_{j j}+\pmb{b}_{ii}) \sum_t \pmb{b}_{i t}\pmb{b}_{t j}\nonumber \\
&+&   
2(\frac{J_1^2}{N-1}-\rho_1^2)^2(\frac{J_1^2\rho_1^2}{N-1}-\frac{J_1^4}{(N-1)^2})\sum_{ij} \delta_{ij}\pmb{b}_{i j}^3(\pmb{b}_{j j}+\pmb{b}_{ii}) \Big) \nonumber \\
\end{eqnarray}
in these last terms it can be seen that if $\rho_1^2\neq J_1^2/N$, for $N\rightarrow \infty$ we are forced to include also higher order overlap distributions.

\end{document}